\documentclass[prd,twocolumn,nofootinbib,10pt,longbibliography,superscriptaddress]{revtex4-2}
\usepackage{textcomp}
\usepackage{alltt}
\usepackage{amsmath}
\usepackage{amsfonts}
\usepackage[bbgreekl]{mathbbol}
\usepackage{calc}
\usepackage{mathrsfs}
\usepackage{amssymb}
\usepackage{amsthm}
\usepackage{array}
\usepackage[dvipsnames]{xcolor}
\usepackage{bbm}
\usepackage{graphicx}
\usepackage[pdftex,breaklinks,colorlinks,
linkcolor=Blue,
citecolor=teal,
anchorcolor=red,
urlcolor=cyan]{hyperref}
\usepackage{mathtools}
\usepackage[T1]{fontenc}
\usepackage{booktabs}
\usepackage{bm}

\renewcommand{\P}{\mathcal{P}}
\newcommand{\res}{\mathcal{R}}

\newcommand{\e}{\epsilon}

\renewcommand{\>}{\rangle}
\newcommand{\varLie}{\mathsterling}

\newcommand{\Lie}{\mathcal{L}}
\newcommand{\beq}{\begin{equation}}
\newcommand{\eeq}{\end{equation}}
\newcommand{\sub}[1]{{}_{{\kern-.165em} #1}}
\newcommand{\dR}{\delta{\kern-.08889em}R}
\newcommand{\ddR}{\delta^2{\kern-.08889em}R}
\newcommand{\dnR}{\delta^n{\kern-.08889em}R}

\newcommand{\pkg}[1]{\textsc{#1}}
\newcommand{\thorn}{{\text{\TH}}}
\newcommand{\edth}{\eth}
\newcommand{\D}[1]{D_{{\kern-.07em} #1}}
\newcommand{\del}[1]{\nabla_{{\kern-.165em} #1}}

\newcommand{\lam}[2]{\lambda_{#1,#2}}

\newcommand{\trab}{\bullet}
\newcommand{\trAB}{\circ}

\renewcommand{\O}{\mathcal{O}}

\newtheorem*{definition*}{Definition}

\begin{document}
\title{Second-order perturbations of the Schwarzschild spacetime: practical, covariant and gauge-invariant formalisms} 
\author{Andrew Spiers}
\affiliation{Nottingham Centre of Gravity, Nottingham NG7 2RD, United Kingdom}
\affiliation{School of Mathematical Sciences, University of Nottingham, University Park, Nottingham NG7 2RD, United Kingdom}
\affiliation{School of Mathematical Sciences and STAG Research Centre, University of Southampton, Southampton, SO17 1BJ, United Kingdom}
\author{Adam Pound}
\affiliation{School of Mathematical Sciences and STAG Research Centre, University of Southampton, Southampton, SO17 1BJ, United Kingdom}
\author{Barry Wardell}
\affiliation{School of Mathematics \& Statistics, University College Dublin, Belfield, Dublin 4, Ireland}
\date{\today}

\begin{abstract}
High-accuracy gravitational-wave modeling demands going beyond linear, first-order perturbation theory. Particularly motivated by the need for second-order perturbative models of extreme-mass-ratio inspirals and black hole ringdowns, we present practical spherical-harmonic decompositions of the Einstein equation, Regge-Wheeler-Zerilli equations, and Teukolsky equation at second perturbative order in a Schwarzschild background. Our formulations are covariant on the $t$--$r$ plane and on the two-sphere, and we express the field equations in terms of gauge-invariant metric perturbations. In a companion \pkg{Mathematica} package, \pkg{PerturbationEquations}, we provide these invariant formulas as well as the analogous formulas in terms of raw, gauge-dependent metric perturbations. Our decomposition of the second-order Einstein equation, when specialized to the Lorenz gauge, was a key ingredient in recent second-order self-force calculations~[Phys. Rev. Lett. 124, 021101 (2020); ibid. 127, 151102 (2021); ibid. 130, 241402 (2023)].
\end{abstract}

\maketitle

\tableofcontents

\section{Introduction}

The first phase of gravitational-wave astronomy~\cite{LIGOScientific:2018mvr, LIGOScientific:2020ibl, LIGOScientific:2021djp} has been a success of experimental physics, data analysis, and the theory of general relativity in its strong-field regime. It also represents a success of perturbation theory. Waveform templates commonly incorporate information from perturbative approximations, particularly from post-Newtonian theory~\cite{Blanchet:2013haa}, which describes the early stages of an inspiral. Many of them additionally incorporate information from black hole perturbation theory in a variety of ways. For example, Effective One Body models are designed to capture the point particle limit~\cite{Buonanno:1998gg}, in which the  motion of a binary reduces to geodesic motion in a black hole spacetime; they can be informed by the associated perturbative fluxes~\cite{Nagar:2006xv,Nagar:2011aa,Taracchini:2014zpa,Albertini:2022dmc,vandeMeent:2023ols} and by perturbative self-force corrections to black hole geodesics~\cite{Damour:2009sm,Antonelli:2019fmq,Nagar:2022fep}; and they often use black hole perturbation theory to describe the final, ringdown phase after merger~\cite{Buonanno:2000ef}.

As detectors are upgraded and new detectors come online, perturbative models must be further improved. In the context of black hole perturbation theory, the overwhelming majority of calculations have been restricted to first, linear perturbative order. There are now at least two prime examples for which gravitational-wave astronomy requires going to second perturbative order, where nonlinear effects first appear. The first example is the ringdown phase of a binary~\cite{Cheung:2022rbm,Mitman:2022qdl,Lagos:2022otp,Green:2022htq}. The second is binaries with small mass ratios, such as extreme-mass-ratio inspirals (EMRIs) in which a stellar-mass compact object orbits a massive black hole~\cite{Hinderer-Flanagan:08,Barack:2018yvs,Pound:2021qin}. 

Ringdowns have, historically, been well modelled as sums of quasinormal modes~\cite{Berti:2009kk}. However, recent work has shown that quadratic coupling between modes can be significant and leave observable signatures in waveforms~\cite{Cheung:2022rbm,Mitman:2022qdl}. Any model using a perturbative treatment of ringdown will likely have to include such nonlinear effects to meet future accuracy requirements. That is especially true for models of massive black hole binaries, which will be observable (with SNRs $\sim 10^3$) by the space-based detector LISA~\cite{Amaro-Seoane:2022rxf}. 

EMRIs, which are also expected to be key sources for LISA~\cite{Amaro-Seoane:2022rxf}, are best modeled by self-force theory, which treats the smaller object as a source of perturbations of the background spacetime of the large black hole~\cite{Barack:2018yvs,Pound:2021qin}. It has been widely accepted for decades that accurately modeling EMRIs necessitates carrying self-force theory to second order~\cite{Rosenthal:06b,Hinderer-Flanagan:08,Detweiler:12,Pound:12a,Gralla:12}. More recently, it has been predicted that second-order self-force calculations can also provide accurate waveforms in the intermediate-mass-ratio regime and even for the mass ratios $\sim$1:10 observable by present-day detectors~\cite{vandeMeent:2020xgc,Ramos-Buades:2022lgf}. That prediction was validated when second-order waveforms were first obtained in 2021~\cite{Wardell:2021fyy,Albertini:2022rfe}. These waveforms were specialized to quasicircular, nonspinning binaries, but their high accuracy across a broad range of mass ratios (and their capacity for rapid waveform generation) provides additional motivation for extending such calculations to more generic binary configurations. 

Unfortunately, although the general formulation of nonlinear perturbation theory on arbitrary backgrounds is well established~\cite{Tomita:74,Tomita-Tajima:76,Bruni-etal:96,Sopuerta-etal:03}, and general formulas, such as the $n$th-order expansion of the Einstein equation, are easily derived at any finite order~\cite{Brizuela-etal:08}, there has been limited development of practical, ready-at-hand tools in black hole spacetimes. Concrete calculations (e.g.,~\cite{Poisson-Vlasov:09,Gleiser-etal:96a,Gleiser-etal:96b,Gleiser-etal:00,Nicasio-etal:00,Ioka-Nakano:07,Nakano-Ioka:07,Lousto-Nakano:08,Yang-etal:15}) have generally been limited to vacuum perturbations and to a small number of harmonic modes. In the simple case of a Schwarzschild background, the most thorough treatment was provided by Brizuela {\em et al.}~\cite{Brizuela-etal:06,Brizuela:2007zza,Brizuela-etal:09,Brizuela-etal:10}, who extended the Regge--Wheeler--Zerilli (RWZ) formalism to second order. This formalism, while useful in many contexts, is limited in that it inherits the sometimes pathological behaviour of the RWZ gauge~\cite{Barack-Ori:01,Hopper-Evans:13,Thompson:2018lgb} and does not provide the entire metric perturbation, missing the $\ell=0$ and $1$ modes that describe the spacetime's mass and momentum. 

There is therefore call for a broader suite of tools. This is especially true for EMRIs, which bring particular complexities. In self-force calculations, the  $\ell=0$ and $1$ modes of the perturbation cannot be ignored; some ingredients are only available in practical form in the Lorenz gauge~\cite{Pound-Miller:14}; and calculations often demand a large number of modes of the first-order perturbation~\cite{vandeMeent:17b} (often above $\ell=50$). 

Our goal with this paper is to provide a comprehensive, practical treatment of second-order calculations on a Schwarzschild background. Our treatment is deliberately modeled on Martel and Poisson's (hereafter MP's) now-standard summary of first-order perturbation theory in Schwarzschild~\cite{Martel-Poisson:05}. Like MP, we include significant review material to make our paper a standalone reference. However, our treatment is more expansive than MP's, covering a wider variety of formulations to make our results useful to the broadest userbase. We also find this larger toolset provides alternative methods that are sometimes more useful than MP's at second order. In all cases, our goal is to decompose the field equations into a set of tensor or spin-weighted spherical harmonics. Our core output is a set of coupling formulas that express harmonic modes of the second-order source as sums of products of first-order field modes. We present these formulas in two forms: in terms of the  gauge-dependent first- and second-order metric perturbations and in terms of gauge-invariant perturbations.

We begin in Sec.~\ref{covariant 2nd order} by reviewing second-order perturbation theory in a generic vacuum background. In Secs.~\ref{2+2 decomposition} and \ref{harmonics} we specialize to a Schwarzschild background and assemble the ingredients required for the harmonic decomposition of the second-order field equations, following MP's description (with some modifications and several extensions)  of the decomposition of four-dimensional covariant quantities into quantities that are separately covariant on the $t$--$r$ plane and on the two-sphere. Section~\ref{gauge} discusses gauge freedom at the level of harmonic modes and the construction of invariant variables. In Sec.~\ref{Decomposed EFE}, we present the decomposition of the second-order Einstein equation. Section~\ref{master equations} then presents the decompositions of the second-order RWZ and Teukolsky equations. We conclude in Sec.~\ref{Conclusion} with a discussion of applications, specifically how our decomposition of the second-order Einstein equation underpinned the recent second-order self-force calculations in Refs.~\cite{Pound:2019lzj,Warburton:2021kwk,Bonetto:2021exn,Wardell:2021fyy}. Table~\ref{table:conventions} and Appendix~\ref{BLS conventions} describe how to translate our formulas into alternative choices of harmonic basis and field variables. 

Alongside our paper, we provide the fully decomposed equations in a companion \pkg{Mathematica} package, \pkg{PerturbationEquations}, which we make available as part of the Black Hole Perturbation Toolkit~\cite{BHPToolkit}. The package provides utilities to work with the second-order Einstein equations, RWZ equations, and Teukolsky equations in a variety of popular harmonic bases and conventions.

\section{Second-order perturbation theory}\label{covariant 2nd order}

Before introducing any decompositions, we begin with the first- and second-order Einstein equations in their covariant, four-dimensional form on an arbitrary vacuum background spacetime, which we will specialize to Schwarzschild in later sections. We keep these formulas generic, but we write them in a form that naturally simplifies in the Lorenz gauge (the gauge used in all second-order self-force calculations to date). At the end of the section we summarize (i) the Bianchi identities that constrain the equations and (ii) the gauge freedom the equations admit.

\subsection{Einstein equations}

We write the exact spacetime metric as ${\sf g}_{\mu\nu}=g_{\mu\nu}+h_{\mu\nu}$, where $g_{\mu\nu}$ is the background metric and $h_{\mu\nu}\sim\e\ll1$ is a small correction, with an associated small stress-energy tensor $T_{\mu\nu}\sim\e$. We will ultimately expand $h_{\mu\nu}$ and $T_{\mu\nu}$ in powers of $\e$, meaning
\begin{align}
h_{\mu\nu} &= \e h^{(1)}_{\mu\nu}+\e^2 h^{(2)}_{\mu\nu}+\O(\e^3),\label{h series}\\
T_{\mu\nu} &= \e T^{(1)}_{\mu\nu}+\e^2 T^{(2)}_{\mu\nu}+\O(\e^3), \label{T series}
\end{align}
and write field equations for $h^{(n)}_{\mu\nu}$. But to organize those equations, we first expand curvature quantities in orders of nonlinearity in $h_{\mu\nu}$.

Explicit perturbative expressions are typically simplest when using the Einstein equation in its trace-reversed form,%
\beq\label{EFE exact}
R_{\mu\nu}[{\sf g}] = 8\pi\left[T_{\mu\nu}-\frac{1}{2}{\sf g}_{\mu\nu}({\sf g}^{-1})^{\alpha\beta}T_{\alpha\beta}\right],
\eeq
where $({\sf g}^{-1})^{\alpha\beta}$ is the inverse of ${\sf g}_{\alpha\beta}$, and where we omit indices on tensorial arguments of functionals. We write the Ricci tensor's expansion in orders of nonlinearity as 
\beq\label{Ricci expansion}
R_{\mu\nu}[g+h] = R_{\mu\nu}[g]+\dR_{\mu\nu}[h] + \ddR_{\mu\nu}[h]+\O(|h_{\mu\nu}|^3), 
\eeq
where $\dnR_{\mu\nu}$ is the (normalized) $n$th functional derivative of $R_{\mu\nu}$, defined by 
\beq
\dnR_{\mu\nu}[\varphi]:=\frac{1}{\lambda!}\frac{d^n}{d\lambda^n}R_{\mu\nu}[g+\lambda\varphi]\bigr|_{\lambda=0}
\eeq
for any rank-2 symmetric tensor $\varphi_{\mu\nu}$. With this definition, $\dnR_{\mu\nu}[h]$ is constructed from the background metric and $n$ copies of $h_{\mu\nu}$. We use the same notation for any quantity constructed from the metric; as a trivial example, $\delta g_{\mu\nu}[h]=h_{\mu\nu}$ and $\delta^n g_{\mu\nu}[h]=0$ for $n>1$.

Concrete formulas for $\dnR_{\mu\nu}$ are found straightforwardly from the spacetime's exact Ricci tensor~\cite{Wald},
\beq\label{exact Ricci}
R_{\mu\nu}[g+h] = R_{\mu\nu}[g] + 2C^\rho{}_{\mu[\nu;\rho]}+2C^\rho{}_{\sigma[\rho}C^\sigma{}_{\nu]\mu},
\eeq
where $C^\alpha{}_{\beta\gamma}$ is the exact difference between the Christoffel symbols of ${\sf g}_{\mu\nu}$ and $g_{\mu\nu}$. Explicitly,
\beq
C^\alpha{}_{\beta\gamma} = \frac{1}{2}({\sf g}^{-1})^{\alpha\delta}\left(2h_{\delta(\beta;\gamma)}-h_{\beta\gamma;\delta}\right);
\eeq
a semicolon and $\del{}$ both denote the covariant derivative compatible with $g_{\alpha\beta}$. The expansion in orders of nonlinearity then immediately follows from the expansion
\beq\label{ginv}
({\sf g}^{-1})^{\alpha\beta} = g^{\alpha\beta} - h^{\alpha\beta}+h^\alpha{}_\gamma h^{\gamma\beta} + \O(|h_{\mu\nu}|^3).
\eeq
Here and throughout this paper, Greek indices are lowered and raised with $g_{\mu\nu}$ and its inverse $g^{\mu\nu}$. 

From Eq.~\eqref{exact Ricci}, \eqref{ginv}, and some simple manipulations (using $R_{\mu\nu}[g]=0$), one finds
\begin{align}
\dR_{\mu\nu}[h] &= -\frac{1}{2}({\cal E}_{\mu\nu}[h] +{\cal F}_{\mu\nu}[h]),\label{dR}\\ 
\ddR_{\mu\nu}[h] &= \frac{1}{2}({\cal A}_{\mu\nu}[h]+{\cal B}_{\mu\nu}[h]+{\cal C}_{\mu\nu}[h]),\label{ddR}
\end{align}
where%
\begin{align}%
{\cal E}_{\mu\nu}[h] &:=\Box h_{\mu\nu}+2R_\mu{}^\alpha{}_\nu{}^\beta h_{\alpha\beta},\label{E}\\
{\cal F}_{\mu\nu}[h] &:=-2\bar h_{\alpha(\mu}{}^{;\alpha}{}_{\nu)},\label{F}
\end{align}
with $\Box:=g^{\mu\nu}\del{\mu}\del{\nu}$, and
\begin{align}
{\cal A}_{\alpha\beta}[h] &:= \tfrac{1}{2}h^{\mu\nu}{}_{;\alpha}h_{\mu\nu;\beta} + h^{\mu}{}_{\beta}{}^{;\nu}\left(h_{\mu\alpha;\nu} - h_{\nu\alpha;\mu}\right),\label{A}\\
{\cal B}_{\alpha\beta}[h] &:= - h^{\mu\nu}\left(2h_{\mu(\alpha;\beta)\nu} - h_{\alpha\beta;\mu\nu} - h_{\mu\nu;\alpha\beta}\right),\label{B}\\
{\cal C}_{\alpha\beta}[h] &:= - \bar h^{\mu\nu}{}_{;\nu}\left(2h_{\mu(\alpha;\beta)}-h_{\alpha\beta;\mu}\right).\label{C}
\end{align}
We use an overbar to denote trace reversal with the background metric, as in $\bar h_{\mu\nu}:=h_{\mu\nu}-\frac{1}{2}g_{\mu\nu}g^{\alpha\beta}h_{\alpha\beta}$. 

If we now substitute Eqs.~\eqref{Ricci expansion} and \eqref{ginv} into the exact Einstein equation~\eqref{EFE exact}, along with the series expansions~\eqref{h series} and \eqref{T series}, then we can equate coefficients of powers of $\e$. The result is the sequence of linear equations
\begin{align}
\dR_{\mu\nu}[h^{(1)}] &= 8\pi {\cal T}^{(1)}_{\mu\nu},\label{EFE1}\\
\dR_{\mu\nu}[h^{(2)}] &= 8\pi {\cal T}^{(2)}_{\mu\nu}- \ddR_{\mu\nu}[h^{(1)}],\label{EFE2}
\end{align}
with matter source terms
\begin{align}
{\cal T}^{(1)}_{\mu\nu} &= \overline{T}^{(1)}_{\mu\nu},\\
{\cal T}^{(2)}_{\mu\nu} &= \overline{T}^{(2)}_{\mu\nu}-\frac{1}{2}\left(h^{(1)}_{\mu\nu}g^{\alpha\beta}-g_{\mu\nu}h^{(1)\alpha\beta}\right)T^{(1)}_{\alpha\beta}.
\end{align}

In the Lorenz gauge, where $\bar h^{\mu\nu}{}_{;\nu}=0$, the quantities ${\cal C}_{\mu\nu}$ and ${\cal F}_{\mu\nu}$ both vanish, simplifying the field equations to 
\begin{align}
{\cal E}_{\mu\nu}[h^{(1)}] &= -16\pi{\cal T}^{(1)}_{\mu\nu},\label{EFE1 Lorenz}\\
{\cal E}_{\mu\nu}[h^{(2)}] &= -16\pi{\cal T}^{(2)}_{\mu\nu}+{\cal A}_{\mu\nu}[h^{(1)}] + {\cal B}_{\mu\nu}[h^{(1)}].\label{EFE2 Lorenz}
\end{align}

Alternatively, we can write the field equations in terms of the perturbed Einstein tensor. The analogues of Eqs.~\eqref{EFE1} and \eqref{EFE2} are
\begin{align}
\delta G_{\mu\nu}[h^{(1)}] &= 8\pi T^{(1)}_{\mu\nu},\label{EFE1 dG}\\
\delta G_{\mu\nu}[h^{(2)}] &= 8\pi T^{(2)}_{\mu\nu}- \delta^2 G_{\mu\nu}[h^{(1)}].\label{EFE2 dG}
\end{align}
The perturbations of $G_{\mu\nu}$ are immediately obtained from those of $R_{\mu\nu}$ using $\delta^n G_{\mu\nu} = \delta^n(R_{\mu\nu} - \frac{1}{2} g_{\mu\nu}g^{\alpha\beta}R_{\alpha\beta})$. In a Ricci-flat background, this simplifies to 
\begin{align}
\delta G_{\mu\nu} &= \overline{\dR}_{\mu\nu},\label{dG=dRbar}\\
\delta^2 G_{\mu\nu} &= \overline{\ddR}_{\mu\nu} - \frac{1}{2}\left(h_{\mu\nu}g^{\alpha\beta}-g_{\mu\nu}h^{\alpha\beta}\right)\dR_{\alpha\beta}\label{ddG=ddRbar}.
\end{align}
In vacuum regions, where $\dR_{\mu\nu}[h^{(1)}]=0=T^{(1)}_{\mu\nu}$, ${\cal T}^{(2)}_{\mu\nu}$ reduces to $\overline{T}^{(2)}_{\mu\nu}$ and $\delta^2 G_{\mu\nu}[h^{(1)}]$ reduces to $\overline{\ddR}_{\mu\nu}[h^{(1)}]$. 

We write the field equations~\eqref{EFE1} and \eqref{EFE2} in generic form as
\beq\label{EFEn dR}
\dR_{\mu\nu}[h^{(n)}] = \bar S^{(n)}_{\mu\nu}.
\eeq
The relations~\eqref{dG=dRbar} and \eqref{ddG=ddRbar} ensure that field equations in the form~\eqref{EFE1} and \eqref{EFE2} can be written in terms of a trace reversal with respect to the background metric,
\beq\label{EFEn dG}
\delta G_{\mu\nu}[h^{(n)}] = S^{(n)}_{\mu\nu},
\eeq
where $S^{(n)}_{\mu\nu} := \bar S^{(n)}_{\mu\nu} - \frac{1}{2}g_{\mu\nu}g^{\alpha\beta}\bar S^{(n)}_{\alpha\beta}$. In the Lorenz gauge,
\begin{align}
{\cal E}_{\mu\nu}[h^{(n)}] &= -2 \bar S^{(n)}_{\mu\nu}\label{EFEn Lorenz}
\end{align}
and
\begin{align}
{\cal E}_{\mu\nu}[\bar h^{(n)}] &= -2 S^{(n)}_{\mu\nu},\label{EFEn Lorenz hbar}
\end{align}
where we have used the fact that $\bar {\cal E}_{\mu\nu}[h]={\cal E}_{\mu\nu}[\bar h]$.

For simplicity, in the paper we only provide the harmonic decompositions of quantities appearing in Eq.~\eqref{EFEn dR} [and therefore also Eq.~\eqref{EFEn Lorenz}]. The companion package \pkg{PerturbationEquations} additionally includes the decompositions of $\delta G_{\mu\nu}$ and $\delta^2 G_{\mu\nu}$.

\subsection{Bianchi identities and conservation equations}

The components of the perturbative Einstein equations are not all independent. They are related by the contracted Bianchi identity $({\sf g}^{-1})^{\beta\gamma}\,{}^{\sf g}\nabla_{\!\gamma}G_{\alpha\beta}[{\sf g}]=0$, where ${}^{\sf g}\nabla_{\!\alpha}$ is the covariant derivative compatible with ${\sf g}_{\alpha\beta}$. Expanding that identity in orders of nonlinearity, we obtain the identities
\begin{align}
g^{\beta\gamma}\del{\gamma}\delta G_{\alpha\beta} &= 0,\label{Bianchi1}\\
g^{\beta\gamma}\del{\gamma}\delta^2 G_{\alpha\beta} &= h^{\beta\gamma}\del{\gamma}\delta G_{\alpha\beta} +2\delta C^{\gamma\beta}{}_{(\alpha}\delta G_{\beta)\gamma}.\label{Bianchi2}
\end{align}
Here $\delta C^\alpha{}_{\beta\gamma}[h]:=\frac{1}{2}g^{\alpha\delta}(2\del{(\beta} h_{\gamma)\delta}-\del{\delta}h_{\beta\gamma})$ is the linear perturbation of the Christoffel symbol. These identities hold for any symmetric rank-two tensor $h_{\alpha\beta}$.

By virtue of the field equations, these identities are equivalent to stress-energy conservation $({\sf g}^{-1})^{\beta\gamma}\,{}^{\sf g}\nabla_{\!\gamma}T_{\alpha\beta}=0$, or
\begin{align}
\nabla^\beta T^{(1)}_{\alpha\beta} &= 0,\\
\nabla^\beta T^{(2)}_{\alpha\beta} &= h^{(1)\beta\gamma}\del{\gamma}T^{(1)}_{\alpha\beta} +2\delta C^{\gamma\beta}{}_{(\alpha}T^{(1)}_{\beta)\gamma}.
\end{align}

Since Eq.~\eqref{Bianchi1} holds for any $h_{\alpha\beta}$, it immediately implies that the sources $S^{(n)}_{\alpha\beta}$ appearing in the field equations~\eqref{EFEn dG} must all be conserved with respect to the background divergence,%
\beq\label{source conservation}
\nabla^\beta S^{(n)}_{\alpha\beta}=0.
\eeq

\subsection{Gauge freedom}

Perturbation theory in GR comes with well-known gauge freedom corresponding to the choice of how to identify points on the exact spacetime with points in the background spacetime~\cite{Geroch:69,Stewart-Walker:74,Bruni-etal:96}; see Sec. IVA of Ref.~\cite{Pound:15b} or Appendix~\ref{transformation of xi tilde} of this paper for a concise summary. To understand the practical consequence of this, let $A=A^{(0)}+\e A^{(1)} + \e^2 A^{(2)}+\O(\e^3)$ be the expansion of a generic tensor of arbitrary rank (in index-free notation). Under a gauge transformation, the terms in this expansion transform as $A^{(n)}\to A^{(n)}+\Delta A^{(n)}$, where%
\begin{subequations}\label{DeltaA}%
\begin{align}
\Delta A^{(1)} &= \Lie_{\xi_{(1)}} A^{(0)},\label{DeltaA1}\\
\Delta A^{(2)} &= \Lie_{\xi_{(2)}} A^{(0)} + \frac{1}{2}\Lie^2_{\xi_{(1)}}A^{(0)} + \Lie_{\xi_{(1)}} A^{(1)}.\label{DeltaA2}
\end{align}
\end{subequations}
Here $\Lie$ denotes a Lie derivative, and the gauge generators $\xi^\mu_{(n)}$ correspond to the small coordinate transformation
\begin{align}
x'^\mu &= x^\mu-\epsilon \xi^\mu_{(1)}(x)-\epsilon^2\!\! \left[\xi^\mu_{(2)}(x)-\frac{1}{2}\xi^\nu_{(1)}(x)\partial_\nu\xi_{(1)}^\mu(x)\right]\nonumber\\
		&\quad +\O(\epsilon^3).\label{coord_transformation}
\end{align}

Applying Eq.~\eqref{DeltaA} to the metric perturbations $h^{(n)}_{\mu\nu}$ yields%
\begin{subequations}\label{Deltah}%
\begin{align}
\Delta h^{(1)}_{\mu\nu} &= \Lie_{\xi_{(1)}} g_{\mu\nu},\label{Deltah1}\\
\Delta h^{(2)}_{\mu\nu} &= \Lie_{\xi_{(2)}} g_{\mu\nu} + \frac{1}{2}\Lie^2_{\xi_{(1)}}g_{\mu\nu} + \Lie_{\xi_{(1)}} h^{(1)}_{\mu\nu}.\label{Deltah2}
\end{align}
\end{subequations}
Applying it to the stress-energy tensor in a vacuum background yields
\begin{subequations}\label{DeltaT}%
\begin{align}
\Delta T^{(1)}_{\mu\nu}&=0,\label{Delta T1}\\
\Delta T^{(2)}_{\mu\nu} &= \Lie_{\xi_{(1)}}T^{(1)}_{\mu\nu}.\label{Delta T2}
\end{align}
\end{subequations}

The field equations~\eqref{EFE1} and \eqref{EFE2} are invariant under a generic gauge transformation, as can be established from the above transformation laws and the identities~\cite{Pound:15b}
\begin{align}
\Delta \dR_{\mu\nu}[h^{(1)}] &= 0,\label{Delta dR1}\\
\Delta \dR_{\mu\nu}[h^{(2)}] &= \dR_{\mu\nu}[\Delta h^{(2)}],\label{Delta dR2}\\
\Delta\ddR_{\mu\nu}[h^{(1)}] &= \Lie_{\xi_{(1)}} \dR_{\mu\nu}[h^{(1)}] -\dR_{\mu\nu}[\tfrac{1}{2}\Lie^2_{\xi_{(1)}}g] \nonumber\\
			&\quad -\dR_{\mu\nu}[\Lie_{\xi_{(1)}}h^{(1)}].\label{Delta ddR1}
\end{align}
Analogous equations apply for the transformation of $\delta^n G_{\mu\nu}$. 

We stress that while the second-order field equation~\eqref{EFE2} is invariant, the individual terms in it are not. In particular, the left-hand side of~\eqref{EFE2} has the nontrivial transformation~\eqref{Delta dR2}, while the sources on the right-hand side have the nontrivial transformations~\eqref{Delta T2} and \eqref{Delta ddR1}. This differs from the situation at first order, where (in a vacuum background) Eqs.~\eqref{Delta T1} and \eqref{Delta dR1} ensure that each side of the field equation is separately invariant.

\section{Tensors and bases on ${\cal M}^2\times S^2$}\label{2+2 decomposition}

When specialized to a Schwarzschild background, the perturbative Einstein equations are fully separable by virtue of the background's stationarity and spherical symmetry. The spherical symmetry allows us to naturally decompose 4D tensorial quantities into $2+2$D quantities. Specifically, we follow MP in writing the spacetime manifold ${\cal M}$ as the Cartesian product ${\cal M}={\cal M}^2\times S^2$, where ${\cal M}^2$ is the ``$t$-$r$ plane'' and $S^2$ is the two-sphere. This method, which is generally attributed to Gerlach and Sengupta~\cite{Gerlach:1979rw}, enables us to work with quantities that are separately covariant on ${\cal M}^2$ and $S^2$. Tensors on $S^2$ are then naturally decomposed into harmonics.

Although we mostly follow MP, we do adopt slightly different notation. Table~\ref{table:MP conversion} provides the conversion between the two.
\begin{table}[tb]
\renewcommand{\arraystretch}{2}
\caption{\label{table:MP conversion}Relationship between our bases and derivatives and those of Martel and Poisson~(MP)~\cite{Martel-Poisson:05}.} 
\begin{ruledtabular}
\begin{tabular}{@{\qquad\qquad\quad} c c @{\qquad\qquad\quad}}
This paper & MP \\
\midrule[.5pt]
$t_a$ & $-f^{-1} t_a^{MP}$\\
$r_a$ & $r_a$\\
$\delta_a$ & $\del{a}$\\
$\D{A}$ & $\D{A}$\\
\end{tabular}
\end{ruledtabular}
\end{table}

\subsection{Covariant decompositions}\label{covariant 2+2 decomposition}

We let $x^a$ be coordinates on ${\cal M}^2$ and give tensors on ${\cal M}^2$ lowercase Latin indices $a, b, c, \ldots$; analogously, we let $\theta^A$ be coordinates on $S^2$ and give tensors on $S^2$ uppercase Latin indices $A, B, C, \ldots$. The background line element can then be written as
\beq
ds^2 = g_{ab}dx^a dx^b + r^2\Omega_{AB}d\theta^Ad\theta^B,
\eeq
where $r$ is the areal radius of a sphere of fixed $x^a$, $g_{ab}$ is the restriction of $g_{\mu\nu}$ to ${\cal M}^2$, and $\Omega_{AB}$ is the metric of the unit sphere. We use $g_{ab}$ and its inverse $g^{ab}$ to lower and raise indices of tensors on ${\cal M}^2$, and $\Omega_{AB}$ and its inverse $\Omega^{AB}$ to lower and raise indices of tensors on $S^2$. We also require the Levi-Civita tensors $\epsilon_{ab}$ and $\epsilon_{AB}$. In standard polar coordinates $\theta^A=(\theta,\phi)$, the tensors on $S^2$ are given by 
\beq
\Omega_{AB}={\rm diag}(1,\sin^2\theta)\quad \text{and} \quad \epsilon_{\theta\phi}=\sin\theta=-\epsilon_{\phi\theta}.
\eeq

Decomposing the field equations~\eqref{EFE1}--\eqref{EFE2} into tensors on ${\cal M}^2$ and $S^2$ requires doing likewise for covariant derivatives. We define $\delta_a$ and $\D{A}$ to be the derivatives compatible with $g_{ab}$ and $\Omega_{AB}$, respectively, with corresponding Christoffel symbols $\Gamma[\delta]^a_{bc}$ and $\Gamma[D]^A_{BC}$. The nonvanishing Christoffel symbols $\Gamma^\mu_{\nu\rho}$ associated with $\del{\alpha}$ are related to these according to $\Gamma^a_{bc}=\Gamma[\delta]^a_{bc}$, $\Gamma^A_{BC}=\Gamma[D]^A_{BC}$, and
\begin{align}\label{Christoffel}
\Gamma^a_{AB} = -r r^a\Omega_{AB},\qquad \Gamma^A_{Bc} = \frac{\delta^A_{\ B} r_c}{r},
\end{align}
where 
\beq\label{ra}
r_a := \partial_a r.
\eeq
This allows us to decompose the components of a derivative $\del{\alpha}v^\beta$ into covariant quantities on ${\cal M}^2$ and $S^2$:%
\begin{subequations}\label{Dv}%
\begin{align}
\del{a} v^b &= \delta_a v^b, \\
\del{a} v^B &= \delta_a v^B+r^{-1}r_av^B,\\
\del{A} v^b &= \D{A} v^b - rr^b \Omega_{AB}v^B,\\
\del{A} v^B &= \D{A} v^B + r^{-1}\delta^B_{\ A}r_cv^c,
\end{align}
\end{subequations}
where $\D{A}$ acts on $v^b$ as it would on a scalar, and $\delta_a$ acts on $v^B$ as it would on a scalar. Similarly, the components of $\del{\alpha}\omega_\beta$ are written as%
\begingroup%
\allowdisplaybreaks
\begin{subequations}\label{Dw}
\begin{align}
\del{a}\omega_b &= \delta_a\omega_b,\\
\del{a}\omega_B &= \delta_a\omega_B-r^{-1}r_a\omega_B,\\
\del{A}\omega_b &= \D{A}\omega_b - r^{-1}\omega_Ar_b,\\
\del{A}\omega_B &= \D{A}\omega_B + r\Omega_{AB}r^c\omega_c.
\end{align}
\end{subequations}
\endgroup
Higher derivatives are expressed in the same manner. 

We will also require the Riemann tensors associated with the derivatives $\delta_a$ and $\D{A}$, $R[\delta]_{abcd}$ and $R[D]_{ABCD}$. They are given by 
\begin{align}
R[\delta]_{abcd} &= \frac{2M}{r^3}(g_{ac}g_{bd}-g_{ad}g_{bc}),\\
R[D]_{ABCD}&=\Omega_{AC}\Omega_{BD}-\Omega_{AD}\Omega_{BC}.
\end{align} 

In concrete calculations, our first step is always to expand contractions into $2+2$D form and then project any free indices onto either ${\cal M}^2$ or ${\cal S}^2$. For example,
\beq
g^{\alpha\beta}\del{\alpha} h_{\beta\gamma} = g^{ab}\del{a} h_{b\gamma} + r^{-2}\Omega^{AB}\del{A} h_{B\gamma}.
\eeq 
Choosing $\gamma=c$ (i.e., projecting onto ${\cal M}^2$) and then using Eq.~\eqref{Dw}, one obtains a fully decomposed expression:
\begin{multline}
g^{\alpha\beta}\del{\alpha} h_{\beta c} = g^{ab}\delta_a h_{bc} + r^{-2}\Omega^{AB}\D{A} h_{Bc} \\ +2r^{-1}r^a h_{ac} - r^{-3}h^{A}{}_{\!A} r_c,
\end{multline}
where $h^A{}_{\!A} =\Omega^{AB}h_{AB}$.

\subsection{Bases on ${\cal M}^2$ and $S^2$}

Most of our results will be fully covariant, without any choice of basis on ${\cal M}^2$ or $S^2$. However, we will on occasion adopt specific bases. 

\subsubsection{Bases on ${\cal M}^2$}

As a coordinate basis for tensors on ${\cal M}^2$, we use $(t_a,r_a)$, where $r_a$ is defined in Eq.~\eqref{ra} and 
\beq
t_a:=\partial_a t.
\eeq
Here $t$ is the usual Schwarzschild time, and we note that MP use the same notation to instead denote the timelike Killing vector; the two are related by $t^a = -f^{-1} t^a_{MP}$, with
\beq
f = (-t^a t_a)^{-1} = r^ar_a = 1-\frac{2M}{r}.
\eeq
In terms of these quantities, we have 
\begin{align}
g_{ab}&=-f t_at_b+f^{-1}r_ar_b,\\
\epsilon_{ab} &= t_a r_b - r_a t_b.
\end{align}

We will also make use of a Newman--Penrose null basis
\begin{align}
l^a &= \frac{\gamma}{\sqrt{2f}}(1,f),\\
n^a &= \frac{1}{\sqrt{2f}\gamma}(1,-f),
\end{align}
where the components are given in $(t,r)$ coordinates and $\gamma=\gamma(r)>0$ is an arbitrary boost factor. This basis satisfies $l^a n_a = -1$ and $l^al_a=0=n^a n_a$, which imply
\begin{align}
g_{ab} &= - l_a n_b - n_al_b,\\
\epsilon_{ab} &= l_a n_b - n_a l_b, 
\end{align}
and
\beq\label{ln identities}
\begin{array}{lr}
l^b\delta_b l^a = l^a\delta_b l^b, &\quad n^b\delta_b n^a = n^a\delta_b n^b, \\
l^b\delta_b n^a = -n^a\delta_b l^b, &\quad n^b\delta_b l^a = - l^a\delta_b n^b. 
\end{array}
\eeq
The divergences that appear in \eqref{ln identities} are given by 
\beq\label{div l and div n}
\delta_al^a = \frac{r^2 f \partial_r\gamma+M \gamma}{r^2\sqrt{2f}} \quad\text{and} \quad \delta_a n^a = \frac{r^2 f\partial_r\gamma - M\gamma}{r^2\sqrt{2f}\gamma^2}.
\eeq

In the definition of the null basis vectors, the boost factor $\gamma$ is commonly chosen to be one of the following:%
\begin{subequations}%
\begin{align}
\text{Carter~\cite{Carter:1987hk,Pound:2021qin}:} &\qquad \gamma=1,\\
\text{Kinnersley~\cite{Kinnersley:1969zza}:} &\qquad \gamma=\sqrt{2/f},\\
\text{Hartle--Hawking~\cite{Hawking:1972hy}:} &\qquad \gamma=\sqrt{f/2}.
\end{align}
\end{subequations}
In the Kinnersley basis, $\delta_a l^a = 0$; in the Hartle--Hawking basis, $\delta_a n^a = 0$; in the Carter basis, $\delta_a l^a = - \delta_a n^a \neq 0$. In the Kinnsersley basis, $l^a$ is tangent to affinely parameterized outgoing null rays, where $r$ is the affine parameter. This makes the Kinnersley basis singular at the future horizon but particularly useful for studying outgoing radiation: in retarded Eddington--Finkelstein coordinates $(u,r)$, 
\beq
l^a_{\rm K}\partial_a=\partial_r \quad \text{and} \quad n^a_{\rm K}\partial_a=\partial_u-(f/2)\partial_r. 
\eeq
In the Hartle--Hawking basis, $n^a$ is tangent to affinely parameterized ingoing null rays, where $r$ is again the affine parameter. This makes the Hartle--Hawking basis singular at the past horizon but particularly useful for studying ingoing radiation: in advanced Eddington--Finkelstein coordinates $(v,r)$,
\beq
l^a_{\rm HH}\partial_a=\partial_v+(f/2)\partial_r\quad \text{and} \quad n^a_{\rm HH}\partial_a=-\partial_r.
\eeq 
The Carter basis is singular at both the past and future horizon, but it has the advantage of maintaining a symmetry between ingoing and outgoing null directions: in double null coordinates $(u,v)$,
\beq 
l^a_{\rm C}\partial_a = \sqrt{\frac{2}{f}}\partial_v \quad\text{and}\quad n^a_{\rm C}\partial_a = \sqrt{\frac{2}{f}}\partial_u.
\eeq

\subsubsection{Bases on $S^2$}

As a basis on $S^2$, we define a complex null vector
\begin{equation}
\tilde m^A= \left(1,\frac{i}{\sin\theta}\right)
\end{equation}
and its complex conjugate, $\tilde m^{A*}$, where the components are given in $(\theta,\phi)$ coordinates. Our definition of $\tilde m^A$ differs by a factor of $\sqrt{2}r$ relative to the traditional Newman--Penrose basis~\cite{Newman-Penrose:66}. With our choice of normalization, the basis vectors satisfy
\begin{equation}\label{m identities}
\begin{array}{ll}
\tilde m^A \tilde m_A=0, &\quad \tilde m^B \D{B} \tilde m^A =  \tilde m^A  \D{B}\tilde m^B, \\
\tilde m^A \tilde m_A^*=2, &\quad \tilde m^{B*} \D{B} \tilde m^A = - \tilde m^A \D{B}\tilde m^{B*}, \\
\end{array}
\end{equation}
and
\begin{align}
\Omega_{AB} &=\frac{1}{2}\left(\tilde m_A \tilde m_{B}^*+\tilde m_{A}^* \tilde m_B\right),\label{Omega to mmb}\\
\epsilon_{AB} &=\frac{i}{2}\left(\tilde m_A \tilde m_B^*-\tilde m_A^* \tilde m_B\right)\label{eps to mmb}.
\end{align}
In $(\theta,\phi)$ coordinates,
\beq
\D{B}\tilde m^{B}=\D{B}\tilde m^{B*}=\cot\theta.
\eeq
Equation~\eqref{eps to mmb} also provides the useful identity 
\beq\label{epsm=im}
\epsilon^A_{\ B} \tilde m^B = i\tilde m^A.
\eeq

It will be useful to also define the Newman--Penrose basis,
\begin{align}
m^A &= \frac{1}{\sqrt{2}r}\tilde m^A,
\end{align}
which satisfies $g_{AB}m^A m^{B*}=1$, and in terms of which
\begin{align}
g_{AB} &= r^4(m_A m^*_B + m^*_A m_B).
\end{align}
The factor of $r^4$ arises from the fact that indices are lowered with $\Omega_{AB}$.

The set of vectors $\{l^\alpha,n^\alpha,m^\alpha,m^{\alpha*}\}$ form a null tetrad on ${\cal M}$, with the natural definitions $l^A=n^A=m^a=0$. A generic symmetric tensor $h_{\alpha\beta}$ can be decomposed into this basis according to
\begin{subequations}%
\begin{align}
h_{ab} &= h_{ll}n_a n_b +h_{l n}(n_a l_b +l_a n_b) + h_{nn} l_a l_b,\\
h_{aA} &= -r^2 h_{l m}n_a m^*_A - r^2 h_{n m}l_a m^*_A + {\rm c.c.},\\
h_{AB} &= r^4 h_{mm}m^*_A m^*_B + r^4 h_{mm^*}m^*_A m_B + {\rm c.c.}
\end{align}
\end{subequations}
Alternatively, we can decompose it in a mixed basis $\{t_a,r_a,\tilde m_A, \tilde m^*_A\}$, according to
\begin{subequations}%
\begin{align}
h_{ab} &= h_{tt}t_a t_b +h_{tr}(t_a r_b + r_a t_b) + h_{rr} r_a r_b,\\
h_{aA} &= \frac{1}{2}\left( h_{t \tilde m}t_a \tilde m^*_A + h_{r \tilde m}r_a \tilde m^*_A + {\rm c.c.}\right),\\
h_{AB} &= \frac{1}{4}\left(h_{\tilde m\tilde m}\tilde m^*_A \tilde m^*_B + h_{\tilde m\tilde m^*}\tilde m^*_A \tilde m_B + {\rm c.c.}\right),
\end{align}
\end{subequations}
or only partially decompose it, according to
\begin{subequations}%
\begin{align}
h_{ab} &= h_{tt}t_a t_b +h_{tr}(t_a r_b + r_a t_b) + h_{rr} r_a r_b,\\
h_{aB} &= h_{tB}t_a + h_{r B}r_a,\\
h_{AB} &= h_{AB}.
\end{align}
\end{subequations}

As a final comment, we observe the main practical advantage of working with the quantities $\{g_{ab},\delta_a,\Omega_{AB},\D{A}\}$. Besides allowing tensor-harmonic decompositions while preserving invariance, these choices enforce that the background quantities on ${\cal M}^2$ commute with those on $S^2$:
\begin{equation}
\D{A} g_{ab} = \delta_a \Omega_{AB} = [\D{A},\delta_a]=0.
\end{equation}
This is not the case if working with $g_{AB}$ and $\del{A}$. Likewise, $\tilde m^A$ often provides a more convenient basis than $m^A$ because 
\beq
\delta_a \tilde m^A=[l,\tilde m]=[n,\tilde m]=0.
\eeq
In this last equation, $[\cdot,\cdot]$ denotes the vector commutator, $[u,v]^\alpha:=u^\beta\partial_\beta v^\alpha-v^\beta\partial_\beta u^\alpha=u^\beta\del{\beta} v^\alpha-v^\beta\del{\beta} u^\alpha={\cal L}_u v^\alpha$. 

By working with trivially commuting quantities, our choices (like MP's) take maximal advantage of Schwarzschild's spherical symmetry.

\section{Decompositions into spin-weighted and tensor spherical harmonics}\label{harmonics}

The literature contains numerous bases of spherical harmonics that can be used to decompose the field equations. With the exception of the Teukolsky formalism, calculations in Schwarzschild spacetime typically use tensor harmonic bases. For that reason, we will decompose the metric perturbation and Einstein equations into tensor harmonics, specifically adopting MP's choice of harmonics. However, for reasons explained below, instead of tensor harmonics we take spin-weighted spherical harmonics ${}_sY_{lm}$ to be the ``base'' harmonics. Our expansions in tensor harmonics will utilize spin-weighted harmonics as an intermediary. This will also allow us to easily connect to the Teukolsky formalism in Sec.~\ref{Teukolsky}. We refer to Brizuela et al. for a treatment that consistently uses tensor harmonics rather than spin-weighted ones~\cite{Brizuela-etal:06,Brizuela:2007zza,Brizuela-etal:09}.

Given the large number of common conventions for harmonic expansions, in Table~\ref{table:conventions} we provide translations between conventions.

\begin{table*}[tb]
\renewcommand{\arraystretch}{2}
\caption{\label{table:conventions}Relationship between harmonic coefficients in various conventions. The relationship between the tensorial and tetrad decompositions is the same for sources (bottom half of table) as for metric perturbations (top half of table).  Additional relations can be found in Eq.~\eqref{tetrad coeffs from tensor coeffs} and Appendix~\ref{BLS conventions} of this paper and Table I of Ref.~\cite{Pound:2021qin}.} 
\begin{ruledtabular}
\begin{tabular}{c c c c}
$\begin{array}{c}\text{This paper}\\[-11pt] \text{(tensorial)}\end{array}$ & $\begin{array}{c}\text{Martel}\\[-11pt]\text{\& Poisson~\cite{Martel-Poisson:05}}\end{array}$ & $\begin{array}{c}\text{This paper}\\[-11pt] \text{(tetrad)}\end{array}$ & Barack--Lousto--Sago~\cite{Barack-Lousto:05,Barack-Sago:07} \\
\midrule[.5pt]
$h^{\ell m}_{ab}$ & $h^{\ell m}_{ab}$ & $l_a l_b h^{\ell m}_{nn} + 2l_{(a} n_{b)} h^{\ell m}_{ln} + n_a n_b h^{\ell m}_{ll}$ & $\begin{array}{r@{}l@{}}\frac{1}{2r}\left[(\bar h_{1\ell m}+f\bar h_{6\ell m})t_at_b+f^{-1}\bar h_{2\ell m}(t_ar_b+r_at_b)\right.\\\left.+f^{-2}(\bar h_{1\ell m}-f \bar h_{6\ell m})r_ar_b\right]\end{array}$ \\
\midrule[.25pt]
$h^{\ell m}_{a+}$ & $j^{\ell m}_{a}$ & $\dfrac{r}{\sqrt{2}\lam{\ell}{1}}\left[l_a (h^{\ell m}_{nm} - h^{\ell m}_{nm^*}) + n_a( h^{\ell m}_{lm}- h^{\ell m}_{lm^*})\right]$ & $\displaystyle\frac{1}{2\lam{\ell}{1}^2}(\bar h_{4\ell m}t_a+f^{-1}\bar h_{5\ell m}r_a)$ \\
\midrule[.25pt]
$h^{\ell m}_{+}$ & $r^2G^{\ell m}$ & $\dfrac{r^2}{\lam{\ell}{2}} \left(h^{\ell m}_{mm}+h^{\ell m}_{m^*m^*}\right)$ & $\displaystyle\frac{r}{\lam{\ell}{2}^2}\bar h_{7\ell m}$ \\
\midrule[.25pt]
$h^{\ell m}_{\trAB}$ & $r^2K^{\ell m}$ & $r^2 h^{\ell m}_{mm^*}$ & $\displaystyle\frac{r}{2}\bar h_{3\ell m}$ \\
\midrule[.25pt]
$h^{\ell m}_{a-}$ & $h^{\ell m}_{a}$ & $-\dfrac{i r}{\sqrt{2}\lam{\ell}{1}}\left[l_a (h^{\ell m}_{nm}+h^{\ell m}_{nm^*}) + n_a(h^{\ell m}_{lm} + h^{\ell m}_{lm^*})\right]$ & $\displaystyle-\frac{1}{2\lam{\ell}{1}^2}(\bar h_{8\ell m}t_a+f^{-1}\bar h_{9\ell m}r_a)$ \\
\midrule[.25pt]
$h^{\ell m}_{-}$ & $h^{\ell m}_{2}$ & $-\dfrac{i r^2}{\lam{\ell}{2}} \left(h^{\ell m}_{mm}-h^{\ell m}_{m^*m^*}\right)$ & $\displaystyle-\frac{r}{\lam{\ell}{2}^2}\bar h_{10\ell m}$ \\
\midrule[.25pt]
$S^{\ell m}_{ab}$ & $Q^{\ell m}_{ab}$ & $\vdots$ & $\begin{array}{r@{}l@{}}\frac{1}{\sqrt{2}}\left[( S_{1\ell m}+f S_{3\ell m})t_at_b+f^{-1} S_{2\ell m}(t_ar_b+r_at_b)\right.\\\left.+f^{-2}( S_{1\ell m}-f  S_{3\ell m})r_ar_b\right]\end{array}$ \\
\midrule[.25pt]
$S^{\ell m}_{a+}$ & $\dfrac{1}{2}Q^{\ell m}_a$ & $\vdots$ & $\dfrac{r}{\sqrt{2}\lam{\ell}{1}}( S_{4\ell m}t_a+f^{-1} S_{5\ell m}r_a)$  \\
\midrule[.25pt]
$S^{\ell m}_{+}$ & $\dfrac{1}{2}Q^\sharp_{\ell m}$  & $\vdots$ & $\dfrac{\sqrt{2}r^2}{\lam{\ell}{2}} S_{7\ell m}$ \\
\midrule[.25pt]
$S^{\ell m}_{\trAB}$ & $\dfrac{r^2}{2}Q^\flat_{\ell m}$ & $\vdots$ & $\dfrac{r^2}{\sqrt{2}} S_{6\ell m}$ \\
\midrule[.25pt]
$S^{\ell m}_{a-}$ & $\dfrac{1}{2}P^{\ell m}_a$ & $\vdots$ & $-\dfrac{r}{\sqrt{2}\lam{\ell}{1}}( S_{8\ell m}t_a+f^{-1} S_{9\ell m}r_a)$ \\
\midrule[.25pt]
$S^{\ell m}_{-}$ & $P^{\ell m}$ & $\vdots$ & $-\dfrac{\sqrt{2}r^2}{\lam{\ell}{2}} S_{10\ell m}$
\vspace{0.35em}
\end{tabular}
\end{ruledtabular}
\end{table*}

\subsection{Spin-weighted harmonics}

A spin-weighted tensor $v$ on $S^2$ is said to have spin weight $s$ if it transforms as $v\to e^{is\varphi}v$ under the complex phase rotation $\tilde m^A\to e^{i\varphi}\tilde m^A$~\cite{Newman-Penrose:66}. In practice, this means $v$'s spin weight is the number of factors of $\tilde m^A$ appearing in it minus the number of factors of $\tilde m^{A*}$ appearing in it. 

We define derivative operators
\begin{subequations}
\label{eq:eth-ethbar}
\begin{align}
\edth &= \tilde m^A\D{A}-s (\D{A} \tilde m^A),\label{eth}\\
\edth' &= \tilde m^{A*}\D{A}+s(\D{A} \tilde m^{A*}),\label{ethbar}
\end{align}
\end{subequations}
that act on tensors of spin-weight $s$.
Our definitions and notation here differ slightly from common conventions in the literature, as summarized in Table~\ref{table:derivatives}. The derivative $\edth $ raises the spin weight by 1, while $\edth'$ lowers it by 1. They satisfy the Leibniz rule [e.g., $\edth (uv) = (\edth u)v + u\edth v$, where $u$ and $v$ can have differing spin weights], and the identities~\eqref{m identities} ensure they annihilate $\tilde m^A$ and $\tilde m^{A*}$:
\beq\label{eth m}
\edth \tilde m^A = \edth' \tilde m^A = \edth \tilde m^{A*} = \edth' \tilde m^{A*} =0.
\eeq
They satisfy the commutation and anti-commutation relations
\begin{subequations}
\begin{align}
\frac{1}{2}(\edth'\edth-\edth\edth') &= i\epsilon^{AB}\D{A}\D{B} + s,\\
\frac{1}{2}(\edth'\edth+\edth\edth') &= \D{A} D^A + s[(\D{B} m^{B*})m^A\D{A}-{\rm c.c.}]\nonumber\\
								&\quad -s^2|\D{A}\tilde m^A|^2.
\end{align}
\end{subequations}
When acting on a spin-weighted scalar (such as a component $h_{am}$), they satisfy $\edth^i\edth'^j=\edth'^j\edth^i$ if $j=i+2s$. Like $\D{A}$, they commute with background quantities on ${\cal M}^2$:
\beq
\edth g_{ab}=\edth' g_{ab}= [\delta_a,\edth] =[\delta_a,\edth']=0.
\eeq

\begin{table}[tb]
\renewcommand{\arraystretch}{2}
\caption{\label{table:derivatives}Relationship between our derivatives and those of Newman \& Penrose~(NP)~\cite{Newman-Penrose:61,Newman-Penrose:66} and Geroch--Held--Penrose (GHP)~\cite{Geroch-Held-Penrose:73}.
We note that Ref.~\cite{Pound:2021qin} adopts GHP conventions. The quantities $\beta$, $\rho$, $\epsilon$, and $\gamma_{\rm NP}$ appearing in the relations are Newman--Penrose spin coefficients, which in our context reduce to $\beta=\frac{1}{2}\D{A} m^A$; $\rho = -l^a r_a/r$; $\epsilon = \frac{1}{2}\delta_a l^a=\frac{1}{2}\partial_r l^r$; and $\gamma_{\rm NP}=-\frac{1}{2}\delta_a n^a=-\frac{1}{2}\partial_r n^r$, or $\gamma_{\rm NP}=-\epsilon'$ in GHP notation. The quantity $b$ is boost weight, defined above Eq.~\eqref{thorn}.} 
\begin{ruledtabular}
\begin{tabular}{c c c c}
This paper & NP~\cite{Newman-Penrose:61} & NP~\cite{Newman-Penrose:66} & GHP \\
\midrule[.5pt]
$\edth$ & $\sqrt{2}r(\bm{\delta}-2s\beta)$ & $-\eth_{\rm NP}$ & $\sqrt{2}r\eth_{\rm GHP}$ \\
$\edth'$ & $\sqrt{2}r(\bar{\bm{\delta}}+2s\beta)$ & $-\bar\eth_{\rm NP}$ & $\sqrt{2}r\eth'_{\rm GHP}$ \\
$\thorn$ & $\bm{D}-2b\epsilon$ & $\vdots$ & $\thorn$ \\
$\thorn'$ & $\bm{\Delta}-2b\gamma_{\rm NP}$ & $\vdots$ & $\thorn'$ 
\end{tabular}
\end{ruledtabular}
\end{table}

A spin-weighted scalar of spin-weight $s$ is conveniently expanded as a sum of spin-weighted spherical harmonics of the same spin weight, defined for $\ell\geq|s|$ as
\begin{equation}\label{sYlm definition}
{}_sY_{\ell m} := \frac{1}{\lam{\ell}{s}}
	\begin{cases}
    (-1)^s\edth^sY_{\ell m}, & 0\leq s\leq \ell,\\
    \edth'^{|s|} Y_{\ell m}, & - \ell\leq s\leq0,
  \end{cases}
\end{equation}
where 
\beq
\lam{\ell}{s}:=\sqrt{\frac{(\ell+|s|)!}{(\ell-|s|)!}} = \sqrt{(l - |s| + 1)_{2|s|}}.
\eeq
We also define for later use the related quantity $\mu_\ell$ by
\beq\label{mu def}
\mu^2_\ell:=(\ell+2)(\ell-1)=(\lam{\ell}{2}/\lam{\ell}{1})^2=\lam{\ell}{1}^2-2.
\eeq
Here we adopt standard definitions; these are precisely the spin-weighted harmonics defined by Newman and Penrose~\cite{Newman-Penrose:66},
simply re-expressed in terms of our convention for the operators $\edth$ and $\edth'$.
These definitions are also consistent with, for example, \pkg{Mathematica}'s \texttt{SphericalHarmonicY} function and with the
\texttt{SpinWeightedSphericalHarmonicY} function in the Black Hole Perturbation toolkit's \pkg{SpinWeightedSpheroidalHarmonics}
package \cite{BHPToolkit,barry_wardell_2023_8112931}.
Note that although Goldberg, et al. \cite{Goldberg-etal:67} is also a standard reference for the spin-weighted spherical harmonics,
their definition includes a non-standard overall factor of $(-1)^m$. 

The spin-weighted harmonics are related to Wigner $D$ matrices (again, following conventions consistent with \pkg{Mathematica}'s \texttt{WignerD} function) according to 
\begin{equation}
{}_sY_{\ell m}(\theta,\phi) = (-1)^s \sqrt{\frac{2\ell+1}{4\pi}}D^\ell_{-sm}(0,\theta,\phi).\label{YtoD}
\end{equation}
They satisfy the orthonormality conditions
\beq
\oint {}_sY^*_{\ell m}\ {}_sY_{\ell'm'} d\Omega = \delta_{\ell\ell'}\delta_{mm'}, \label{sYlm-orthonormal}
\eeq
where $d\Omega=\sin\theta d\theta d\phi$, and they have the properties~\cite{Goldberg-etal:67}%
\begin{subequations}\label{sYlm identities}%
\begin{align}
{}_sY^*_{\ell m} &= (-1)^{m+s}{}_{-s}Y_{\ell,-m},\label{sYlm-conj}\\
\edth{}_sY_{\ell m} &= -\sqrt{(\ell-s)(\ell+s+1)}\,{}_{s+1}Y_{\ell m},\label{spin raising}\\
\edth'{}_sY_{\ell m} &= \sqrt{(\ell+s)(\ell-s+1)}\,{}_{s-1}Y_{\ell m},\label{spin lowering}\\
\edth'\edth{}_sY_{\ell m} &= - (\ell-s)(\ell+s+1)\,{}_sY_{\ell m}.\label{edthbar-edth-sYlm}
\end{align}
\end{subequations}
Due to our sign convention for $\edth$, Eqs.~\eqref{spin raising} and \eqref{spin lowering} differ by an overall sign relative to the analogous formulas in Ref.~\cite{Goldberg-etal:67}. The identity~\eqref{edthbar-edth-sYlm} is an eigenvalue equation that can equivalently be written as 
\beq
\frac{1}{2}\left(\edth\edth'+\edth'\edth\right){}_sY_{\ell m} = - \left[\ell(\ell+1)-s^2\right]\,{}_sY_{\ell m}.
\eeq

Spin-weighted harmonics are convenient for two key reasons. First, Eq.~\eqref{EFE2} involves many derivatives, and any number of covariant derivatives of $Y_{\ell m}$ can be easily written in terms of ${}_sY_{\ell m}$. For example, using $\D{A}Y_{\ell m}=\frac{1}{2}(\tilde m_A \tilde m^{B*}+\tilde m^*_A\tilde m^B)\D{B}Y_{\ell m}$ together with Eqs.~\eqref{eq:eth-ethbar} and \eqref{sYlm identities}, one finds  
\begin{equation}\label{DY}
\D{A} Y_{\ell m} = \frac{\lam{\ell}{1}}{2}\left({}_{-1}Y_{\ell m}\tilde m_A-{}_1Y_{\ell m}\tilde m^*_A\right).
\end{equation}
Doing the same for $\D{A}\D{B}Y_{\ell m}$ and making use of Eqs.~\eqref{eq:eth-ethbar}, \eqref{sYlm identities} and \eqref{m identities}, one finds
\begin{align}
\D{A}\D{B}Y_{\ell m} &= \frac{\lam{\ell}{2}}{4}\left({}_{-2}Y_{\ell m}\tilde m_A \tilde m_B+{}_2Y_{\ell m}\tilde m^*_A \tilde m^*_B\right) \nonumber\\
							&\quad -\frac{\lam{\ell}{1}^2}{2} Y_{\ell m}\Omega_{AB}.\label{DDY}
\end{align}
Higher derivatives are given in Eqs.~\eqref{DDDY} and \eqref{DDDDY}. 

The second reason spin-weighted harmonics are useful is that when one expands the first-order field in a basis of harmonics, the sources in Eq.~\eqref{EFE2} involve products of  those harmonics, and decomposing that product into a sum of single harmonics requires the integral of three harmonics. With spin-weighted harmonics, that integral is easily found. We define the desired integral as
\begin{equation}\label{Cdef}
C^{ \ell m s}_{\ell'm's'\ell''m''s''}:=\oint {}_{s}Y^*_{\ell m}{}_{s'}Y_{\ell' m'}{}_{s''}Y_{\ell'' m''}d\Omega. 
\end{equation}
For $s=s'+s''$, one can explicitly evaluate the integral using Eq.~\eqref{YtoD} and then following Sec. 30B of Ref.~\cite{Hecht:00} to derive the integral of three Wigner $D$ matrices. The result is~\cite{Shiraishi:12}%
\begin{align}\label{coupling}
C^{\ell m s}_{\ell'm's'\ell''m''s''} &= (-1)^{m+s}\sqrt{\frac{(2\ell+1)(2\ell'+1)(2\ell''+1)}{4\pi}}\nonumber\\
									&\quad	\times \begin{pmatrix} \ell &  \ell' &  \ell'' \\ s & -s' & -s''\end{pmatrix}
														  \begin{pmatrix} \ell &  \ell' &  \ell'' \\ -m & m' & m''\end{pmatrix}\!,\!
\end{align}
where the arrays are $3j$ symbols. It follows from the symmetries of the $3j$ symbol that%
\begin{subequations}\label{C symmetries}%
\begin{align}
C^{\ell m s}_{\ell'm's'\ell''m''s''} &= (-1)^{\ell+\ell'+\ell''}C^{\ell m -s}_{\ell'm'-s'\ell''m''-s''},\\
C^{\ell m s}_{\ell'm's'\ell''m''s''} &= (-1)^{\ell+\ell'+\ell''}C^{\ell -m s}_{\ell'-m's'\ell''-m''s''},\\
C^{\ell m s}_{\ell'm's'\ell''m''s''} &= C^{\ell m s}_{\ell''m''s''\ell'm's'}.
\end{align}
\end{subequations}
It also follows that the usual rules associated with coupling of angular momenta are enforced, since the $C^{\ell m s}_{\ell'm's'\ell''m''s''}$ are zero unless:
\begin{subequations}
  \label{Crules}
\begin{align}
m &= m'+m'',\\
\left|\ell'-\ell''\right|&\leq \ell \leq \ell'+\ell''.\label{triangle inequality}
\end{align}
\end{subequations}
Finally, we note that for $\ell=m=s=0$ and $s''=-s'$, the result collapses to
\beq\label{ell=0 coupling C}
C^{000}_{\ell'm's'\ell''m''s''} = \frac{(-1)^{m'+s'}}{\sqrt{4\pi}}\delta_{\ell'\ell''}\delta_{m',-m''}.
\eeq

To decompose Eq.~\eqref{EFE2} in tensor harmonics, we will express all quantities in terms of spin-weighted harmonics. Equation~\eqref{coupling} then becomes the essential tool in the decomposition. To the best of our knowledge, this strategy has not appeared in prior literature.

\subsection{Tensor harmonics}

Tensor harmonics of rank $s$ are constructed from symmetric and trace-free combinations of covariant derivatives of ordinary scalar harmonics $Y_{\ell m}$~\cite{Brizuela-etal:06}:\footnote{To maintain compatibility with MP, we have introduced a minus sign into Ref.~\cite{Brizuela-etal:06}'s definition of $X^{\ell m}_{A_1\cdots A_s}$.}%
\begin{subequations}\label{rank-s harmonics}%
\begin{align}
Y^{\ell m}_{A_1\cdots A_s}&:=\D{\langle A_1}\cdots \D{A_s\rangle}Y_{\ell m},\\
X^{\ell m}_{A_1\cdots A_s}&:=-\epsilon_{\langle A_1}{}^C \D{A_2}\cdots \D{A_s\rangle}\D{C} Y_{\ell m},
\end{align}
\end{subequations}
where angular brackets denote the symmetric-trace-free part of a tensor, with traces defined using $\Omega^{AB}$. These harmonics are defined only for $\ell\geq s$, as they identically vanish for $0\leq\ell<s$. They are related to spin-weighted harmonics by the simple formulas%
\begin{subequations}\label{tensor vs spin-weighted harmonics}%
\begin{align}
Y^{\ell m}_{A_1\cdots A_s} &= \frac{\lam{\ell}{s}}{2^s}\big[{}_{-s}Y_{\ell m}\tilde m_{A_1}\cdots \tilde m_{A_s}\nonumber\\
	&\qquad\qquad+(-1)^s{}_sY_{\ell m}\tilde m^*_{A_1}\cdots \tilde m^*_{A_s}\big],\label{tensor Y vs sYlm}\\
X^{\ell m}_{A_1\cdots A_s} &= -\frac{i\lam{\ell}{s}}{2^s}\big[{}_{-s}Y_{\ell m}\tilde m_{A_1}\cdots \tilde m_{A_s}\nonumber\\
&\qquad\qquad-(-1)^s{}_sY_{\ell m}\tilde m^*_{A_1}\cdots \tilde m^*_{A_s}\big]; \label{tensor X vs sYlm}
\end{align}
\end{subequations}
see Appendix~\ref{Covariant derivs of Y}. The harmonics $Y_{\ell m}$ and $Y^{\ell m}_{A_1\cdots A_s}$ are said to have even parity, transforming as $Y^{\ell m}_{A_1\cdots A_s}\to \displaystyle(-1)^\ell Y^{\ell m}_{A_1\cdots A_s}$ under the parity inversion $(\theta,\phi)\to(\pi-\theta,\phi+\pi)$, while $X^{\ell m}_{A_1\cdots A_s}$ are said to have odd parity, transforming as $X^{\ell m}_{A_1\cdots A_s}\to\displaystyle (-1)^{\ell+1}X^{\ell m}_{A_1\cdots A_s}$. In the linearized field equations, the even- and odd-parity sectors decouple. However, at second order they couple through the source terms in the field equation~\eqref{EFE2}. 

Brizuela et al.~\cite{Brizuela-etal:06,Brizuela:2007zza,Brizuela-etal:09} worked consistently with tensor harmonics rather than spin-weighted ones, motivating their use of rank-$s>2$ harmonics. However, in our case we will only require rank-1 (vector) and rank-2 tensor harmonics. Specializing Eqs.~\eqref{rank-s harmonics} to these cases, we see that the vector harmonics, defined for $\ell\geq1$, are given by%
\begin{subequations}\label{vector harmonics def}%
\begin{align}
Y^{\ell m}_A &:= \D{A}Y_{\ell m},\\
X^{\ell m}_A &:= -\epsilon_A{}^C\D{C}Y_{\ell m},
\end{align}
\end{subequations}
and the tensor harmonics, defined for $\ell\geq2$, are given by%
\begin{subequations}\label{tensor harmonics def}%
\begin{align}
Y^{\ell m}_{AB} &:= \left[\D{A}\D{B}+\frac{1}{2} \ell(\ell+1)\Omega_{AB}\right]Y_{\ell m},\\
X^{\ell m}_{AB} &:= -\epsilon_{(A}{}^C\D{B)}\D{C}Y_{\ell m}.
\end{align}
\end{subequations}
In the formula for $Y^{\ell m}_{AB}$, we have used the eigenvalue equation 
\beq
\D{A}D^AY_{\ell m}=- \ell(\ell+1)Y_{\ell m}.
\eeq

By construction, the tensor harmonics are trace-free:
\begin{align}\label{MP tracefree}
\Omega^{AB}Y^{\ell m}_{AB} = 0 = \Omega^{AB}X^{\ell m}_{AB}.
\end{align}
From Eq.~\eqref{tensor vs spin-weighted harmonics}, they are related to spin-weighted harmonics as%
\begin{subequations}\label{MP to sYlm}%
\begin{align}
Y^{\ell m}_{A} &=\frac{\lam{\ell}{1}}{2}
							\left({}_{-1}Y_{\ell m}\tilde m_A-{}_1Y_{\ell m}\tilde m^*_A\right),\\
X^{\ell m}_{A} &= -\frac{\lam{\ell}{1}}{2}i
							\left({}_{-1}Y_{\ell m}\tilde m_A+{}_1Y_{\ell m}\tilde m^*_A\right),\\
Y^{\ell m}_{AB} &= \frac{\lam{\ell}{2}}{4}
							\left({}_{-2}Y_{\ell m}\tilde m_A \tilde m_B+{}_2Y_{\ell m}\tilde m^*_A \tilde m^*_B\right),\\
X^{\ell m}_{AB} &= -\frac{\lam{\ell}{2}}{4}i
							\left({}_{-2}Y_{\ell m}\tilde m_A \tilde m_B-{}_2Y_{\ell m}\tilde m^*_A \tilde m^*_B\right).
\end{align}
\end{subequations}
They are orthogonal with respect to the natural inner product on $S^2$, but they are not orthonormal:%
\begin{subequations}\label{MP orthogonality}%
\begin{align}
\oint Y_{\ell m}^{A*}Y^{\ell' m'}_A d\Omega &=  \lam{\ell}{1}^2\delta_{\ell\ell'}\delta_{mm'},\\
\oint X_{\ell m}^{A*}X^{\ell' m'}_A d\Omega &=  \lam{\ell}{1}^2\delta_{\ell\ell'}\delta_{mm'},\\
\oint Y_{\ell m}^{AB*}Y^{\ell' m'}_{AB} d\Omega &= \frac{\lam{\ell}{2}^2}{2}\delta_{\ell\ell'}\delta_{mm'},\\
\oint X_{\ell m}^{AB*}X^{\ell' m'}_{AB} d\Omega &= \frac{\lam{\ell}{2}^2}{2}\delta_{\ell\ell'}\delta_{mm'},
\end{align}
\end{subequations}
and
\beq
\oint X_{\ell m}^{A*}Y^{\ell' m'}_A d\Omega = 0 = \oint X_{\ell m}^{AB*}Y^{\ell' m'}_{AB} d\Omega.\\
\eeq

\subsection{Harmonic expansions}

In terms of the MP harmonics, any symmetric tensor $v_{\mu\nu}$ can be expanded as%
\begin{subequations}\label{harmonic-expansion-generic}%
\begin{align}
v_{ab} &= \sum_{\ell m}v^{\ell m}_{ab}Y_{\ell m},\\
v_{aA} &= \sum_{\ell m}\left(v^{\ell m}_{a+}Y_A^{\ell m}+v^{\ell m}_{a-}X_A^{\ell m}\right)\!,\\
\!\!v_{AB} &= \sum_{\ell m}\left(v_\trAB^{\ell m}\Omega_{AB}Y^{\ell m}+v^{\ell m}_{+}Y_{AB}^{\ell m}+v^{\ell m}_{-}X_{AB}^{\ell m}\right)\!,\!\!
\end{align}
\end{subequations}
where the coefficients are functions of $x^a$. Here and throughout this paper, sums range over all allowed values of $\ell$ and $m$.  If $v_{\mu\nu}$ is real-valued, then all of its harmonic coefficients satisfy 
\beq\label{conjugation}
v^{\ell,-m}_{\boldsymbol{\cdot}}=(-1)^m v^{\ell m*}_{\boldsymbol{\cdot}}.
\eeq
Here and below we use the following shorthand:
\begin{definition*}\label{dot}
A dot, as in $v^{\ell m}_{\boldsymbol{\cdot}}$, is used to denote a generic tensor-harmonic coefficient, in this case any of $v^{\ell m}_{ab}$, $v^{\ell m}_{a\pm}$, $v^{\ell m}_{\pm}$, or $v^{\ell m}_{\trAB}$.
\end{definition*}

Our convention in Eq.~\eqref{harmonic-expansion-generic} differs slightly from that of MP, who followed tradition~\cite{Regge-Wheeler:57} by introducing a factor of $r^2$ in front of $v^{\ell m}_\trAB$ and $v^{\ell m}_{+}$. Our notation also differs from tradition in that we  uniformly use a ``$+$'' sign to denote the coefficient of an even-parity vector or tensor harmonic, a ``$-$'' sign to denote the coefficient of an odd-parity vector or tensor harmonic, and a ``$\trAB$'' to denote one-half the angular trace of a tensor.

Using the orthogonality of the harmonics, each of the coefficients in Eq.~\eqref{harmonic-expansion-generic} can be written as an integral against the appropriate harmonic:
\begingroup%
\allowdisplaybreaks%
\begin{subequations}\label{coefficient integrals}%
\begin{align}
v^{lm}_{ab} &= \oint v_{ab}Y^*_{lm}d\Omega,\\
v^{lm}_{a+} &= \frac{1}{\lam{\ell}{1}^2}\oint v_{aA}Y^{A*}_{lm}d\Omega,\\
v^{lm}_{a-} &= \frac{1}{\lam{\ell}{1}^2}\oint v_{aA}X^{A*}_{lm}d\Omega,\\
v^{lm}_{\trAB} &= \frac{1}{2}\oint v_{AB}\Omega^{AB}Y^*_{lm}d\Omega,\\
v^{lm}_+ &= \frac{2}{\lam{\ell}{2}^2}\oint v_{AB}Y^{AB*}_{lm}d\Omega,\\
v^{lm}_- &= \frac{2}{\lam{\ell}{2}^2}\oint v_{AB}X^{AB*}_{lm}d\Omega.
\end{align}
\end{subequations}
\endgroup%
To facilitate use of Eq.~\eqref{coupling}, in practice we express these as integrals against spin-weighted harmonics using the relations~\eqref{MP to sYlm}.

These expansions in tensor harmonics are covariant; they do not depend on any choice of basis vectors on $S^2$. If we adopt the null basis $\{m^A,m^{A*}\}$ on $S^2$, then components of a symmetric tensor $v_{\mu\nu}$ can instead be expanded in spin-weighted harmonics according to%
\begingroup
\allowdisplaybreaks
\begin{subequations}
\begin{align}
v_{ab} &= \sum_{\ell m} v^{\ell m}_{ab}Y_{\ell m},\\
v_{am} &= \sum_{\ell m} v^{\ell m}_{am}\,{}_{1}Y_{\ell m},\\
v_{am^*} &= \sum_{\ell m} v^{\ell m}_{am^*}\,{}_{-1}Y_{\ell m},\\
v_{mm} &= \sum_{\ell m} v^{\ell m}_{mm}\,{}_{2}Y_{\ell m},\\
v_{m^*m^*} &= \sum_{\ell m} v^{\ell m}_{m^*m^*}\,{}_{-2}Y_{\ell m},\\
v_{mm^*} &= \sum_{\ell m} v^{\ell m}_{mm^*}Y_{\ell m}.
\end{align}
\end{subequations}
\endgroup
In these expansions, the spin weights are carried by the harmonics; the coefficients have spin weight 0. 

If $v_{\mu\nu}$ is real, then $v_{am^*}=(v_{am})^*$ and $v_{m^*m^*}=(v_{mm})^*$.  Together with Eq.~\eqref{sYlm-conj}, this implies
\begin{subequations}
\begin{align}
v^{\ell,-m}_{am} &= -(-1)^m (v^{\ell m}_{am^*})^*,\\
v^{\ell,- m}_{mm} &= (-1)^m (v^{\ell m}_{m^*m^*})^*.
\end{align}
\end{subequations}
More generally, the modes of a spin-weight $s$ scalar, $v = \sum_{\ell m}v_{\ell m}\,{}_s Y_{\ell m}$, are related to the modes of its complex conjugate, $v^* = \sum_{\ell m}v^*_{\ell m}\,{}_{-s} Y_{\ell m}$, by
\beq\label{v conjugate spin-weighted}
v^*_{\ell m} = (-1)^{m+s}(v^{\ell,-m})^*.
\eeq

The coefficients in the spin-weighted harmonic decomposition are easily related to those in Eq.~\eqref{harmonic-expansion-generic}:
\begin{subequations}\label{tetrad coeffs from tensor coeffs}
\begin{align}
v^{\ell m}_{am} &= -\frac{\lam{\ell}{1}}{\sqrt{2}r} (v^{\ell m}_{a+} + i v^{\ell m}_{a-}),\\
v^{\ell m}_{am^*} &= \frac{\lam{\ell}{1}}{\sqrt{2}r} (v^{\ell m}_{a+} - i v^{\ell m}_{a-}),\\
v^{\ell m}_{mm} &= \frac{\lam{\ell}{2}}{2r^2} (v^{\ell m}_+ + i v^{\ell m}_-),\\
v^{\ell m}_{m^*m^*} &= \frac{\lam{\ell}{2}}{2r^2} (v^{\ell m}_+ - i v^{\ell m}_-),\\
v^{\ell m}_{mm^*} &= \frac{1}{r^2}v^{\ell m}_\trAB.
\end{align}
\end{subequations}

We conclude with the explicit expansion of our main quantity of interest: the metric perturbation. Its expansion reads
\begin{subequations}\label{h-Yilm}
\begin{align}
h_{ab} &= \sum_{\ell m}h^{\ell m}_{ab}Y^{\ell m},\\
h_{aA} &= \sum_{\ell m}\left(h^{\ell m}_{a+}Y_A^{\ell m}+h^{\ell m}_{a-}X_A^{\ell m}\right)\!,\\
h_{AB} &= \sum_{\ell m}\left(h_\trAB^{\ell m}\Omega_{AB}Y^{\ell m}+h^{\ell m}_{+}Y_{AB}^{\ell m}+h^{\ell m}_{-}X_{AB}^{\ell m}\right)\!.
\end{align}
\end{subequations}
We will likewise write the decomposition of a generic source term in the Einstein equations~\eqref{EFEn dG} as
\begin{subequations}\label{S-Yilm}
\begin{align}
S_{ab} &= \sum_{\ell m}S^{\ell m}_{ab}Y^{\ell m},\\
S_{aA} &= \sum_{\ell m}\left(S^{\ell m}_{a+}Y_A^{\ell m}+S^{\ell m}_{a-}X_A^{\ell m}\right)\!,\\
S_{AB} &= \sum_{\ell m}\left(S_\trAB^{\ell m}\Omega_{AB}Y^{\ell m}+S^{\ell m}_{+}Y_{AB}^{\ell m}+S^{\ell m}_{-}X_{AB}^{\ell m}\right)\!.
\end{align}
\end{subequations}
The field equations also often involve the trace-reversal of these fields, $\bar h_{\mu\nu}$ and $\bar S_{\mu\nu}$. To facilitate trace reversals at the level of harmonic coefficients, for a generic field $v_{\mu\nu}$ we introduce 
\beq
v^{\ell m}_\trab := \frac{1}{2}g^{ab}v^{\ell m}_{ab}
\eeq
in analogy with $v^{\ell m}_\trAB$. The coefficients in the tensor-harmonic expansion of $\bar v_{\mu\nu}:=v_{\mu\nu}-\frac{1}{2}g_{\mu\nu}g^{\alpha\beta}v_{\alpha\beta}$ are then related to those in the expansion of $v_{\mu\nu}$ by%
\begin{subequations}\label{trace reverse modes}%
\begin{align}
\bar v^{\ell m}_{ab} &= v^{\ell m}_{ab} - g_{ab}(v^{\ell m}_\trab + r^{-2}v^{\ell m}_\trAB),\\
\bar v^{\ell m}_\trAB &= - r^2 v^{\ell m}_\trab,\\
\bar v^{\ell m}_{a\pm} &= v^{\ell m}_{a\pm} \quad\text{and} \quad \bar v^{\ell m}_{\pm} = v^{\ell m}_{\pm}.
\end{align}
\end{subequations}
If instead we expand $v_{\mu\nu}$ and $\bar v_{\mu\nu}$ in spin-weighted harmonics, then
\beq
\bar v^{\ell m}_{ln} = v^{\ell m}_{mm^*}\quad\text{and}\quad \bar v^{\ell m}_{mm^*} = v^{\ell m}_{ln},
\eeq
and all other coefficients satisfy $\bar v^{\ell m}_{\boldsymbol{\cdot}}=v^{\ell m}_{\boldsymbol{\cdot}}$.

\section{Gauge transformations and invariant perturbations}\label{gauge}

In Ref.~\cite{Martel-Poisson:05}, MP wrote the first-order Einstein equation in terms of a set of gauge-invariant metric perturbations. Here we extend that approach to second order. In the accompanying \pkg{PerturbationEquations} package we also provide the second-order field equations in terms of the original, gauge-dependent perturbations. We discuss the relative merits of each approach at the end of the section.

\subsection{Gauge transformations of harmonic coefficients}

We first examine how tensor-harmonic coefficients transform under a change of gauge. That requires decomposing Eq.~\eqref{Deltah}, which in turn requires the decompositions of Lie derivatives. Consider the Lie derivative $\Lie_\xi v_{\mu\nu}$ of a symmetric tensor $v_{\mu\nu}$ along a vector $\xi^\mu=(\zeta^a,Z^A)$. It has components
\begin{subequations}\label{LieS}
\begin{align}
(\Lie_\xi v)_{ab} &=  l_\zeta v_{ab} +\varLie_Z v_{ab} + 2 v_{C(a} \delta_{b)}Z^C,\\
(\Lie_\xi v)_{aB} &=  l_\zeta v_{aB} +\varLie_Z v_{aB} + v_{BC} \delta_{a}Z^C\nonumber\\
						&\quad + v_{ac} \D{B}\zeta^c,\\
(\Lie_\xi v)_{AB} &=  l_\zeta v_{AB} +\varLie_Z v_{AB} + 2 v_{c(A} \D{B)}\zeta^c,
\end{align}
\end{subequations}
where $ l_\zeta$ is a Lie derivative on ${\cal M}^2$, and $\varLie_Z$ is a Lie derivative on $S^2$. We can use the decomposition~\eqref{LieS} to calculate the components of $\Delta h^{(1)}_{\mu\nu}$ straightforwardly from Eq.~\eqref{Deltah1}. Given that result, we may then use the decomposition~\eqref{LieS} a second time to calculate the components of $\Delta h^{(2)}_{\mu\nu}$, after rewriting Eq.~\eqref{Deltah2} as  
\beq
\Delta h^{(2)}_{\mu\nu}=\Lie_{\xi_{(2)}}g_{\mu\nu}+H_{\mu\nu}\label{Dh2}
\eeq
with
\beq
H_{\mu\nu} := \Lie_{\xi_{(1)}}\!\!\left(h^{(1)}_{\mu\nu}+\tfrac{1}{2}\Delta h^{(1)}_{\mu\nu}\right).\label{H}
\eeq
To obtain the harmonic expansion of the result, we expand $\xi^\mu_{(n)}$ in vector harmonics as
\begin{subequations}\label{xilm}
\begin{align}
\zeta^a_{(n)} &= \sum_{\ell m}\zeta^a_{(n)\ell m}Y_{\ell m},\\
Z^A_{(n)} &= \sum_{\ell m}(Z^+_{(n)\ell m}Y^A_{\ell m}+Z^-_{(n)\ell m}X^A_{\ell m}),
\end{align}
\end{subequations}
where, recall, $Y^A_{\ell m}:=\Omega^{AB}Y^{\ell m}_B$ and $X^A_{\ell m}:=\Omega^{AB}X^{\ell m}_B$.

For the first-order transformation $\Delta h^{(1)}_{\mu\nu}={\cal L}_{\xi_{(1)}}g_{\mu\nu}$, Eq.~\eqref{LieS} reduces to
\begin{subequations}\label{Dh1}
\begin{align}
\Delta h^{(1)}_{ab} &=  2\delta_{(a}\zeta^{(1)}_{b)},\\
\Delta h^{(1)}_{aB} &=  r^2\delta_{a}Z^{(1)}_{B}+\D{B}\zeta^{(1)}_a,\\
\Delta h^{(1)}_{AB} &=  2r r_c\zeta^c_{(1)} \Omega_{AB} +2r^2\D{(A}Z^{(1)}_{B)}.
\end{align}
\end{subequations}
Substituting the harmonic expansion \eqref{xilm} into Eq.~\eqref{Dh1}, one finds%
\begingroup
\allowdisplaybreaks
\begin{subequations}\label{Dh1lm}
\begin{align}
\Delta h^{(1)\ell m}_{ab} &= 2\delta_{(a}\zeta^{(1)\ell m}_{b)},\label{Delta h1lmab}\\ 
\Delta h^{(1)\ell m}_{a+} &= r^2\delta_aZ^{(1)\ell m}_+ + \zeta_a^{(1)\ell m},\label{Delta h1lmap}\\
\Delta h^{(1)\ell m}_{a-} &= r^2\delta_aZ^{(1)\ell m}_-,\label{Delta h1lmam}\\
\Delta h^{(1)\ell m}_{\trAB} &= 2rr_c\zeta^c_{(1)\ell m}-\ell (\ell +1)r^2 Z^{(1)\ell m}_+,\label{Delta h1lmO}\\
\Delta h^{(1)\ell m}_{\pm} &= 2r^2 Z^{(1)\ell m}_\pm.\label{Delta h1lmpm}
\end{align}
\end{subequations}
\endgroup

In the same way, the harmonic decomposition of Eq.~\eqref{Dh2} reads%
\begingroup
\allowdisplaybreaks
\begin{subequations}\label{Dh2lm}
\begin{align}
\Delta  h^{(2)\ell m}_{ab} &= 2\delta_{(a}\zeta^{(2)\ell m}_{b)} + H^{\ell m}_{ab},\label{Delta h2lmab}\\
\Delta h^{(2)lm}_{a+} &=  r^2\delta_aZ^{(2)\ell m}_+ + \zeta_a^{(2)\ell m}+ H^{\ell m}_{a+},\label{Delta h2lmap}\\
\Delta h^{(2)lm}_{a-} &=  r^2\delta_aZ^{(2)\ell m}_- + H^{\ell m}_{a-},\label{Delta h2lmam}\\
\Delta h^{(2)lm}_{\trAB} &= 2rr_c\zeta^c_{(2)\ell m}-\ell (\ell +1)r^2 Z^{(2)\ell m}_+ \nonumber\\
&\quad + H^{\ell m}_\trAB,\label{Delta h2lmO}\\
\Delta h^{(2)lm}_{\pm} &= 2r^2 Z^{(2)\ell m}_\pm + H^{\ell m}_{\pm}.\label{Delta h2lmpm}
\end{align}
\end{subequations}
\endgroup
The mode decompositions of the quadratic quantity $H_{\alpha\beta}$ is obtained through the following steps: 
\begin{enumerate}
\item Write Eq.~\eqref{H} in $2+2$D form using Eq.~\eqref{LieS}
\item Substitute the harmonic expansions of $h_{\alpha\beta}^{(1)}$ and $\xi^\alpha_{(1)}$
\item Use Eqs.~\eqref{vector harmonics def}, \eqref{tensor harmonics def}, \eqref{DY}, \eqref{DDY}, \eqref{DDDY}, and \eqref{DDDDY} to express tensor harmonics and their covariant derivatives as sums of spin-weighted harmonics
\item Use Eqs.~\eqref{coefficient integrals} to pick out the tensor-harmonic coefficients of the result 
\item Use Eq.~\eqref{Cdef} to express the result in terms of the constants $C^{\ell m s}_{\ell'm's'\ell''m''s''}$ 
\item Use the symmetries \eqref{C symmetries} and relabel summation indices to minimize the number of constants $C^{\ell m s}_{\ell'm's'\ell''m''s''}$. 
\end{enumerate}
This gives us the final expressions %
\begingroup
\begin{subequations}\label{H decomposition}
\begin{align}
H^{\ell m}_{ab} &= \sum_{\substack{\ell'm'\ell''m''\\s'=0,1}}\!\!\lam{\ell^{\prime}}{s'} \lam{\ell^{\prime \prime}}{s'}\, C^{\ell m 0}_{\ell'm's'\ell''m''-s'}\nonumber\\[-15pt] &\qquad\qquad\qquad\qquad\qquad\times H^{\ell'm's'\ell''m''-s'}_{ab},\\[5pt]
H^{\ell m}_{a\pm} &= \sum_{\substack{\ell'm'\ell''m''\\s'=1,2}}\!\! \frac{\lam{\ell^{\prime}}{s'} \lam{\ell^{\prime \prime}}{1-s'}}{\lam{\ell}{1}}\, C^{\ell m 1}_{\ell'm's'\ell''m'',1-s'}\nonumber\\[-15pt] &\qquad\qquad\qquad\qquad\qquad\times H^{\ell'm's'\ell''m'',1-s'}_{a\pm},\\[5pt]
H^{\ell m}_\trAB &= \sum_{\substack{\ell'm'\ell''m''\\s'=0,1,2}}\!\!\lam{\ell^{\prime}}{s'} \lam{\ell^{\prime \prime}}{s'}\,  C^{\ell m0}_{\ell'm' s'\ell''m''-s'}\nonumber\\[-15pt] &\qquad\qquad\qquad\qquad\qquad\times H^{\ell'm's'\ell''m''-s'}_\trAB,\\[5pt]
H^{\ell m}_{\pm} &= \sum_{\substack{\ell'm'\ell''m''\\s'=1,2,3}}\!\!\frac{\lam{\ell^{\prime}}{s'} \lam{\ell^{\prime \prime}}{2-s'}}{\lam{\ell}{2}}\,C^{\ell m 2}_{\ell'm' s'\ell''m'',2 -s'}\nonumber\\[-15pt] &\qquad\qquad\qquad\qquad\qquad\times H^{\ell'm' s'\ell''m'',2 -s'}_\pm,
\end{align}
\end{subequations}
\endgroup
where the quantities $H^{\ell'm's'\ell''m''s''}_{\boldsymbol{\cdot}}$ are made up of quadratic products of first-order mode coefficients $h^{(1)\ell' m'}_{\boldsymbol{\cdot}}$, $h^{(1)\ell'' m''}_{\boldsymbol{\cdot}}$, $\zeta^a_{(1)\ell' m'}$, $\zeta^a_{(1)\ell'' m''}$, $Z^\pm_{(1)\ell' m'}$, and $Z^\pm_{(1)\ell'' m''}$. We give those products, which we refer to as coupling functions, explicitly in Eq.~\eqref{H coupling functions}.

Equation~\eqref{H decomposition} has the appearance of a quintuple sum 
\beq
\sum_{\ell'=0}^\infty\ \sum_{\ell''=0}^\infty\ \sum_{m'=-\ell'}^{+\ell'}\ \sum_{m''=-\ell''}^{+\ell''}\ \sum_{s'=0}^{s'_{\rm max}}.
\eeq
However, these summation ranges are restricted by the factors $C^{\ell m s}_{\ell'm' s'\ell''m''s''}$, which enforce the conditions in Eq.~\eqref{Crules}.
The definitions of tensor harmonics also automatically enforce $\ell'\geq s'$ and we additionally require $s=s'+s''$. Together these restrictions reduce the sums to
\beq\label{restricted sums}
\sum_{s'=0}^{s'_{\rm max}}\ \sum_{\ell'=s'}^\infty\ \sum_{\substack{\ell''=\,\text{max}(|s-s'|,\\ \qquad\qquad|\ell-\ell'|)}}^{\ell+\ell'}\ \sum_{m'=-\ell'}^{+\ell'}\delta_{m'',m-m'}.
\eeq

\subsection{Common gauge choices}

There are several common gauge conditions in Schwarzschild spacetime. These include the RWZ gauge~\cite{Regge-Wheeler:57,Zerilli:70}, the ingoing and outgoing radiation gauges (IRG and ORG), and the Lorenz gauge:
\begin{subequations}%
\begin{align}
\text{RWZ:} &\qquad h_{mm}=0=(h_{am} - h_{am^*}).\\
\text{IRG:} &\qquad h_{\alpha\beta}l^\beta=0=g^{\alpha\beta}h_{\alpha\beta}.\\
\text{ORG:} &\qquad h_{\alpha\beta}n^\beta=0=g^{\alpha\beta}h_{\alpha\beta}.\\
\text{Lorenz:} &\qquad \nabla^\beta \bar h_{\alpha\beta} = 0.
\end{align}
\end{subequations}
Here and below, we do not distinguish between $h^{(1)}_{\alpha\beta}$ and $h^{(2)}_{\alpha\beta}$. The gauge conditions can be applied to either or both of them, and one can ``mix and match'' by adopting a different condition for $h^{(2)}_{\alpha\beta}$ than for $h^{(1)}_{\alpha\beta}$. 

The RWZ gauge is the most common because it greatly simplifies the field equations. Following MP, we will make extensive use of it in the sections below.

The radiation gauges, which reduce to so-called lightcone gauges~\cite{Poisson:05,Poisson-Vlasov:09} for specific choices of null basis, are particularly useful for studying ingoing or outgoing radiation. If the Kinnersley tetrad is used, then the IRG condition ensures that radially outgoing null cones have the same coordinate description in the perturbed spacetime as in the background spacetime: surfaces of constant retarded time $u$ are null cones, and $r$ is an affine parameter along the generators of the cones. If the Hartle--Hawking tetrad is used, then the ORG condition ensures that the analogous statements apply to ingoing null cones.\footnote{The traditional names and geometrical features of the radiation gauge conditions may appear antithetical: the {\em outgoing} radiation gauge preserves {\em ingoing} null cones, which should make it ideal for studying ingoing waves, while the {\em ingoing} radiation gauge preserves {\em outgoing} null cones, which should make it ideal for studying outgoing waves. This clash stems from the particular metric reconstruction method traditionally used to obtain the metric perturbation in these gauges, reviewed in Sec.~\ref{Teukolsky} below. Despite the geometrical features of the gauge conditions, the reconstruction method yields metric perturbations that match the gauges' names: outgoing (ingoing) radiation is asymptotically regular in the ORG (IRG), while ingoing (outgoing) radiation is asymptotically irregular in the ORG (IRG)~\cite{Shah-etal:11}.} The radiation gauges also ensure that $r$ remains an areal radius in the perturbed spacetime. This follows from the fact that  if $h_{\alpha\beta}l^\beta=0$ or $h_{\alpha\beta}n^\beta=0$, then the traceless condition $g^{\alpha\beta}h_{\alpha\beta}=0$ is equivalent to $\Omega^{AB}h_{AB}=0$; since $\Omega^{AB}h_{AB}$ is proportional to the perturbation of the area element on the sphere of constant $r$, the surface area of the sphere remains $4\pi r^2$. 

Finally, the Lorenz gauge condition is useful for putting the perturbative field equations in the symmetric hyperbolic form~\eqref{EFE1 Lorenz}--\eqref{EFE2 Lorenz}.

In mode-decomposed form, the gauge conditions become
\begin{subequations}%
\begin{align}
\text{RWZ:} &\quad\ \;  h^{\ell m}_{\pm} = h^{\ell m}_{a+} = 0.\label{RWZ condition}\\
\text{IRG:} &\ \begin{cases}\ h^{\ell m}_{la} = h^{\ell m}_{l\pm} = 0,\\[.1em]
				\ h^{\ell m}_{\trAB} = 0.\end{cases}\\
\text{ORG:} &\ \begin{cases}\ h^{\ell m}_{na} = h^{\ell m}_{n\pm} = 0,\\[.1em]
				\ h^{\ell m}_{\trAB} = 0.\end{cases}\\
\text{Lorenz:} &\ \begin{cases}\ \delta^b\bar h^{\ell m}_{ab} = \displaystyle\frac{1}{r^3}\!\left(\lam{\ell}{1}^2 r \bar h^{\ell m}_{a+}+2r_a\bar h^{\ell m}_\trAB-2r^2r^b\bar h^{\ell m}_{ab}\right)\!,\\[.65em]
\ \delta^b\bar h^{\ell m}_{b+} = \displaystyle \frac{1}{2r^2}\!\left(\mu^2_\ell \bar h^{\ell m}_+ -4rr^b \bar h^{\ell m}_{b+}-2\bar h^{\ell m}_\trAB\right)\!,\\[.65em]
\ \delta^b\bar h^{\ell m}_{b-} = \displaystyle\frac{1}{2r^2}\!\left(\mu^2_\ell \bar h^{\ell m}_- - 4rr^b \bar h^{\ell m}_{b-}\right)\!.
\end{cases}
\end{align}
\end{subequations}
where $\mu_\ell$ is defined in Eq.~\eqref{mu def}. Note that MP denote this same quantity as $\mu$, meaning $\mu_{\rm MP} = \mu^2_\ell$.

All four conditions leave residual gauge freedom, meaning in each case one can find gauge perturbations $\del{(\alpha}\xi_{\beta)}$ that satisfy the relevant gauge conditions. Specifically, the RWZ gauge condition does not constrain $\ell=0$ modes or the $\ell=1$ mode $h^{1 m}_{a-}$; the radiation gauges can be altered by a gauge vector satisfying $l^\beta\del{(\alpha}\xi_{\beta)}=0$ (IRG) or $n^\beta\del{(\alpha}\xi_{\beta)}=0$ (ORG) and $\del\alpha\xi^\alpha=0$; and the Lorenz gauge can be altered by a gauge vector satisfying $\nabla^\beta \overline{\del{(\alpha}\xi_{\beta)}}=0$ (which is equivalent to $\Box\xi^\alpha=0$). In the next sections, we will specifically analyse (and remove) the gauge freedom in the $\ell=0,1$ modes.

\subsection{Gauge-fixing procedure and residual gauge freedom}\label{gauge fixing}

It is common in perturbation theory to construct gauge-invariant metric perturbations using a gauge-fixing procedure; see, for example, Refs.~\cite{Nakamura:2003wk,Nakamura:2004wr,Nakamura:2006rk,Nakamura:07,Martel-Poisson:05,Brizuela:2007zza,Brizuela-etal:09,Merlin:2016boc} and Nakamura's recent series of papers exploring this method in Schwarzschild spacetime~\cite{Nakamura:2021mfv,Nakamura:2021ftr,Nakamura:2021hki,Nakamura:2021onm,Nakamura:2021jlc}. The idea is to identify gauge conditions that completely fix the gauge, leaving no residual gauge freedom. The gauge vectors, call them $\tilde \xi^\mu_{(n)}$, that transform from a generic gauge to the fixed gauge are then determined by the perturbations in the generic gauge, call them $h^{(n)}_{\mu\nu}$. Referring to the transformation rule~\eqref{Deltah}, we can then construct gauge-invariant perturbations $\widetilde{h}^{(n)}_{\mu\nu}$ that are simply the metric perturbations in the fixed gauge expressed in terms of the perturbations $h^{(n)}_{\mu\nu}$ in the generic gauge:
\begin{subequations}\label{tilde h}%
\begin{align}
\widetilde{h}^{(1)}_{\mu\nu} &= h^{(1)}_{\mu\nu} + \Lie_{\tilde\xi_{(1)}}g_{\mu\nu},\\
\widetilde{h}^{(2)}_{\mu\nu} &= h^{(2)}_{\mu\nu} + \Lie_{\tilde\xi_{(2)}}g_{\mu\nu} + \widetilde{H}_{\mu\nu},
\end{align}
\end{subequations}
where
\begin{align}
\widetilde{H}_{\mu\nu} &:= \Lie_{\tilde \xi_{(1)}}\!\!\left(h^{(1)}_{\mu\nu}+\tfrac{1}{2}\Lie_{\tilde \xi_{(1)}}g_{\mu\nu}\right).
\end{align}
Analogously, referring to the transformation rule~\eqref{DeltaT}, we construct invariant stress-energy perturbations,
\begin{subequations}\label{tilde T}%
\begin{align}
\widetilde{T}^{(1)}_{\mu\nu} &= T^{(1)}_{\mu\nu},\\
\widetilde{T}^{(2)}_{\mu\nu} &= T^{(2)}_{\mu\nu} +\Lie_{\tilde \xi_{(1)}}T^{(1)}_{\mu\nu}.\label{tilde T2}
\end{align}
\end{subequations}
The most obvious example of such invariants are the variables used by MP, which use the RWZ gauge conditions to specify $\widetilde{h}^{(1)}_{\mu\nu}$. We review those RWZ-based invariant variables in Sec.~\ref{gauge fixing ell>1} below.

To our knowledge, this procedure has always specified the fixed gauge through conditions on $\widetilde{h}^{(n)}_{\mu\nu}$, which then determines $\tilde\xi^\mu_{(1)}=\tilde\xi^\mu_{(1)}[h^{(1)}]$ and $\tilde\xi^\mu_{(2)}=\tilde\xi^\mu_{(2)}[h^{(1)},h^{(2)}]$ via Eq.~\eqref{tilde h}. If $h^{(n)}_{\mu\nu}$ happens to already be in the fixed gauge, then $\tilde\xi^\mu_{(n)}=0$ and $\widetilde{h}^{(n)}_{\mu\nu}=h^{(n)}_{\mu\nu}$. But if $h^{(n)}_{\mu\nu}$ is in any other gauge, then the quantities $\widetilde{h}^{(n)}_{\mu\nu}$ are invariants constructed from $h^{(n)}_{\mu\nu}$; no matter the choice of gauge used to calculate $h^{(n)}_{\mu\nu}$, $\widetilde{h}^{(n)}_{\mu\nu}$ take the value of the perturbations in the fixed gauge. 

However, such a procedure is necessarily incomplete because conditions on the metric perturbation cannot fully specify the gauge. This is because of the Killing symmetries of the background. If $\xi^\mu_{(1)}$ is a Killing vector of the background, then the gauge transformation $\Delta h^{(1)}_{\mu\nu}=\Lie_{\xi_{(1)}}g_{\mu\nu}$ vanishes. This means any gauge condition on $\widetilde{h}^{(1)}_{\mu\nu}$ can only fix the gauge up to infinitesimal isometries of the background. 

In linear perturbation theory, this incomplete gauge fixing is not problematic. Since the $\ell m$ modes decouple from each other, one can fully fix the $\ell>1$ gauge freedom through conditions on the $\ell>1$ pieces of $\widetilde{h}^{(1)}_{\mu\nu}$. The gauge ambiguity is then confined to the $\ell=0$ and $\ell=1$ perturbations. Those perturbations are very often simply ignored because in vacuum they are only perturbations toward another stationary black hole solution (specifically, a linear-in-spin Kerr solution).

However, at second order the residual gauge ambiguity does manifest itself in the metric perturbation. If $\xi^\mu_{(1)}$ is a Killing vector of the background, then it induces a nontrivial transformation $\Delta h^{(2)}_{\mu\nu} = \Lie_{\xi_{(1)}}h^{(1)}_{\mu\nu}$. This implies that if $\tilde \xi^\mu_{(1)}$ is only determined up to the addition of a background Killing vector, then $\widetilde h^{(2)}_{\mu\nu}$ is not invariant.

Appendix~\ref{transformation of xi tilde} analyzes the general transformation properties of $\tilde\xi^\mu_{(n)}$ and $\widetilde h^{(n)}_{\mu\nu}$ and the implications of residual gauge freedom. In the body of the paper, we outline a specific type of gauge-fixing prescription that eliminates the residual freedom. Our prescription differs from others by enforcing a condition on $\widetilde{T}^{(2)}_{\mu\nu}$; through Eq.~\eqref{tilde T}, this imposes additional conditions on $\tilde \xi^\mu_{(1)}$. We then obtain vectors $\tilde\xi^\mu_{(1)}=\tilde\xi^\mu_{(1)}[h^{(1)},T^{(2)}]$ and $\tilde\xi^\mu_{(2)}=\tilde\xi^\mu_{(2)}[h^{(1)},h^{(2)}]$ and fully invariant perturbations $\widetilde h^{(n)}_{\mu\nu}$ and $\widetilde{T}^{(n)}_{\mu\nu}$. This restricts our prescription to nonvacuum perturbations. In global vacuum, without a matter distribution to refer to, fixing the residual gauge freedom would require specifying a value of time and angular position at some physically identifiable event in the perturbed spacetime.

We detail a particular gauge-fixing scheme in the remainder of this section. Our procedure for  $\ell>1$ follows tradition, while our procedure for $\ell=0$ and $\ell=1$ appears here for the first time.

\subsection{Gauge fixing for $\ell>1$}\label{gauge fixing ell>1}

We follow MP and Brizuela et al.~\cite{Brizuela-etal:09} by putting the $\ell>1$ pieces of $\widetilde{h}^{(n)}_{\mu\nu}$ in the RWZ gauge, setting 
\beq
\widetilde{h}^{(n)\ell m}_{\pm}=0=\widetilde{h}^{(n)\ell m}_{a+}. 
\eeq

At first order, the analog of Eqs.~\eqref{Delta h1lmap} and \eqref{Delta h1lmpm} then imply that the vector $\tilde\xi^\mu_{(1)}$ has $\ell>1$ modes given by%
\begin{subequations}%
\begin{align}
\tilde \zeta^{(1)\ell m}_a &= -h^{(1)\ell m}_{a+}- r^2\delta_a\tilde Z^{(1)\ell m}_+,\\
\tilde Z^{(1)\ell m}_\pm &= - \frac{h^{(1)\ell m}_\pm}{2r^2}.
\end{align}
\end{subequations}
Substituting these formulas into the analogues of Eqs.~\eqref{Delta h1lmab}, \eqref{Delta h1lmam}, and \eqref{Delta h1lmO}, we find the nonzero $\ell>1$ pieces of $\widetilde{h}^{(1)}_{\mu\nu}$ are%
\begin{subequations}\label{h1tilde ell>1}%
\begin{align}
\widetilde{h}^{(1)\ell m}_{ab} &:= h^{(1)\ell m}_{ab} + 2\delta_{(a}\tilde\zeta^{(1)\ell m}_{b)},\\
\widetilde{h}^{(1)\ell m}_{a-} &:=  h^{(1)\ell m}_{a-} + r^2\delta_a\tilde Z^{(1)\ell m}_-,\\
\widetilde{h}^{(1)\ell m}_\trAB &:= h^{(1)\ell m}_\trAB +2rr_c\tilde\zeta^c_{(1)\ell m} \nonumber\\
&\quad\;  -\ell(\ell+1)r^2\tilde Z^{(1)\ell m}_+.
\end{align}
\end{subequations}
These are the equivalent of MP's Eqs.~(4.10), (4.11), and (5.7). The fields $\widetilde{h}^{(1)\ell m}_{\boldsymbol{\cdot}}$ are invariant regardless of what we do with $\ell=0,1$ modes.

At second order, the analogues of Eqs.~\eqref{Delta h2lmap} and \eqref{Delta h2lmpm} imply that the vector $\tilde\xi^\mu_{(2)}$ has $\ell>1$ modes given by
\begin{subequations}%
\begin{align}
\tilde \zeta^{(2)\ell m}_a &= - h^{(2)\ell m}_{a+} -\widetilde H^{\ell m}_{a+} - r^2\delta_a\tilde Z^{(2)\ell m}_+,\\
\tilde Z^{(2)\ell m}_\pm &= - \frac{h^{(2)\ell m}_\pm +\widetilde H^{\ell m}_\pm}{2r^2},
\end{align}
\end{subequations}
and the analogues of Eqs.~\eqref{Delta h2lmab}, \eqref{Delta h2lmam}, and \eqref{Delta h2lmO} imply that the nonzero $\ell>1$ modes of $\widetilde{h}^{(2)}_{\mu\nu}$ are
\begin{subequations}%
\begin{align}
\widetilde{h}^{(2)\ell m}_{ab} &:= h^{(2)\ell m}_{ab} +\widetilde H^{\ell m}_{ab} + 2\delta_{(a}\tilde\zeta^{(2)\ell m}_{b)},\\
\widetilde{h}^{(2)\ell m}_{a-} &:=  h^{(2)\ell m}_{a-} +\widetilde H^{\ell m}_{a-} + r^2\delta_a\tilde Z^{(2)\ell m}_-,\\
\widetilde{h}^{(2)\ell m}_\trAB &:= h^{(2)\ell m}_\trAB +\widetilde H^{\ell m}_{\trAB}  +2rr_c\tilde\zeta^c_{(2)\ell m} \nonumber\\
				&\quad\; -\ell(\ell+1)r^2\tilde Z^{(2)\ell m}_+.
\end{align}
\end{subequations}
As per the discussion in the preceding section and Appendix~\ref{transformation of xi tilde}, these fields are not yet invariants. They only become invariant once we fix the $\ell=0,1$ modes of the first-order field $\widetilde h^{(1)}_{\mu\nu}$.

\subsection{Gauge fixing for $\ell=0$}

For $\ell=0$, the only nonvanishing pieces of $h^{(n)}_{\mu\nu}$ are the scalar-harmonic modes $h^{(n)00}_{ab}$ and $h^{(n)00}_{\trAB}$. Equation~\eqref{tilde h} reduces to%
\begin{subequations}\label{ell=0 transformation}%
\begin{align}
\widetilde{h}^{(1)00}_{ab} &= {h}^{(1)00}_{ab}+\tilde\zeta^c_{(1)00}\partial_c g_{ab} + 2\partial_{(a}\tilde\zeta^c_{(1)00}g_{b)c},\\ 
\widetilde{h}^{(1)00}_{\trAB} &= {h}^{(1)00}_{\trAB}+2rr_c\tilde\zeta^c_{(1)00}\label{ell=0 trace transformation}
\end{align}
\end{subequations}
at first order. It reduces to the same equations at second order with the replacement ${h}^{(1)00}_{\boldsymbol{\cdot}}\to{h}^{(2)00}_{\boldsymbol{\cdot}}+\widetilde H^{00}_{\boldsymbol{\cdot}}$ on the right-hand side. Given this simple replacement, we only list results at first order in this and the next two sections below.

In these sections, we streamline the analysis by restricting ourselves to perturbations that are asymptotically flat at spatial infinity.\footnote{Lifting this restriction is straightforward. If perturbations are not asymptotically flat, or are in a gauge that does not manifest the asymptotic flatness, then integrals such as the one in Eq.~\eqref{zeta t tilde 100} become ill defined. One can then take the Hadamard finite part of such integrals~\cite{Blanchet-Damour:86}. As an example, suppose an integrand of the form $\sum_{j=0}^{k+1}a_j r^{k-j}+\gamma(t,r)$, where $\gamma(t,r)$ falls off faster than $1/r$. We can define the integral as $\sum_{j=0}^{k}\frac{a_j}{k-j+1} r^{k-j+1}+a_{k+1}\ln r+\int_{\infty}^r\!\!f^{-1}(r')\gamma(t,r') dr'$. Any length scale in the logarithm can be absorbed into $\varrho^{(1)}(t)$.} For the monopole mode, this implies
\begin{align}
h^{(n)00}_{ab} &= \frac{c^{(n)}_{ab}}{r}+\O(r^{-2}),\\
h^{(n)00}_{\trAB} &= r^2\left[\frac{c^{(n)}_{\trAB}}{r}+\O(r^{-2})\right]
\end{align}
for some $t$-independent constants $c^{(n)}_{ab}$ and $c^{(n)}_\trAB$. Note that this restricts the gauge of $h^{(n)}_{\mu\nu}$ in addition to restricting the asymptotic geometry. Moreover, we make the simplifying assumption that the gauge of $h^{(n)}_{\mu\nu}$ satisfies $c^{(n)}_{tr}=0$. 

We are not aware of any work that has constructed gauge invariants $\widetilde h^{(n)00}_{ab}$ and $\widetilde h^{(n)00}_{\trAB}$ that are local functions of $(t,r)$. We will instead allow one of the components to be a \emph{nonlocal} function that takes the form of a radial integral. However, at least at first order we are able to construct a local invariant from it through differentiation.

We adopt the following gauge-fixing conditions:%
\begin{subequations}\label{ell=0 gauge fixing}%
\begin{align}
\widetilde{h}^{(n)00}_\trAB &= 0,\label{ell=0 trace condition}\\
\widetilde{h}^{(n)00}_{tr} &= 0,\label{ell=0 tr condition}\\
\lim_{r\to\infty}\widetilde{h}^{(n)00}_{tt} &= 0,\label{ell=0 tt condition}
\end{align}
\end{subequations}
where the limit is taken at fixed $t$. 

Now, the trace condition~\eqref{ell=0 trace condition}, with Eq.~\eqref{ell=0 trace transformation}, fully determines $\tilde \zeta^r_{(1)00}$ to be
\begin{align}
\tilde \zeta^r_{(1)00} &= - \frac{1}{2r}h^{(1)00}_{\trAB}.
\end{align}
With Eq.~\eqref{ell=0 transformation}, this in turn determines
\begin{align}\label{h00rr tilde}
\widetilde{h}^{(1)00}_{rr} &= h^{(1)00}_{rr} +\frac{M}{r^3f^2}h^{(1)00}_{\trAB} - \frac{1}{f}\partial_{r}\left(\frac{1}{r}h^{(1)00}_{\trAB}\right).
\end{align}
The quantity $\widetilde{h}^{(1)00}_{rr}$ is invariant even without specifying the remaining component $\tilde\zeta^{t}_{(1)00}$. The quantity $\widetilde{h}^{(2)00}_{rr}$ is given by the same formula with $h^{(1)00}_{\boldsymbol{\cdot}}\to h^{(2)00}_{\boldsymbol{\cdot}}+H^{00}_{\boldsymbol{\cdot}}$, but it is not invariant until  $\tilde\zeta^{t}_{(1)00}$ is specified.

Next, the condition~\eqref{ell=0 tr condition}, with the $t$--$r$ component of Eq.~\eqref{ell=0 transformation}, restricts $\tilde\zeta^{t}_{(n)00}$ to be 
\begin{align}\label{zeta t tilde 100}
\tilde\zeta^t_{(1)00} &= \varrho^{(1)}(t) \nonumber\\
			&\quad +\int_{\infty}^r\!\!\left(f^{-1}{h}^{(1)00}_{tr} + f^{-2}\partial_{t}\tilde\zeta^r_{(1)00}\right)dr',
\end{align}
where $\varrho^{(1)}$ is an arbitrary function of $t$. 

The condition~\eqref{ell=0 tt condition}, with the $t$--$t$ component of Eq.~\eqref{ell=0 transformation}, then implies
\begin{align}
\varrho^{(1)} &= \varrho^{(1)}_0
\end{align}
for some constant $\varrho^{(1)}_0$. This represents a time translation; it is a timelike Killing vector of the background, which we cannot fix using conditions on $\widetilde h^{(1)}_{\mu\nu}$. To fix this remaining freedom, we examine the transformation of the stress-energy tensor. We defer that procedure to Sec.~\ref{residual Killing freedom}.

The remaining nonzero component of $\widetilde{h}^{(1)00}_{ab}$ is now a nonlocal invariant given by
\begin{align}\label{h00tt tilde}
\widetilde{h}^{(1)00}_{tt} &= {h}^{(1)00}_{tt}-\frac{2M}{r^2}\tilde\zeta^r_{(1)00} - 2f\partial_{t}\tilde\zeta^t_{(1)00}.
\end{align}
It is nonlocal because of the radial integral in $\tilde\zeta^t_{(1)00}$. However, we can immediately construct a local invariant from it: 
\beq
\varphi^{(1)}:=\partial_r\!\left(f^{-1}\widetilde{h}^{(1)00}_{tt}\right),
\eeq
or more explicitly,
\begin{multline}\label{ell=0 invariant}
\varphi^{(1)} = \partial_r(f^{-1}{h}^{(1)00}_{tt})-2f^{-1}\partial_{t}{h}^{(1)00}_{tr}\\
 +\partial_r\!\left(\frac{M}{r^3f}h^{(1)00}_\trAB\right) + \frac{1}{rf^{2}}\partial^2_{t}h^{(1)00}_\trAB.
\end{multline}

Our local invariants $\widetilde h^{(1)00}_{rr}$ and $\varphi^{(1)}$ are related to the quantities $\psi_0$ and $o_0$ in Ref.~\cite{Thompson:2016fxe} by $\widetilde h^{(1)00}_{rr}=2\psi_0$ and $\varphi^{(1)}=2f^{-1}o_0$.

At second order, the above formulas remain valid if we replace ${h}^{(1)00}_{\boldsymbol{\cdot}}$ with ${h}^{(2)00}_{\boldsymbol{\cdot}}+\widetilde H^{00}_{\boldsymbol{\cdot}}$. However, the invariant $\varphi^{(2)}$ is not manifestly local because $\widetilde H^{00}_{\boldsymbol{\cdot}}$ depends on the nonlocal quantity $\tilde \zeta^t_{(1)00}$. It might be possible to express $\widetilde H^{00}_{\boldsymbol{\cdot}}$ in terms of local quantities, or to construct alternative second-order invariants that are manifestly local, but we leave this for interested readers to explore.

\subsection{Gauge fixing for $\ell=1$: even parity}

For even-parity $\ell=1$ perturbations, the only nonvanishing pieces of $h^{(n)}_{\mu\nu}$ are the scalar- and vector-harmonic modes $h^{(n)1m}_{ab}$, $h^{(n)1m}_{\trAB}$, and $h^{(n)1m}_{a+}$. Equation~\eqref{tilde h} reduces to%
\begin{subequations}\label{ell=1 even-parity transformation}
\begin{align}
\widetilde{h}^{(1)1m}_{ab} &= {h}^{(1)1m}_{ab}+ \tilde\zeta^c_{(1)1m}\partial_c g_{ab} + 2\partial_{(a}\tilde\zeta^c_{(1)1m}g_{b)c},\\ 
\widetilde{h}^{(1)1m}_{a+} &= {h}^{(1)1m}_{a+}+\tilde\zeta_a^{(1)1m}+r^2\delta_a \tilde Z^{(1)1m}_+,\\
\widetilde{h}^{(1)1m}_{\trAB} &= {h}^{(1)1m}_{\trAB}+ 2rr_c\tilde\zeta^c_{(1)1m}-2r^2 \tilde Z^{(1)1m}_+,
\end{align}
\end{subequations}
at first order and to the same equations at second order with the replacement ${h}^{(1)1m}_{\boldsymbol{\cdot}}\to{h}^{(2)1m}_{\boldsymbol{\cdot}}+H^{1m}_{\boldsymbol{\cdot}}$. Our assumption of asymptotic flatness implies
\begin{subequations}
\begin{align}
h^{(n)1m}_{ab} &= \frac{c^{(n)m}_{ab}}{r^2}+\O(r^{-3}),\\
{h}^{(n)1m}_{a+} &= r\left[\frac{c^{(n)m}_{a+}}{r^2}+\O(r^{-3})\right],\\ 
h^{(n)1m}_{\trAB} &= r^2\left[\frac{c^{(n)m}_{\trAB}}{r^2}+\O(r^{-3})\right]
\end{align}
\end{subequations}
for $r\to\infty$ at fixed $t$, where $c^{(n)m}_{ab}$, $c^{(n)m}_{a+}$, and $c^{(n)m}_{\trAB}$ are constants.

A convenient set of gauge-fixing conditions are
\begin{subequations}
\begin{align}
\widetilde{h}^{(n)1m}_{a+} &= 0,\\
\widetilde{h}^{(n)1m}_\trAB &= 0,\\
\lim_{r\to\infty}\left(\widetilde{h}^{(n)1m}_{tt}\right) &= 0,\label{h1mtt falloff}\\ 
\lim_{r\to\infty}\left(r\widetilde{h}^{(n)1m}_{tr}\right) &= 0, \label{h1mtr falloff}\\
\lim_{r\to\infty}\left(r^2\widetilde{h}^{(n)1m}_{rr}\right)&=0. \label{h1mrr falloff}
\end{align}
\end{subequations}
The first of these determines $\tilde\zeta_a^{(1)1m}$ in terms of $\tilde Z^{(1)1m}_+$,
\beq
\tilde\zeta_a^{(1)1m} = - h^{(1)1m}_{a+} - r^2\delta_a \tilde Z^{(1)1m}_+,
\eeq
and the second determines a radial differential equation for $\tilde Z^{(1)1m}_+$,
\beq
2r^2\partial_r\!\left(rf\tilde Z^{(1)1m}_+\right) = {h}^{(1)1m}_{\trAB} - 2rf h^{(1)1m}_{r+}.
\eeq
The solution to this equation is
\beq
\tilde Z^{(1)1m}_+ = \frac{\kappa^{(1)m}(t)}{rf} +\frac{1}{rf}\int_\infty^r \left(\frac{{h}^{(1)1m}_{\trAB}}{2r'^2} - \frac{f  h^{(1)1m}_{r+}}{r'}\right)dr',
\eeq
where $\kappa$ is an arbitrary function of $t$.

$\kappa$ represents an asymptotic translation of the coordinate system. The condition~\eqref{h1mtt falloff} imposes $\partial^2_t\kappa=0$, which enforces that the fixed gauge is not asymptotically accelerating. The condition~\eqref{h1mtr falloff} imposes $\partial_t\kappa=0$; the fixed gauge is asymptotically stationary with respect to the asymptotic frame of $h^{(n)}_{\mu\nu}$. Finally, the condition~\eqref{h1mrr falloff} imposes 
\beq
\kappa^{(1)m} = \frac{1}{3M}\left(c^{(1)m}_\trAB-2c^{(1)m}_{r+}\right).
\eeq

The nonzero invariants are now
\begin{subequations}\label{h1mp tilde}
\begin{align}
\widetilde{h}^{(1)1m}_{tt} &= {h}^{(1)1m}_{tt} - \frac{2M}{r^2} \tilde\zeta^r_{(1)1m} - 2f\partial_t\tilde\zeta^t_{(1)1m},\\
\widetilde{h}^{(1)1m}_{tr} &= {h}^{(1)1m}_{tr} + f^{-1}\partial_t\tilde\zeta^r_{(1)1m}-f\partial_r\tilde\zeta^t_{(1)1m},\\
\widetilde{h}^{(1)1m}_{rr} &= {h}^{(1)1m}_{rr} - \frac{2M}{r^2f^2}\tilde\zeta^r_{(1)1m} + 2f^{-1}\partial_r\tilde\zeta^r_{(1)1m}.
\end{align}
\end{subequations}
Again these are nonlocal. They depend on a radial integral, and through $\kappa^{(1)m}$ they depend explicitly on the values $h^{(1)1m}_{\boldsymbol{\cdot}}$ at spatial infinity. But again we can construct a set of local invariants through differentiation:
\begin{subequations}
\begin{align}
\varphi^{(1)m}_{tt} &= \partial_r\!\left[r^4\partial_r\!\left(r^{-1} f\widetilde{h}^{(1)1m}_{tt}\right)\right]\!,\\
\varphi^{(1)m}_{tr} &= \partial_r\!\left(r f^2\widetilde{h}^{(1)1m}_{tr}\right)\!,\\
\varphi^{(1)m}_{rr} &= \partial_r\!\left(r^2f^3\widetilde{h}^{(1)1m}_{rr}\right)\!.
\end{align}
\end{subequations}
Explicitly, in terms of $h^{(1)1m}_{\boldsymbol\cdot}$, 
\begin{subequations}
\allowdisplaybreaks
\begin{align}
\varphi^{(1)m}_{tt} &=\frac{1}{r^2} \bigl[Mr^2f \partial_r^2{h_\trAB}
   -M(5r-12 M)\partial_r{h_\trAB}\nonumber\\*
   &\quad 
   -r^4\partial_t^2\partial_r {h_\trAB}
   -2 r^3\partial_t^2{h_\trAB}
   +2Mr^2f\partial_r {h_{r+}}\nonumber\\*
   &\quad +2r^5 f \partial_t^2\partial_r{h_{r+}}
   +2r^3(3r-4M)\partial_t^2{h_{r+}}\nonumber\\*
   &\quad
   -2 M (3r-8M) {h_{r+}}
   -2r^5f\partial_t\partial_r^2 {h_{t+}}\nonumber\\*
   &\quad
   +4 r^3f\partial_t{h_{t+}}
   -4 r^4\partial_t\partial_r{h_{t+}}
   +r^5f\partial_r^2 {h_{tt}}\nonumber\\*
   &\quad
   +2r^4\partial_r {h_{tt}}
   -2 r^3f{h_{tt}}\bigr]\nonumber\\
   &\quad
   +\frac{4 M}{r^3} (2 r-5 M) {h_\trAB},\\
\varphi^{(1)m}_{tr} &= \frac{1}{r^2} \bigl[-r^2f\partial_t\partial_r {h_\trAB}
   +M\partial_t{h_\trAB}
   +rf(r-4M)\partial_t{h_{r+}}\nonumber\\*
   &\quad
   +r^3f^2\partial_t\partial_r {h_{r+}}
   -r^3f^2\partial_r^2 {h_{t+}}
   -r^2f\partial_r{h_{t+}}\nonumber\\*
   &\quad
   +r^3f^2\partial_r{h_{tr}}
   +rf(r+2M){h_{tr}}\bigr]\nonumber\\*
   &\quad
   -\frac{2 M}{r^3} (r-4 M) {h_{t+}},\\
\varphi^{(1)m}_{rr} &= -\frac{f}{r}\bigl[r^2 f\partial_r^2 h_\trAB-(r-3M) \partial_r h_\trAB 
   +2r^2f\partial_r h_{r+} \nonumber\\*
   &\quad + 2(r-M) h_{r+} - r^3f^2\partial_r h_{rr} - 2 rf (r+M)
   h_{rr}\bigr] \nonumber\\*
   &\quad +\frac{2Mf}{r^2}h_\trAB.
\end{align}
\end{subequations}
Here, for readability, we have omitted superscript ``$(1)\ell m$'' labels on the right-hand side.

Again, at second order we replace ${h}^{(1)1m}_{\boldsymbol{\cdot}}$ with ${h}^{(2)1m}_{\boldsymbol{\cdot}}+\widetilde H^{1m}_{\boldsymbol{\cdot}}$.

\subsection{Gauge fixing for $\ell=1$: odd parity}

For odd-parity $\ell=1$ perturbations, the only nonvanishing piece of $h^{(n)}_{\mu\nu}$ is the vector-harmonic mode $h^{(n)1m}_{a-}$. At first order, Eq.~\eqref{tilde h} reduces to%
\beq\label{ell=1 odd-parity transformation}
\widetilde{h}^{(1)1m}_{a-} = {h}^{(1)1m}_{a-}+r^2\delta_a \tilde Z^{(1)1m}_-.
\eeq
Our assumption of asymptotic flatness at spatial infinity implies the falloff condition 
\beq
{h}^{(n)1m}_{a-}=r\left[\frac{c^{(n)}_{a-}}{r^2}+\O(r^{-3})\right] 
\eeq
for some $t$-independent constants $c^{(n)}_{a-}$.

We impose conditions
\begin{subequations}
\begin{align}
\widetilde{h}^{(n)1m}_{r-} &= 0,\\
\lim_{r\to\infty}\left(\widetilde{h}^{(n)1m}_{t-}\right) &= 0.
\end{align}
\end{subequations}
The first implies
\beq\label{Z11m tilde}
\tilde Z^{(1)1m}_- = \varpi^{(1)m}(t) -\int^r_\infty \frac{h^{(1)1m}_{r-}}{r'^{2}}dr',
\eeq
and the second implies
\beq
\varpi^{(1)m} = \varpi^{(1)m}_0
\eeq
for some constant $\varpi^{(1)m}_0$. This constant represents the rotational Killing vector of the background, and once again we are unable to determine it through conditions on $\widetilde h^{(1)}_{\mu\nu}$.

The nonzero invariant component is
\beq\label{h1mm tilde}
\widetilde{h}^{(1)1m}_{t-} = {h}^{(1)1m}_{t-}+r^2\partial_t \tilde Z^{(1)1m}_-
\eeq
This is a nonlocal invariant, but we can construct a local invariant from it:
\beq\label{ell=1 odd invariant}
\varphi^{(1)m}_- := \partial_r\left(r^{-2}\widetilde{h}^{(1)1m}_{t-}\right).
\eeq
Explicitly,
\beq
\varphi^{(1)m}_- = \partial_r\bigl(r^{-2}h^{(1)1m}_{t-}\bigr) -\partial_t \bigl(r^{-2}h^{(1)1m}_{r-}\bigr).
\eeq
This local invariant is related to the quantity $W_1$ in Ref.~\cite{Thompson:2016fxe} by $\varphi^{(1)m}_- = -W_1/r^4$.

At second order, we replace ${h}^{(1)1m}_{a-}$ with ${h}^{(2)1m}_{a-}+\widetilde H^{1m}_{a-}$ in these formulas.

\subsection{Residual Killing freedom and comments on gauge fixing}\label{residual Killing freedom}

We have now fully fixed the gauge freedom at first and second order, up to the Killing vectors represented by the constants $\varrho^{(1)}_0$ and $\varpi^{(1)m}_0$ in Eqs.~\eqref{zeta t tilde 100} and \eqref{Z11m tilde}. To fix those remaining constants, we can impose conditions on the stress-energy tensor.

We decompose $\tilde\xi^\mu_{(1)}$ into the Killing and non-Killing pieces, denoting the former by $K^\mu_{(t)}$ (for the timelike Killing vector) and $K^\mu$ (for the rotational Killing vector), and denoting the latter by $\hat\xi^\mu_{(1)}$. The invariant $\widetilde T^{(2)}_{\mu\nu}$ defined in Eq.~\eqref{tilde T2} is then
\beq
\widetilde T^{(2)}_{\mu\nu} = T^{(2)}_{\mu\nu} + K^\alpha_{(t)}\partial_\alpha T^{(1)}_{\mu\nu} +\Lie_K T^{(1)}_{\mu\nu} + \Lie_{\hat\xi_{(1)}} T^{(1)}_{\mu\nu}. 
\eeq
Imposing conditions on $\widetilde T^{(2)}_{\mu\nu}$ allows us to rearrange this to obtain equations for the constants in $K^\mu_{(t)}$ and $K^\mu$. 

The full list of now fully fixed invariants is~\eqref{h1tilde ell>1}, \eqref{h00rr tilde}, \eqref{h00tt tilde}, \eqref{h1mp tilde}, and \eqref{h1mm tilde}.

Having now completed our gauge-fixing procedure, we consider its merits relative to the obvious alternative: simply adopting a convenient gauge and solving the perturbative Einstein equations in that gauge. We first enumerate some merits of the gauge-fixing scheme.
\begin{enumerate}
\item It fully elucidates and isolates the gauge-invariant degrees of freedom in the metric perturbation. By removing all gauge degrees of freedom, we have reduced the metric perturbation to the set of invariant fields $\widetilde{h}^{(n)00}_{tt}$ and $\widetilde{h}^{(n)00}_{rr}$ for $\ell=0$; $\widetilde{h}^{(n)1 m}_{ab}$ and $\widetilde{h}^{(n)1 m}_{t-}$ for $\ell=1$; and $\widetilde{h}^{(n)\ell m}_{ab}$, $\widetilde{h}^{(n)\ell m}_{a-}$, and $\widetilde{h}^{(n)\ell m}_{\trAB}$ for $\ell>1$.
\item Although the invariants are, in general, nonlocal functionals of the metric perturbation, the method also provides a simple recipe for deriving local invariants, at least at first order in perturbation theory. It might be possible to write the field equations entirely in terms of these local invariants (although we leave that possibility unexplored).
\item Even if we wish to work with the gauge-dependent metric perturbations and choose some convenient gauge, it can be expedient to derive the field equations for the invariant variables. From them, field equations for the gauge-dependent metric perturbations can be obtained simply by substituting the definitions of $\widetilde h^{(n)\ell m}_{\boldsymbol\cdot}$ in terms of $h^{(n)\ell m}_{\boldsymbol\cdot}$.
\end{enumerate}
Contrast this with a clear disadvantage:
\begin{enumerate}
\item Writing the Einstein equations in terms of the invariant metric perturbation components is equivalent to simply adopting the fixed gauge as one's working gauge. This might very well be an unfortunate choice, as it is often useful to choose a gauge that is well adapted to one's particular problem.
\end{enumerate}
Note that this last point does not mean that working in a convenient gauge is always equivalent to working in some fully fixed gauge. If one works in a gauge with residual gauge freedom, such as the Lorenz gauge, then the residual freedom is eliminated through a choice of boundary conditions rather than at the level of the field equations.

In this paper, part of our motivation for presenting a gauge-fixing formalism is to remain in the tradition of MP. However, we recognize the merits of both approaches and therefore provide field equations both for invariant variables and for raw, gauge-dependent metric perturbations.

\section{Mode decomposition of the first- and second-order Einstein equations}\label{Decomposed EFE}

Having assembled the necessary tools, we now apply them to the Einstein field equations~\eqref{EFE1} and~\eqref{EFE2}. Sections~\ref{Yilm-decomposition - linear terms} and~\ref{Yilm-decomposition - quadratic terms} present the harmonic expansions of the quantities appearing in the field equations, relegating lengthy expressions to Appendices~\ref{M2S2-decomposition} and \ref{coupling functions}. Section~\ref{summary} then summarizes the field equations in various forms, and Sec.~\ref{Bianchi decomposed} presents the mode-decomposed conservation equations that constrain the field equations.

We decompose the quantities in Eqs.~\eqref{EFE1} and~\eqref{EFE2} following the steps outlined above Eq.~\eqref{H decomposition}. In the body of the paper we present mode-decomposed formulas for the linear quantities ${\cal E}_{\mu\nu}[h]$ and ${\cal F}_{\mu\nu}[h]$ and the quadratic quantities ${\cal A}_{\mu\nu}[h]$, ${\cal B}_{\mu\nu}[h]$, and ${\cal C}_{\mu\nu}[h]$; the companion package \pkg{PerturbationEquations} includes also the decompositions of the field equations in the form~\eqref{EFE1 dG}--\eqref{EFE2 dG}.

We present our results in terms of the invariant fields~$\widetilde{h}^{(n)}_{\mu\nu}$. However, in \pkg{PerturbationEquations} we also provide the raw results, without the gauge-fixing procedure. As mentioned in the Introduction, to the best of our knowledge, this is the first time a complete mode decomposition has been presented for the second-order Einstein equations, with arbitrary first-order mode content and in an arbitrary gauge.

\subsection{Linear curvature terms}\label{Yilm-decomposition - linear terms}

Equation~\eqref{dR} expresses the linearized Ricci tensor $\dR_{\mu\nu}[h]$ in terms of the quantities ${\cal E}_{\mu\nu}[h]$ and ${\cal F}_{\mu\nu}[h]$ defined in Eqs.~\eqref{E} and \eqref{F}. The $2+2$D decomposition of those quantities is given in Eqs.~\eqref{E-M2S2} and \eqref{F-M2S2}. Substituting the mode expansion~\eqref{h-Yilm}, we obtain the modes ${\cal E}^{\ell m}_{\boldsymbol{\cdot}}[h^{\ell m}_{\boldsymbol{\cdot}}]$ and ${\cal F}^{\ell m}_{\boldsymbol{\cdot}}[h^{\ell m}_{\boldsymbol{\cdot}}]$. We then make the replacement $h_{\mu\nu}\to\widetilde{h}_{\mu\nu}$, with $\widetilde{h}^{\ell m}_{\pm}=0=\widetilde{h}^{\ell m}_{a+}$.

The results are%
\begin{subequations}\label{Elm}%
\allowdisplaybreaks%
\begin{align}
{\cal E}^{\ell m}_{ab} &= \Box_{{\cal M}^2}\widetilde{h}^{\ell m}_{ab} 
								+ \frac{4M}{r^5}g_{ab}\left(2r^2\widetilde{h}_{\trab}^{\ell m}-\widetilde{h}_{\trAB}^{\ell m}\right) \nonumber\\*
							&\quad - \frac{4}{r^2} \widetilde{h}^{\ell m}_{c(a}r_{b)}r^c + \frac{4}{r^4}\widetilde h_{\trAB}^{\ell m} r_{a} r_{b} 
								 \nonumber\\*
							&\quad - \frac{1}{r^2}\widetilde{h}^{\ell m}_{ab} \left(\lam{\ell}{1}^2+\frac{4 M}{r}\right)  + \frac{2}{r} r^{c}\delta_{c}\widetilde{h}^{\ell m}_{ab},\\
{\cal E}^{\ell m}_{a+} &=  -\frac{2}{r^3}r_a\widetilde{h}^{\ell m}_{\trAB} +\frac{2}{r}\widetilde{h}^{\ell m}_{ab}r^b,\\
{\cal E}^{\ell m}_{a-} &= \Box_{{\cal M}^2}\widetilde{h}_{a-}^{\ell m} -\frac{1}{r^2}\widetilde{h}_{a-}^{\ell m}\left(\lam{\ell}{1}^2-\frac{2M}{r}\right) \nonumber\\*
&\quad - \frac{4}{r^2}\widetilde{h}_{b-}^{\ell m} r_{a} r^{b},\\
{\cal E}^{\ell m}_\trAB &=\Box_{{\cal M}^2}\widetilde{h}_{\trAB}^{\ell m} - \frac{\lambda_1^2}{r^2}\widetilde{h}_{\trAB}^{\ell m} - \frac{4M}{r}\widetilde{h}_{\trab}^{\ell m} \nonumber\\*
							&\quad +2\widetilde{h}^{\ell m}_{ab} r^{a} r^{b} -\frac{2}{r} r^{a} \delta_{a}\widetilde{h}_{\trAB}^{\ell m},\\
{\cal E}^{\ell m}_{\pm} &= \frac{4}{r}  r^{a}\widetilde{h}_{a\pm}^{\ell m},
\end{align}
\end{subequations}
where
\beq
\Box_{{\cal M}^2}:=g^{ab}\delta_a\delta_b,
\eeq
and
\begin{subequations}\label{Flm}%
\allowdisplaybreaks%
\begin{align}
{\cal F}^{\ell m}_{ab} &= -  2\delta_{(a}\delta^{c}\widetilde h^{\ell m}_{b)c} + 2 \delta_{a}\delta_{b}\widetilde h_{\trab}^{\ell m} + \frac{2 }{r^2}\delta_{a}\delta_{b}\widetilde h_{\trAB}^{\ell m} \nonumber\\*
&\quad -  \frac{4}{r} r^{c} \delta_{(a}\widetilde h^{\ell m}_{b)c} -  \frac{2}{r^3} r_{(a} \delta_{b)}\widetilde h_{\trAB}^{\ell m} - \frac{4 M }{r^3}\widetilde h^{\ell m}_{ab} \nonumber\\*
&\quad + \frac{4}{r^2} r^{c}r_{(a}\widetilde h^{\ell m}_{b)c},\\
{\cal F}^{\ell m}_{a+} &=  - \delta^{b}\widetilde h^{\ell m}_{ab}+ 2\delta_{a}\widetilde h_{\trab}^{\ell m} + \frac{1}{r^2}\delta_{a}\widetilde h_{\trAB}^{\ell m} - \frac{2}{r} r_{a}\widetilde h_{\trab}^{\ell m} \nonumber\\
&\quad -  \frac{2}{r} r^{b}\widetilde h^{\ell m}_{ab},\\
{\cal F}^{\ell m}_{a-} &=  - \delta_{a}\delta^{b}\widetilde h_{b-}^{\ell m} + \frac{2}{r} r_{a} \delta^{b}\widetilde h_{b-}^{\ell m} -  \frac{2}{r} r^{b}\delta_{a}\widetilde h_{b-}^{\ell m} \nonumber\\*
&\quad - \frac{2 M}{r^3}\widetilde h_{a-}^{\ell m} + \frac{6}{r^2}  r_{a} r^{b} \widetilde h_{b-}^{\ell m},\\
{\cal F}^{\ell m}_\trAB &= - 2 rr^{a}\delta^{b}\widetilde{h}^{\ell m}_{ab}+2 rr^{a} \delta_{a}\widetilde{h}_{\trab}^{\ell m} - 4 \widetilde{h}^{\ell m}_{ab} r^{a} r^{b} \nonumber\\*
&\quad - \lam{\ell}{1}^2 \widetilde{h}_{\trab}^{\ell m}  + \frac{2}{r} r^{a}\delta_{a}\widetilde{h}_{\trAB}^{\ell m},\\
{\cal F}^{\ell m}_+ &= 2\widetilde h_{\trab}^{\ell m}, \\
{\cal F}^{\ell m}_- &= - 2\delta^{a}\widetilde h_{a-}^{\ell m} - \frac{4}{r}r^{a}\widetilde h_{a-}^{\ell m}.
\end{align}
\end{subequations}
Here we have written the expressions for a generic symmetric tensor $\widetilde{h}_{\mu\nu}$, which can be either $\widetilde{h}^{(1)}_{\mu\nu}$ or $\widetilde{h}^{(2)}_{\mu\nu}$. 

At first order, these expressions are valid in all gauges since (i) they are valid in at least one gauge (the gauge in which $h^{(1)}_{\mu\nu}=\widetilde{h}^{(1)}_{\mu\nu}$), and (ii) they express the invariant quantity $\dR_{\mu\nu}[h^{(1)}]$ in terms of invariant fields. If desired, we can express these quantities in terms of $h^{(1)}_{\mu\nu}$ in any gauge by substituting the explicit expressions~\eqref{h1tilde ell>1}, \eqref{h00rr tilde}, \eqref{h00tt tilde}, \eqref{h1mp tilde}, and \eqref{h1mm tilde} for $\widetilde{h}^{(1)}_{\mu\nu}$ in terms of $h^{(1)}_{\mu\nu}$. Alternatively, we can solve the field equations directly for the invariant fields.

At second order, $\dR_{\mu\nu}[h^{(2)}]$ is not invariant, and the above expressions are valid \emph{only} in the gauge for which $h^{(2)}_{\mu\nu}=\widetilde{h}^{(2)}_{\mu\nu}$. However, the second-order field equation~\eqref{EFE2}, taken as a whole, \emph{is} invariant, meaning it will remain valid in all gauges after the replacements $h^{(n)}_{\mu\nu}\to\widetilde{h}^{(n)}_{\mu\nu}$ and $T^{(2)}_{\mu\nu}\to\widetilde{T}^{(2)}_{\mu\nu}$.

\subsection{Quadratic curvature terms}\label{Yilm-decomposition - quadratic terms}

Equation~\eqref{ddR} expresses the second-order Ricci tensor $\ddR_{\mu\nu}[h]$ in terms of the quantities ${\cal A}_{\mu\nu}[h]$, ${\cal B}_{\mu\nu}[h]$, and ${\cal C}_{\mu\nu}[h]$ defined in Eqs.~\eqref{A}--\eqref{C}. The $2+2$D decomposition of those quantities is given in Eqs.~\eqref{A-M2S2}, \eqref{B-M2S2}, \eqref{C-M2S2}. Substituting the mode expansion~\eqref{h-Yilm}, and following the same steps that led to Eq.~\eqref{H decomposition}, for ${\cal A}_{\mu\nu}$ we obtain %
\begin{subequations}\label{Ailm}%
\begin{align}
{\cal A}^{\ell m}_{ab} &= \sum_{\substack{\ell'm'\ell''m''\\s'=0,1}}\!\!\lam{\ell^{\prime}}{s'} \lam{\ell^{\prime \prime}}{s'}\, C^{\ell m 0}_{\ell'm's'\ell''m''-s'}\nonumber\\[-15pt] &\qquad\qquad\qquad\qquad\qquad\times {\cal A}^{\ell'm's'\ell''m''-s'}_{ab},\\[5pt]
{\cal A}^{\ell m}_{a\pm} &= \sum_{\substack{\ell'm'\ell''m''\\s'=1,2}}\!\! \frac{\lam{\ell^{\prime}}{s'} \lam{\ell^{\prime \prime}}{1-s'}}{\lam{\ell}{1}}\, C^{\ell m 1}_{\ell'm's'\ell''m'',1-s'}\nonumber\\[-15pt] &\qquad\qquad\qquad\qquad\qquad\times {\cal A}^{\ell'm's'\ell''m'',1-s'}_{a\pm},\\[5pt]
{\cal A}^{\ell m}_\trAB &= \sum_{\substack{\ell'm'\ell''m''\\s'=0,1,2}}\!\!\lam{\ell^{\prime}}{s'} \lam{\ell^{\prime \prime}}{s'}\,  C^{\ell m0}_{\ell'm' s'\ell''m''-s'}\nonumber\\[-15pt] &\qquad\qquad\qquad\qquad\qquad\times {\cal A}^{\ell'm's'\ell''m''-s'}_\trAB,\\[5pt]
{\cal A}^{\ell m}_{\pm} &= \sum_{\substack{\ell'm'\ell''m''\\s'=1,2}}\!\!\frac{\lam{\ell^{\prime}}{s'} \lam{\ell^{\prime \prime}}{2-s'}}{\lam{\ell}{2}}\,C^{\ell m 2}_{\ell'm' s'\ell''m'',2 -s'}\nonumber\\[-15pt] &\qquad\qquad\qquad\qquad\qquad\times {\cal A}^{\ell'm' s'\ell''m'',2 -s'}_\pm,
\end{align}
\end{subequations}
where the quantities ${\cal A}^{\ell'm's'\ell''m''s''}_{\boldsymbol{\cdot}}$ are made up of products of $h^{(1)\ell'm'}_{\boldsymbol{\cdot}}$ and $h^{(1)\ell''m''}_{\boldsymbol{\cdot}}$. We display these quantities in Eq.~\eqref{A coupling functions} in terms of the invariants $\widetilde h^{(1)}_{\mu\nu}$. If ${\cal A}_{\mu\nu}$ is calculated in a generic gauge in terms of $h^{(1)}_{\mu\nu}$, then $s'_{\rm max}$ in the above sums is increased by one because of the involvement of the tensor modes $h^{(1)\ell' m'}_\pm$ and $h^{(1)\ell'' m''}_\pm$; in the invariant form of the field equations, those higher-spin terms appear instead on the left-hand side of the field equations, hidden within $\widetilde h^{(2)}_{\mu\nu}$. 

Similarly, ${\cal B}^{\ell m}_{\boldsymbol{\cdot}}$ and ${\cal C}^{\ell m}_{\boldsymbol{\cdot}}$ are given by the sums
\begin{subequations}\label{Bilm}%
\begin{align}
{\cal B}^{\ell m}_{ab} &= \sum_{\substack{\ell'm'\ell''m''\\s'=0,1}}\!\!\lam{\ell^{\prime}}{s'} \lam{\ell^{\prime \prime}}{s'}\, C^{\ell m 0}_{\ell'm's'\ell''m''-s'}\nonumber\\[-15pt] &\qquad\qquad\qquad\qquad\qquad\times {\cal B}^{\ell'm's'\ell''m''-s'}_{ab},\\[5pt]
{\cal B}^{\ell m}_{a\pm} &= \sum_{\substack{\ell'm'\ell''m''\\s'=1,2}}\!\! \frac{\lam{\ell^{\prime}}{s'} \lam{\ell^{\prime \prime}}{1-s'}}{\lam{\ell}{1}}\, C^{\ell m 1}_{\ell'm's'\ell''m'',1-s'}\nonumber\\[-15pt] &\qquad\qquad\qquad\qquad\qquad\times {\cal B}^{\ell'm's'\ell''m'',1-s'}_{a\pm},\\[5pt]
{\cal B}^{\ell m}_\trAB &= \sum_{\substack{\ell'm'\ell''m''\\s'=0,1}}\!\!\lam{\ell^{\prime}}{s'} \lam{\ell^{\prime \prime}}{s'}\,  C^{\ell m0}_{\ell'm' s'\ell''m''-s'}\nonumber\\[-15pt] &\qquad\qquad\qquad\qquad\qquad\times {\cal B}^{\ell'm's'\ell''m''-s'}_\trAB,\\[5pt]
{\cal B}^{\ell m}_{\pm} &= \sum_{\substack{\ell'm'\ell''m''\\s'=1,2}}\!\!\frac{\lam{\ell^{\prime}}{s'} \lam{\ell^{\prime \prime}}{2-s'}}{\lam{\ell}{2}}\,C^{\ell m 2}_{\ell'm' s'\ell''m'',2 -s'}\nonumber\\[-15pt] &\qquad\qquad\qquad\qquad\qquad\times {\cal B}^{\ell'm' s'\ell''m'',2 -s'}_\pm,
\end{align}
\end{subequations}
\begin{subequations}\label{Cilm}%
\begin{align}
{\cal C}^{\ell m}_{ab} &= \sum_{\substack{\ell'm'\ell''m''\\s'=0,1}}\!\!\lam{\ell^{\prime}}{s'} \lam{\ell^{\prime \prime}}{s'}\, C^{\ell m 0}_{\ell'm's'\ell''m''-s'}\nonumber\\[-15pt] &\qquad\qquad\qquad\qquad\qquad\times {\cal C}^{\ell'm's'\ell''m''-s'}_{ab},\\[5pt]
{\cal C}^{\ell m}_{a\pm} &= \sum_{\substack{\ell'm'\ell''m''\\s'=1}}\!\! \frac{\lam{\ell^{\prime}}{s'} \lam{\ell^{\prime \prime}}{1-s'}}{\lam{\ell}{1}}\, C^{\ell m 1}_{\ell'm's'\ell''m'',1-s'}\nonumber\\[-15pt] &\qquad\qquad\qquad\qquad\qquad\times {\cal C}^{\ell'm's'\ell''m'',1-s'}_{a\pm},\\[5pt]
{\cal C}^{\ell m}_\trAB &= \sum_{\substack{\ell'm'\ell''m''\\s'=0,1}}\!\!\lam{\ell^{\prime}}{s'} \lam{\ell^{\prime \prime}}{s'}\,  C^{\ell m0}_{\ell'm' s'\ell''m''-s'}\nonumber\\[-15pt] &\qquad\qquad\qquad\qquad\qquad\times {\cal C}^{\ell'm's'\ell''m''-s'}_\trAB,\\[5pt]
{\cal C}^{\ell m}_{\pm} &= \sum_{\substack{\ell'm'\ell''m''\\s'=1,2}}\!\!\frac{\lam{\ell^{\prime}}{s'} \lam{\ell^{\prime \prime}}{2-s'}}{\lam{\ell}{2}}\,C^{\ell m 2}_{\ell'm' s'\ell''m'',2 -s'}\nonumber\\[-15pt] &\qquad\qquad\qquad\qquad\qquad\times {\cal C}^{\ell'm' s'\ell''m'',2 -s'}_\pm,
\end{align}
\end{subequations}
where the quantities ${\cal B}^{\ell'm's'\ell''m''s''}_{\boldsymbol{\cdot}}$ and ${\cal C}^{\ell'm's'\ell''m''s''}_{\boldsymbol{\cdot}}$ are made up of products of $h^{(1)\ell'm'}_{\boldsymbol{\cdot}}$ and $h^{(1)\ell''m''}_{\boldsymbol{\cdot}}$. We display these quantities in Eqs.~\eqref{B coupling functions} and \eqref{C coupling functions} in terms of the invariants $\widetilde h^{(1)}_{\mu\nu}$.
In all cases, the sums run over the restricted range of mode numbers displayed in Eq.~\eqref{restricted sums}. 

\subsection{Stress-energy terms}

The stress-energy terms in the field equations~\eqref{EFEn dR} are more straightforwardly decomposed. In the first-order equation we have
\beq\label{calT1 modes}
\widetilde{\cal T}^{(1)\ell m}_{\boldsymbol{\cdot}} = \overline{T}^{(1)\ell m}_{\boldsymbol{\cdot}},
\eeq
where $\overline{T}^{(1)\ell m}_{\boldsymbol{\cdot}}$ are related to $T^{(1)\ell m}_{\boldsymbol{\cdot}}$ by Eq.~\eqref{trace reverse modes}.  In the second-order equation we have the harmonic modes of the invariant $\widetilde{\cal T}^{(2)}_{\mu\nu} = {\cal T}^{(2)}_{\mu\nu}+ \Lie_{\tilde\xi_{(1)}}\overline{T}^{(1)}_{\mu\nu}$. Expressed in terms of the invariants $\widetilde{T}^{(n)}_{\mu\nu}$ and $\widetilde{h}^{(1)}_{\mu\nu}$, this quantity reads
\begin{multline}\label{calT2}
\widetilde{\cal T}^{(2)}_{\mu\nu} = \widetilde{T}^{(2)}_{\mu\nu} - \frac{1}{2}g_{\mu\nu}g^{\alpha\beta}\widetilde{T}^{(2)}_{\alpha\beta} \\
+ \frac{1}{2}\left(g_{\mu\nu}\widetilde{h}^{(1)}_{\alpha\beta} - \widetilde{h}^{(1)}_{\mu\nu}g_{\alpha\beta}\right)T^{\alpha\beta}_{(1)}.
\end{multline}
Its mode expansion is given in Eq.~\eqref{calT2 modes}.

\subsection{Summary of the decomposed field equations}\label{summary}

\subsubsection{Field equations in terms of invariant variables}

In summary, we can write our covariant, gauge-invariant, tensor-harmonic decomposition of the Einstein equations~\eqref{EFE1} and \eqref{EFE2} as%
\begin{subequations}\label{EFEnlm fixed}%
\begin{align}
\dR^{\ell m}_{\boldsymbol\cdot}[\widetilde{h}^{(1)\ell m}_{\boldsymbol\cdot}] &= 8\pi{\cal T}^{(1)\ell m}_{\boldsymbol\cdot},\\
\dR^{\ell m}_{\boldsymbol\cdot}[\widetilde{h}^{(2)\ell m}_{\boldsymbol\cdot}] &=8\pi\widetilde{\cal T}^{(2)\ell m}_{\boldsymbol\cdot}-\ddR^{\ell m}_{\boldsymbol\cdot}[\widetilde{h}^{(1)}],\label{EFE2lm fixed}
\end{align}
\end{subequations}
using the shorthand introduced below Eq.~\eqref{conjugation}. The mode-decomposed operator on the left-hand side is
\beq
\dR^{\ell m}_{\boldsymbol\cdot}[\widetilde{h}^{(n)\ell m}_{\boldsymbol\cdot}] = -\frac{1}{2}({\cal E}^{\ell m}_{\boldsymbol\cdot}[\widetilde{h}^{(n)\ell m}_{\boldsymbol\cdot}]+{\cal F}^{\ell m}_{\boldsymbol\cdot}[\widetilde{h}^{(n)\ell m}_{\boldsymbol\cdot}]),
\eeq
with ${\cal E}^{\ell m}_{\boldsymbol\cdot}[\widetilde{h}^{(n)\ell m}_{\boldsymbol\cdot}]$ and ${\cal F}^{\ell m}_{\boldsymbol\cdot}[\widetilde{h}^{(n)\ell m}_{\boldsymbol\cdot}]$ as given in Eqs.~\eqref{Elm} and~\eqref{Flm}. 

The stress-energy source terms on the right-hand side are given in Eqs.~\eqref{calT1 modes} and \eqref{calT2 modes}. The quadratic source term is 
\beq
\ddR^{\ell m}_{\boldsymbol\cdot}[\widetilde{h}^{(1)}] = \frac{1}{2}({\cal A}^{\ell m}_{\boldsymbol\cdot}[\widetilde{h}^{(1)}]+{\cal B}^{\ell m}_{\boldsymbol\cdot}[\widetilde{h}^{(1)}]+{\cal C}^{\ell m}_{\boldsymbol\cdot}[\widetilde{h}^{(1)}]),
\eeq
where the quantities ${\cal A}^{\ell m}_{\boldsymbol\cdot}$, ${\cal B}^{\ell m}_{\boldsymbol\cdot}$, and ${\cal C}^{\ell m}_{\boldsymbol\cdot}$ are infinite sums of products of modes of $\widetilde{h}^{(1)}_{\mu\nu}$. These sums take the form~\eqref{Ailm} for all three calligraphic quantities (with differing $s'_{\rm max}$), where (i) the sums run over the range in Eq.~\eqref{restricted sums}, (ii) $C^{\ell m s}_{\ell'm's'\ell''m''s''}$ is given in Eq.~\eqref{coupling}, and (iii) (as an example) the quadratic coupling functions ${\cal A}^{\ell'm's'\ell''m''s''}_{\boldsymbol{\cdot}}$ are given in Eqs.~\eqref{A coupling functions}.



The even- and odd-parity sectors of these field equations decouple at each order. This is because on the left-hand side of the field equations, the even-parity terms $\dR^{\ell m}_{ab}$, $\dR^{\ell m}_{\trAB}$, $\dR^{\ell m}_{a+}$ and $\dR^{\ell m}_{+}$ only depend on the even-parity perturbations $\widetilde{h}^{\ell m}_{ab}$ and $\widetilde{h}^{\ell m}_{\trAB}$; and odd-parity terms $\dR^{\ell m}_{a-}$ and $\dR^{\ell m}_{-}$ only depend on the odd-parity perturbations $\widetilde{h}^{\ell m}_{a-}$.

However,  due to the quadratic source term, the even- and odd-parity first-order fields do become coupled in the second-order field equation~\eqref{EFE2lm fixed}. The even-parity fields $\widetilde{h}^{(2)\ell m}_{ab}$ and $\widetilde{h}^{(2)\ell m}_{\trAB}$ are sourced by ``even $\times$ even'' products ($\widetilde{h}^{(1)\ell' m'}_{ab}$ and $\widetilde{h}^{(1)\ell' m'}_{\trAB}$ multiplying $\widetilde{h}^{(1)\ell'' m''}_{ab}$ and $\widetilde{h}^{(1)\ell'' m''}_{\trAB}$) as well as by ``odd $\times$ odd'' products ($\widetilde{h}^{(1)\ell' m'}_{a-}$ multiplying $\widetilde{h}^{(1)\ell'' m''}_{a-}$). The odd-parity fields $\widetilde{h}^{(2)\ell m}_{a-}$ are sourced by ``even $\times$ odd'' products ($\widetilde{h}^{(1)\ell' m'}_{ab}$ and $\widetilde{h}^{(1)\ell' m'}_{\trAB}$ multiplying $\widetilde{h}^{(1)\ell'' m''}_{a-}$).

Similarly, the $\ell m$ modes decouple from one another at each order, but the first-order modes couple to one another in the second-order source. A second-order mode $\widetilde{h}^{(2)\ell m}_{\boldsymbol{\cdot}}$ with any given $\ell$ value is generically sourced by \emph{all} first-order modes $\widetilde{h}^{(1)\ell' m'}_{\boldsymbol{\cdot}}$ from $\ell'=0$ to $\infty$.

The modes of the trace-reversed field equations~\eqref{EFEn dG} can be obtained from these modes using the relations~\eqref{trace reverse modes}.

\subsubsection{Field equations in a generic gauge}

If we do not make the replacement $h^{(n)\ell m}_{\boldsymbol\cdot}\to\widetilde{h}^{(n)\ell m}_{\boldsymbol\cdot}$, then we arrive at the raw field equations
\beq\label{EFEnlm generic}
\dR^{\ell m}_{\boldsymbol\cdot}[h^{(n)\ell m}_{\boldsymbol\cdot}] = \bar S^{(n)\ell m}_{\boldsymbol\cdot},
\eeq
which have all the same features as Eq.~\eqref{EFEnlm fixed} but are substantially more complicated due to the nonvanishing $h^{(n)\ell m}_{a\pm}$ and $h^{(n)\ell m}_{\pm}$. The sources are 
\begin{subequations}
\begin{align}
\bar S^{(1)\ell m}_{\boldsymbol\cdot} &= 8\pi{\cal T}^{(1)\ell m}_{\boldsymbol\cdot}= 8\pi\overline{T}^{(1)\ell m}_{\boldsymbol\cdot},\\
\bar S^{(2)\ell m}_{\boldsymbol\cdot}&= 8\pi{\cal T}^{(2)\ell m}_{\boldsymbol\cdot}-\ddR^{\ell m}_{\boldsymbol\cdot}[h^{(1)}].\label{S2lm generic}
\end{align}
\end{subequations}

We can also obtain these equations by starting from Eq.~\eqref{EFEnlm fixed} and substituting the expressions~\eqref{h1tilde ell>1}, \eqref{h00rr tilde}, \eqref{h00tt tilde}, \eqref{h1mp tilde}, and \eqref{h1mm tilde} for $\widetilde{h}^{(n)\ell m}_{\boldsymbol\cdot}$ in terms of $h^{(n)\ell m}_{\boldsymbol\cdot}$. Additional manipulations (involving the Bianchi identities, for example) are required to put the result in precisely the form of Eq.~\eqref{EFEnlm generic}, but the equations are necessarily equivalent.

\subsubsection{Field equations in the Lorenz gauge}

In the Lorenz gauge, where the fields ${\cal F}_{\mu\nu}$ and ${\cal C}_{\mu\nu}$ vanish, the field equations~\eqref{EFEnlm generic} reduce to
\beq\label{EFEnlm Lorenz}
{\cal E}^{\ell m}_{\boldsymbol\cdot}[h^{(n)\ell m}_{\boldsymbol\cdot}] = -2\bar S^{(n)\ell m}_{\boldsymbol\cdot},
\eeq
with the same first-order source $\bar S^{(1)\ell m}_{\boldsymbol\cdot}= 8\pi\overline{T}^{(1)\ell m}_{\boldsymbol\cdot}$ and with
\begin{align}
\bar S^{(2)\ell m}_{\boldsymbol\cdot} &= 8\pi{\cal T}^{(2)\ell m}_{\boldsymbol\cdot} -\frac{1}{2}\!\left( {\cal A}^{\ell m}_{\boldsymbol\cdot}[h^{(1)}]+{\cal B}^{\ell m}_{\boldsymbol\cdot}[h^{(1)}]\right)\!.\label{S2lm Lorenz}
\end{align}

The Lorenz-gauge field equations (at first order) in the MP harmonic basis are described in detail in Ref.~\cite{Osburn-etal:14}. Most self-force calculations in the Lorenz gauge have instead been in the closely related Barack--Lousto--Sago harmonic basis; see Table~\ref{table:conventions} and Appendix~\ref{BLS conventions}.

\subsection{Conservation equations}\label{Bianchi decomposed}

Due to stress-energy conservation, the source terms $S^{(n)\ell m}_{\boldsymbol{\cdot}}$ in the field equations (and therefore the field equations themselves) are not all independent. They are related by the  mode decomposition of the conservation equation~\eqref{source conservation}, which divides into the three equations%
\begin{subequations}%
\begin{align}%
\delta^b S^{\ell m}_{ab} &= \frac{1}{r^3}\!\left(\lam{\ell}{1}^2 r  S^{\ell m}_{a+}+2r_a S^{\ell m}_\trAB-2r^2r^b S^{\ell m}_{ab}\right)\!,\\
\delta^b S^{\ell m}_{b+} &= \frac{1}{2r^2}\!\left(\mu^2_\ell  S^{\ell m}_+ -4rr^b  S^{\ell m}_{b+}-2 S^{\ell m}_\trAB\right)\!,\\
\delta^b S^{\ell m}_{b-} &= \frac{1}{2r^2}\!\left(\mu^2_\ell  S^{\ell m}_- - 4rr^b  S^{\ell m}_{b-}\right).
\end{align}
\end{subequations}
The quantity $\mu^2_\ell$ appearing here is defined in Eq.~\eqref{mu def}.

\section{Master scalars and metric reconstruction}\label{master equations}

As an alternative to directly solving the Einstein equations, a common approach in black hole perturbation theory is to instead solve one or more scalar field equations for master scalar variables. The metric perturbation is then reconstructed from the master scalar(s). 

Here we summarize the formulation of this approach at second order. We specifically describe the most common variants of the approach: the RWZ formalism and the Teukolsky formalism~\cite{Teukolsky:1972my,Teukolsky:73}. Both of these are intimately related to the Weyl scalars of the perturbed spacetime, although the RWZ formalism is less often described in those terms.

The formalisms in this section are broadly identical at first and second order, with the only difference being the source terms. We therefore omit the label $n$ that indicates a quantity's perturbative order. However, we do emphasise some particular features that distinguish the second-order problem from the first-order one. 

\subsection{Regge--Wheeler--Zerilli formalism}\label{RWZ}

For the RWZ formalism we adopt the conventions of MP and Ref.~\cite{Hopper:2010uv}. Our treatment at second order differs from that of Brizuela et al.~\cite{Brizuela-etal:09} through our choice of master scalars and our inclusion of low ($\ell=0,1$) modes in $h^{(1)}_{\mu\nu}$. 

\subsubsection{Master functions}

The RWZ formalism has two master functions, one for even-parity perturbations and one for odd-parity perturbations, each of them only defined for $\ell>1$. We specifically adopt the Zerilli--Moncrief function $\Psi^{\ell m}_{\rm even}$~\cite{Zerilli:70,Moncrief:1974am} in the even-parity sector and the Cunningham--Price--Moncrief function $\Psi^{\ell m}_{\rm odd}$~\cite{Cunningham:1978zfa} in the odd-parity sector, with MP's choice of normalizations. They are closely related to the real and imaginary parts, respectively, of the linearized Weyl scalar $\delta\psi_2$~\cite{Aksteiner:2010rh}.

In terms of our invariant metric perturbations, these functions are
\begin{multline}
\Psi^{\ell m}_{\rm even} = \frac{2r}{(\lam{\ell}{1})^2}\bigg[r^{-2}\widetilde{h}^{\ell m}_\trAB \\+ \frac{2}{\Lambda_\ell}\left(r^a r^b\widetilde{h}^{\ell m}_{ab}-rr^a\delta_a(r^{-2}\widetilde{h}^{\ell m}_\trAB) \right)\bigg] 
\end{multline}
and
\beq
\Psi^{\ell m}_{\rm odd} = \frac{2r}{\mu^2_\ell}\epsilon^{ab}\left(\delta_a\widetilde{h}^{\ell m}_{b-}-\frac{2}{r}r_a\widetilde{h}^{\ell m}_{b-}\right)\!,
\eeq
with 
\beq
\Lambda_\ell := \mu^2_\ell + \frac{6M}{r}.
\eeq

The even-parity master function satisfies the 2D scalar wave equation
\beq
(\Box_{{\cal M}^2}-V^\ell_{\rm even})\Psi^{\ell m}_{\rm even} = S^{\ell m}_{\rm even},
\eeq
with the potential
\beq
V^{\ell}_{\rm even} = \frac{1}{\Lambda_\ell^2}\left[\frac{\mu^4_\ell}{r^2}\left(\lam{\ell}{1}^2+\frac{6M}{r}\right)+\frac{36M^2}{r^4}\!\left(\mu^2_\ell+\frac{2M}{r}\right)\right].
\eeq
The source term is constructed from the source in the Einstein equation according to
\begin{multline}\label{S even}
S^{\ell m}_{\rm even} = \frac{8}{\Lambda_\ell}r^a \widetilde S^{\ell m}_{a+}-\frac{2}{r}\widetilde S^{\ell m}_{+}\\
+\frac{2}{\lam{\ell}{1}^2\Lambda_\ell}\bigg\{\frac{24M}{\Lambda_\ell}r^a r^b \widetilde S^{\ell m}_{ab}-4r^2r^a\delta_a \widetilde S^{\ell m}_\trab\\
+\frac{4f}{r} \widetilde S^{\ell m}_{\trAB}+\frac{2r}{\Lambda_\ell}\bigg[\mu^2_\ell(\mu^2_\ell-2)\\
+\frac{12M}{r}(\mu^2_\ell-3)+\frac{84M^2}{r^2}\bigg]\widetilde S^{\ell m}_{\trab}\bigg\}.
\end{multline}

The odd-parity master function likewise satisfies a 2D scalar wave equation,
\beq
(\Box_{{\cal M}^2}-V^\ell_{\rm odd})\Psi^{\ell m}_{\rm odd} = S^{\ell m}_{\rm odd},
\eeq
with the potential
\beq
V^{\ell}_{\rm odd} = \frac{\ell(\ell+1)}{r^2}-\frac{6M}{r^3}
\eeq
and a source term
\beq
S^{\ell m}_{\rm odd} = -\frac{4r}{\mu^2_\ell}\epsilon^{ab}\delta_a \widetilde S^{\ell m}_{b-}.
\eeq

\subsubsection{Metric reconstruction}

From $\Psi^{\ell m}_{\rm even}$ and $\Psi^{\ell m}_{\rm odd}$, we can reconstruct the invariants $\widetilde{h}^{\ell m}_{\boldsymbol{\cdot}}$ for $\ell>1$. They are given by~\cite{Hopper:2010uv}
\begin{subequations}\label{RWZ reconstruction}
\begin{align}
\widetilde{h}^{\ell m}_{tt} &= f^2\widetilde{h}^{\ell m}_{rr}+2f\tilde S^{\ell m}_+,\\
\widetilde{h}^{\ell m}_{tr} &= r\partial_t\partial_r\Psi^{\ell m}_{\rm even}+r B_\ell\partial_t\Psi^{\ell m}_{\rm even}\nonumber\\
&\qquad\qquad\quad +\frac{2r^2}{\lam{\ell}{1}^2}\left(\tilde S^{\ell m}_{tr}-\frac{2r}{\Lambda_\ell f}\partial_t \tilde S^{\ell m}_{tt}\right)\!,\\
\widetilde{h}^{\ell m}_{rr} &= \frac{1}{4r^2f^2}\left[\Lambda_\ell\left(\lam{\ell}{1}^2 r\Psi^{\ell m}_{\rm even}-2\widetilde{h}^{\ell m}_\trAB\right)\right.\nonumber\\
&\qquad\qquad\quad\left. +4r^3r^a\delta_a(r^{-2}\widetilde{h}^{\ell m}_\trAB)\right]\!,\\
\widetilde{h}^{\ell m}_{a-} &= \frac{1}{2}\epsilon_{a}{}^b\delta_b\left(r\Psi^{\ell m}_{\rm odd}\right)+\frac{2r^2}{\mu^2_\ell}\tilde S^{\ell m}_{a-},\\
\widetilde{h}^{\ell m}_\trAB &= r^2r^a\delta_a\Psi^{\ell m}_{\rm even}+r^2A_\ell\Psi^{\ell m}_{\rm even}-\frac{4r^4}{\lam{\ell}{1}^2\Lambda_\ell} \tilde S^{\ell m}_{tt},
\end{align}
\end{subequations}
where
\begin{align}
A_\ell &:= \frac{1}{2r\Lambda_\ell}\left[\lam{\ell}{2}^2+\frac{6M}{r}\left(\mu^2_\ell+\frac{4M}{r}\right)\right]\!,\\
B_\ell &:= \frac{1}{rf\Lambda_\ell}\left[\mu^2_\ell\left(1-\frac{3M}{r}\right)-\frac{6M^2}{r^2}\right]\!.
\end{align}

As pointed out by Brizuela et al.~\cite{Brizuela-etal:09}, this metric reconstruction is problematic at large $r$. For a linear metric perturbation that is geometrically asymptotically flat, the quantities $\widetilde h^{(1)}_{\mu\nu}$ blow up at large $r$. If the second-order source is constructed from those quantities, it is also asymptotically singular. Dealing with such a source is problematic numerically but also makes the choice of physical, retarded boundary conditions unclear. We can trace the emergence of this poor behaviour starting from $\Psi^{(1)\ell m}_{\rm even}$. If the first-order source is spatially bounded and we impose retarded boundary conditions, then an outgoing mode with frequency $\omega$ behaves as $\Psi^{(1)\ell m}_{\rm even}\sim e^{-i\omega u}$ at large $r$.  The reconstruction formula~\eqref{RWZ reconstruction} then implies $\widetilde h^{(1)\ell m}_\trAB\sim r^2 e^{-i\omega u}$ and $\widetilde h^{(1)\ell m}_{ab}\sim r e^{-i\omega u}$; this contrasts with the natural behavior $h^{(1)\ell m}_\trAB\sim r e^{-i\omega u}$ and $h^{(1)\ell m}_{ab}\sim r^{-1} e^{-i\omega u}$ in a well-behaved gauge. The second-order Einstein tensor then behaves as $\delta^2G_{\mu\nu}[\widetilde h^{(1)},\widetilde h^{(1)}]\sim r^2$, and the source~\eqref{S even} constructed from it blows up even more rapidly.

One possible route around this is to work with alternative master variables at second order. Another route, suggested by an analysis in Ref.~\cite{Spiers_etal1}, is to work with alternative variables at \emph{first} order (or equivalently, work with $h^{(1)}_{\mu\nu}$ in a particular, nice gauge, rather than working with $\widetilde h^{(1)}_{\mu\nu}$). We defer further discussion of asymptotics to the Conclusion and to Ref.~\cite{Spiers_etal}, where we will provide a thorough treatment of the problem.

\subsection{Teukolsky formalism}\label{Teukolsky}

For the Teukolsky formalism we follow the conventions of Ref.~\cite{Pound:2021qin}; most equations in this section are mode-decomposed, Schwarzschild specializations of equations in that reference. Our treatment also incorporates recent work on nonvacuum metric reconstruction by Green, Hollands, and Zimmerman (GHZ)~\cite{Green:2019nam}. Although the overarching formalism here was detailed in our earlier Ref.~\cite{Spiers_etal1}, this is the first time (to our knowledge) that a second-order Teukolsky equation has appeared in mode-decomposed form with generic first-order mode content.

\subsubsection{Master scalars}

We first introduce some additional tools from Geroch, Held, and Penrose (GHP)~\cite{Geroch-Held-Penrose:73}. In analogy with spin-weighted quantities, a tensor $v$ is said to have boost weight $b$ if it transforms as $v\to \gamma^b v$ under the boost $(l^a,n^a)\to (\gamma l^a,\gamma^{-1}n^a)$. In practice, this means $v$'s boost weight is the number of factors of $l^a$ appearing in it minus the number of factors of $n^a$ appearing in it. We next define derivatives $\thorn$ and $\thorn'$ that act on boost-weighted tensors just as $\edth$ and $\edth'$ act on spin-weighted ones, meaning
\begin{subequations}
\begin{align}
\thorn v &= (l^a\delta_a - b\, \delta_a l^a)v,\label{thorn}\\
\thorn' v &= (n^a\delta_a + b\, \delta_a n^a)v.
\end{align}
\end{subequations}
Here and below we simplify definitions using the GHP prime operation: 
\beq
{}':\ m^A\leftrightarrow m^{A*},\ l^a\leftrightarrow n^a. 
\eeq
$\thorn$ raises the boost weight by 1, while $\thorn'$ lowers it by 1. They satisfy the Leibniz rule [e.g., $\thorn(u v) = v\thorn u+u\thorn v$ even for $u$ and $v$ of differing boost weights], and Eq.~\eqref{ln identities} ensures they annihilate $l^a$ and $n^a$:
\beq
\thorn l^a = \thorn' l^a = \thorn n^a = \thorn' n^a = 0.
\eeq
They satisfy the commutation relation
\begin{align}
(\thorn'\thorn - \thorn\thorn') &= -\epsilon^{ab}\delta_a\delta_b -\frac{2Mb}{r^3},
\end{align}
and the anti-commutation relation
\begin{align}
(\thorn'\thorn + \thorn\thorn') &= - \Box_{{\cal M}^2} - 2 b^2(\delta_{a} l^a)(\delta_b n^b) \nonumber\\
								&\quad  + b[2(\delta_{b} n^{b})l^a\delta_{a} - 2(\delta_{b} l^{b})n^a\delta_{a} \nonumber\\
								&\qquad \quad
  + (l^a\delta_{a}\delta_{b} n^{b}) - (n^a\delta_{a}\delta_{b} l^{b})].
\end{align}
They commute with $\edth$ and $\edth'$ and with all other background quantities on $S^2$:
\beq
\thorn\Omega_{AB} =\thorn'\Omega_{AB} = [\thorn,\D{A}] = [\thorn',\D{A}] = 0. 
\eeq
In the Kinnersley basis in retarded coordinates $(u,r)$, $\thorn=\partial_r$ when acting on a scalar; in the Hartle--Hawking basis in advanced coordinates $(v,r)$, $\thorn'=-\partial_r$ when acting on a scalar.

The field variable in this formalism can be the linear perturbation of either of the Weyl scalars $\psi_0$ or $\psi_4$,%
\begin{subequations}%
\begin{align}
\delta\psi_0[h] &:= \delta C_{\alpha\mu\beta\nu}[h]l^\alpha m^\mu l^\beta m^\nu,\\
\delta\psi_4[h] &:= \delta C_{\alpha\mu\beta\nu}[h]n^\alpha m^{\mu*}n^\beta m^{\nu*},
\end{align}
\end{subequations}
where $\delta C_{\alpha\mu\beta\nu}[h]$ is the linearized Weyl tensor. These variables are related by the GHP prime operation $\delta\psi_4 = \delta\psi'_0$. At first order, they are the first-order perturbations of the Weyl scalars, $\psi^{(1)}_0=\delta \psi_0[h^{(1)}]$ and $\psi^{(1)}_4=\delta \psi_4[h^{(1)}]$. However, at second order they form only part of the second-order perturbations of the Weyl scalars,%
\begin{subequations}%
\begin{align}
\psi^{(2)}_0 &= \delta \psi_0[h^{(2)}]+\delta^2\psi_0[h^{(1)},e_{(1)}],\\
\psi^{(2)}_4 &= \delta \psi_4[h^{(2)}]+\delta^2\psi_4[h^{(1)},e_{(1)}]. 
\end{align}
\end{subequations}
Here $e^{\alpha}_{(1)}$ represents the first-order perturbations of the tetrad legs, $\{l^\alpha_{(1)},n^\alpha_{(1)},m^\alpha_{(1)},m^{\alpha*}_{(1)}\}$. The quantities $\delta^2\psi_0$ and $\delta^2\psi_4$ are quadratic in the tetrad perturbations and in $h^{(1)}_{\alpha\beta}$. In Ref.~\cite{Spiers_etal1}, we discuss some of the merits of working with $\delta\psi_0[h^{(2)}]$ or $\delta\psi_4[h^{(2)}]$ rather than with $\psi^{(2)}_{0}$ or $\psi^{(2)}_4$.

For easy reference in the following sections, it will be convenient to write the four-dimensional Teukolsky equations prior to mode decomposition. Written in compact form, they read%
\begin{subequations}\label{4D Teukolsky}%
\begin{align}
\hat O \delta\psi_0 &= \hat S^{\alpha\beta}S_{\alpha\beta},\\
\hat O' \delta\psi_4 &= \hat S'^{\alpha\beta}S_{\alpha\beta},
\end{align}
\end{subequations}
where we use a hat to denote linear differential operators, $S_{\alpha\beta}$ is the source in the Einstein equation $\delta G_{\alpha\beta}[h]=S_{\alpha\beta}$, and we again use the GHP prime to show the manifest symmetry between the two equations. The operators $\hat O$ and $\hat S^{\alpha\beta}$ are given explicitly in four-dimensional form in Eqs.~(58) and~(59) of Ref.~\cite{Pound:2021qin} (with the conversions in Table~\ref{table:derivatives} and $\tau=\tau'=0$ in Schwarzschild). When acting on $\delta\psi_{0}$ and $\delta\psi_4$, respectively, these reduce in the Schwarzschild case to
\begin{subequations}
\begin{align}
  \label{Ohat}
\hat O = (\thorn-5\rho)(\thorn'-\rho') - \frac{1}{2r^2}\edth\edth' +\frac{3M}{r^3},\\
\hat O' = (\thorn'-5\rho')(\thorn-\rho) - \frac{1}{2r^2}\edth'\edth +\frac{3M}{r^3},
\end{align}
\end{subequations}
where we have introduced
\beq\label{rho and rho'}
\rho = -\frac{r_al^a}{r} \quad \text{and} \quad \rho' = -\frac{r_an^a}{r}.
\eeq
We will also refer to their adjoints, $\hat O^\dagger$ and $\hat S^\dagger_{\alpha\beta}$ and their GHP primes, adopting Wald's definition
\beq
\int A\,\hat U B \,dV = \int (\hat U^\dagger\! A) B \,dV
\eeq
for any linear differential operator $\hat U$ and tensor test fields $A$ and $B$ of appropriate ranks. The operator $\hat O^\dagger$ is related to $\hat O'$ by
\beq
\hat O^\dagger = r^4\hat O' r^{-4},
\eeq
and  $\hat S^\dagger_{\alpha\beta}$ is given explicitly in Eq.~(60) of Ref.~\cite{Pound:2021qin} (again with the conversions in Table~\ref{table:derivatives} and $\tau=\tau'=0$).

Expanded in spin-weighted harmonics, $\delta\psi_0$ and $\delta\psi_4$ read
\begin{subequations}
\begin{align}
\delta\psi_0 &= \sum_{\ell m}\delta\psi^{\ell m}_0\, {}_2Y_{\ell m},\\
\delta\psi_4 &= \sum_{\ell m}\delta\psi^{\ell m}_4\, {}_{-2}Y_{\ell m},
\end{align}
\end{subequations}
with coefficients
\begin{multline}
\delta\psi^{\ell m}_0 = -\frac{1}{4r^2}\Big[\lam{\ell}{2} h^{\ell m}_{ll} + 2^{3/2}\mu_\ell \,r (\thorn - \rho) h^{\ell m}_{lm}\\
  +2r^2(\thorn-\rho)^2h^{\ell m}_{mm} \Big]
\end{multline}
and
\begin{multline}
\delta\psi^{\ell m}_4 = -\frac{1}{4r^2}\Big[\lam{\ell}{2} h^{\ell m}_{nn} - 2^{3/2}\mu_\ell \,r (\thorn' - \rho') h^{\ell m}_{nm^*}\\
  +2r^2(\thorn'-\rho')^2h^{\ell m}_{m^*m^*} \Big].
\end{multline}
Note that the formulas for $\delta\psi^{\ell m}_0$ and $\delta\psi^{\ell m}_4$ differ only by an application of the GHP prime operation and a change in sign of the term proportional to $\mu_\ell$; the latter change stems from the sign difference between Eqs.~\eqref{spin raising} and \eqref{spin lowering}. This pattern carries over to the equations for the source modes below.

At the level of these $\ell m$ modes, the Teukolsky equations~\eqref{4D Teukolsky} become
\begin{subequations}
\begin{align}
\left[(\thorn-5\rho)(\thorn'-\rho')+\frac{\mu^2_\ell}{2r^2}+\frac{3M}{r^3}\right]\delta\psi^{\ell m}_{0} &= S^{\ell m}_{0},\\
\left[(\thorn'-5\rho')(\thorn-\rho)+\frac{\mu^2_\ell}{2r^2}+\frac{3M}{r^3}\right]\delta\psi^{\ell m}_{4} &= S^{\ell m}_{4},
\end{align}
\end{subequations}
with source terms
\begin{subequations}
\begin{align}
S^{\ell m}_0 &= -\frac{\lam{\ell}{2}}{4r^2} S^{\ell m}_{ll}-\frac{1}{\sqrt{2}r}\mu_\ell(\thorn-3\rho)S^{\ell m}_{lm}\nonumber\\
		&\quad -\frac{1}{2}(\thorn-5\rho)(\thorn-\rho)S^{\ell m}_{mm},\\
S^{\ell m}_4 &= -\frac{\lam{\ell}{2}}{4r^2} S^{\ell m}_{nn}+\frac{1}{\sqrt{2}r}\mu_\ell(\thorn'-3\rho')S^{\ell m}_{nm^*}\nonumber\\
		&\quad -\frac{1}{2}(\thorn'-5\rho')(\thorn'-\rho')S^{\ell m}_{m^*m^*}.
\end{align}
\end{subequations}

\subsubsection{Green--Hollands--Zimmerman metric reconstruction}

Metric reconstruction from $\delta\psi_0$ or $\delta\psi_4$ has traditionally followed a method due to Chrzanowski, Cohen, and Kegeles~\cite{Chrzanowski:75,Cohen-Kegeles:75,Cohen-Kegeles:79}, neatly explained by Wald~\cite{Wald:78}. That method was specialized to homogeneous solutions with $S_{\alpha\beta}=0$, but it has recently been extended to generic sourced perturbations by GHZ.

The method writes the metric perturbation in a radiation gauge in terms of a Hertz potential $\Phi$ and a ``corrector tensor'' $x_{\alpha\beta}$. We first focus on reconstruction in the IRG, in which case the perturbation reads
\beq\label{hIRG GHZ}
h^{\rm IRG}_{\alpha\beta} = 2{\rm Re}\left(\hat S^\dagger_{\alpha\beta}\Phi_{\rm IRG}\right) + x^{\rm IRG}_{\alpha\beta}.
\eeq
The traditional, source-free reconstruction method uses only the first term; the corrector tensor then corrects for the failure of that method in the presence of sources.

For convenience below, we define 
\beq
k^{\rm IRG}_{\alpha\beta}:=2{\rm Re}\left(\hat S^\dagger_{\alpha\beta}\Phi_{\rm IRG}\right). 
\eeq
It satisfies 
\beq
k^{\rm IRG}_{\alpha\beta}l^\beta=0=k^{\rm IRG}_{\alpha\beta}g^{\alpha\beta}, 
\eeq
while $x^{\rm IRG}_{\alpha\beta}$ satisfies 
\beq
x^{\rm IRG}_{\alpha\beta}l^\beta=0 \quad \text{but}\quad x^{\rm IRG}_{\alpha\beta}g^{\alpha\beta}\neq0. 
\eeq
$x^{\rm IRG}_{\alpha\beta}$ satisfies the Einstein equation 
\beq\label{xIRG EFE}
\delta G_{\alpha\beta}[x^{\rm IRG}]l^\beta=S_{\alpha\beta}l^\beta, 
\eeq
and it is only nonzero if $S_{\alpha\beta}l^\beta\neq0$. $k^{\rm IRG}_{\alpha\beta}$ then satisfies the remainder of the Einstein equation,
\beq
\delta G_{\alpha\beta}[k^{\rm IRG}] = S_{\alpha\beta}-\delta G_{\alpha\beta}[x^{\rm IRG}]
\eeq
(implying $\delta G_{\alpha\beta}[k^{\rm IRG}]l^\beta = 0$). Unlike $x^{\rm IRG}_{\alpha\beta}$, $k^{\rm IRG}_{\alpha\beta}$ is nonzero so long as $\delta\psi_0$ is nonzero, even if $S_{\alpha\beta}$ vanishes. The Hertz potential itself satisfies the adjoint Teukolsky equation $\hat O^\dagger\Phi_{\rm IRG}=\eta_{\rm IRG}$, where the source $\eta_{\rm IRG}$ vanishes if $S_{\alpha\beta}=0$, but we will not require $\eta_{\rm IRG}$ explicitly. 

At the level of modes, the nonzero components of $k^{\rm IRG}_{\alpha\beta}$ are given by
\begin{subequations}\label{k IRG}
\begin{align}
k^{\ell m}_{nn} &= -\frac{\lam{\ell}{2}}{4r^2}\left(\Phi^{\ell m}_{\rm IRG}+\bar\Phi^{\ell m}_{\rm IRG} \right),\\
k^{\ell m}_{nm^*} &= \frac{\mu_\ell}{2\sqrt{2}r}(\thorn+2\rho)\Phi^{\ell m}_{\rm IRG},\\
k^{\ell m}_{m^*m^*} &= -\frac{1}{2}(\thorn-\rho)(\thorn+3\rho)\Phi^{\ell m}_{\rm IRG}
\end{align}
\end{subequations}
together with
\begin{subequations}
\begin{align}
k^{\ell m}_{nm}&=(-1)^{m+1}\left(k^{\ell,-m}_{nm^*}\right)^* ,\\
k^{\ell m}_{mm}&=(-1)^{m}\left(k^{\ell,-m}_{m^*m^*}\right)^*.
\end{align}
\end{subequations}
We suppress the IRG label on the left-hand side to avoid overcrowded notation. The modes of the Hertz potential can be found by solving the inversion relation
\beq\label{Hertz IRG}
\frac{1}{4}\thorn^4\bar\Phi_{\rm IRG}^{\ell m} = \delta\psi^{\ell m}_0.
\eeq
Note that for the Hertz potential (and no other quantity) we use a bar to denote the complex conjugate; modes of $\Phi$ can be obtained from modes of $\bar\Phi$ using Eq.~\eqref{v conjugate spin-weighted}, which in this case implies
\beq
\Phi_{\rm IRG}^{\ell m} = (-1)^m(\bar\Phi^{\ell,- m}_{\rm IRG})^*.
\eeq
Because $\thorn$ is a derivative along $l^a$, Eq.~\eqref{Hertz IRG} is a fourth-order ordinary differential equation along outgoing null rays. In the Kinnersley tetrad in retarded coordinates, it reduces to $\frac{1}{4}\partial_r^4\bar\Phi_{\rm IRG}^{\ell m} = \delta\psi^{\ell m}_0$. 

Like the Hertz potential, the corrector tensor can be obtained by solving ordinary differential equations along outgoing null rays. The Einstein equation~\eqref{xIRG EFE} reduces to the following hierarchical sequence of differential equations for the nonzero components of $x^{\rm IRG}_{\alpha\beta}$:
\begin{subequations}\label{x IRG}
\begin{align}
\rho^2\thorn\!\left(\rho^{-2}\thorn x^{\ell m}_{mm^*}\right) &= -S^{\ell m}_{ll},\\
\thorn\!\left[\rho^2\thorn(\rho^{-2}x^{\ell m}_{nm})\right] &= -2S^{\ell m}_{lm} + \frac{\lam{\ell}{1}}{\sqrt{2}r}\thorn x^{\ell m}_{mm^*},\\
\rho^2\thorn(\rho^{-1}x^{\ell m}_{nn}) &= -S^{\ell m}_{ln} +\bigg[\frac{\lam{\ell}{1}^2}{2r^2} - \frac{2M}{r^3}+2\rho\rho' \nonumber\\
&\qquad\quad -2(\rho'\thorn+\rho\thorn')+\thorn'\thorn\bigg]x^{\ell m}_{mm^*} \nonumber\\
&\quad -\frac{\lam{\ell}{1}}{2^{3/2}r}\left(\thorn - 3\rho\right)(x^{\ell m}_{nm} - x^{\ell m}_{nm^*}),
\end{align}
\end{subequations}
together with
\begin{subequations}
\begin{align}
x^{\ell m}_{nm^*}&=(-1)^{m+1}\left(x^{\ell,-m}_{nm}\right)^* ,
\end{align}
\end{subequations}
again suppressing the IRG label. We again recall that $\thorn=\partial_r$ in the Kinnersley tetrad in retarded coordinates.

Reconstruction in the ORG is precisely analogous. All of its formulas can be obtained from those above by applying the GHP prime together with $\mu_\ell\to-\mu_\ell$ and $\lam{\ell}{1}\to-\lam{\ell}{1}$, beginning with the prime of Eq.~\eqref{hIRG GHZ},
\beq
h^{\rm ORG}_{\alpha\beta} = 2{\rm Re}\left(\hat S'^\dagger_{\alpha\beta}\Phi_{\rm ORG}\right) + x^{\rm ORG}_{\alpha\beta}.
\eeq
The primed analogues of Eqs.~\eqref{Hertz IRG} and~\eqref{x IRG} are ordinary differential equations along ingoing null rays. In this case the differential equations simplify in the Hartle--Hawking tetrad in ingoing null coordinates, for which $\thorn'=-\partial_r$.

%

\section{Conclusion}\label{Conclusion}

In this paper we have attempted a comprehensive treatment of the second-order perturbative field equations in a Schwarzschild background.

With tensor spherical harmonics defined in Eqs.~\eqref{vector harmonics def} and~\eqref{tensor harmonics def} and first- and second-order metric perturbations $h^{(1)}_{\mu\nu}$ and $h^{(2)}_{\mu\nu}$ expanded as in Eq.~\eqref{h-Yilm}, the harmonic coefficients $h^{(n)\ell m}_{\boldsymbol\cdot}$ satisfy the first- and second-order Einstein equations~\eqref{EFEnlm generic}. Those equations apply in all gauges, but the left-hand side of the equations, as well as the source terms in the second-order field equation, will take different values depending on the choice of gauge. If we specialize to the Lorenz gauge, the field equations reduce to Eq.~\eqref{EFEnlm Lorenz}. If we adopt gauge-invariant field variables, as defined in Sec.~\ref{gauge fixing}, then the field equations instead reduce to Eq.~\eqref{EFEnlm fixed}.

As an alternative to directly solving the Einstein equations, one can solve master equations for scalar variables and then reconstruct the metric perturbation. This is described in Sec.~\ref{RWZ} within a RWZ formalism and in Sec.~\ref{Teukolsky} within a Teukolsky formalism.

Regardless of which formulation is adopted, the essential ingredient in each of the second-order field equations is a coupling formula: a formula for each mode of the second-order source as an infinite sum over products of modes of the first-order field. In previous literature, such coupling formulas were presented for the Regge--Wheeler and Zerilli equations, omitting $\ell=0,1$ modes from both the first- and second-order fields and restricting $\ell\geq2$ modes to the RWZ gauge (or equivalently, adopting  gauge-fixed invariant RWZ variables for $\ell\geq2$). We have presented a number of extensions and generalizations: (i) the sources in the Einstein equations in ``raw'' form without any gauge fixing and with arbitrary mode content, (ii) the sources in the Einstein equations in terms of invariant fields, including invariant $\ell=0,1$ modes derived using a novel gauge-fixing method, (iii) the second-order RWZ sources including $\ell=0,1$ first-order input modes, (iv) the source in the second-order Teukolsky equation in a convenient ``reduced'' form. We have also, as far as possible, attempted to cast all of these in a unified framework. Most importantly, we have created the package \pkg{PerturbationEquations} to work with these sources in a variety of conventions.

A crucial question in all these formulations is how quickly, if at all, the coupling formulas converge. The answer, as analyzed in Ref.~\cite{Miller-Wardell-Pound:16}, is that the convergence is dictated by the smoothness of the first-order field. If the first-order field contains a singularity, then its harmonic modes decay slowly in a neighborhood of the singularity, and evaluating the sum of products in the coupling formulas becomes infeasible. This challenge is critical in the self-force context, where the convergence becomes arbitrarily slow at points arbitrarily close to the particle. This has been overcome in practice using the strategy described in Ref.~\cite{Miller-Wardell-Pound:16}, which requires knowledge of the four-dimensional singularity structure. More efficient strategies are likely possible. 

Another important question is how well the source terms behave at large distances and near the horizon. Poor behaviour there will represent an obstacle to numerical integration of the field equations and difficulties in establishing physically correct boundary conditions. This has posed a problem in most second-order calculations. As reviewed in Sec.~\ref{RWZ}, the RWZ metric variables are not asymptotically flat, which causes poor behaviour of the second-order sources in the RWZ equations. This can be ameliorated by working with modified master functions. However, even in asymptotically well-behaved gauges, such as the Lorenz gauge, practical implementations can encounter nonconvergent retarded integrals~\cite{Pound:15c}. This problem has been addressed by developing post-Minkowskian and near-horizon expansions that can be used to derive physical boundary conditions in the Lorenz gauge, as described in Ref.~\cite{Pound:15c} and in forthcoming work. A superior method of eliminating nonconvergent integrals, explained in Refs.~\cite{Spiers_etal1, Spiers_etal}, is to work with variables adapted to the physical lightcone structure of the perturbed spacetime.

Followup papers will detail how the second-order self-force results in Refs.~\cite{Pound:2019lzj,Warburton:2021kwk,Bonetto:2021exn,Wardell:2021fyy} were obtained by combining (i) the coupling formulas derived in this paper, (ii) the two-timescale expansion of the Lorenz-gauge field equations in Ref.~\cite{Miller:2020bft}, (iii) the strategies developed in Refs.~\cite{Miller-Wardell-Pound:16,Pound:15c} to overcome slow convergence of the coupling formulas and nonconvergence of two-timescale retarded integrals, (iv) an extension of the ``puncture scheme'' in Ref.~\cite{Wardell-Warburton:15}, and (v) the punctures in Ref.~\cite{Pound-Miller:14}.

\acknowledgments
We are grateful to Jonathan Thompson for comparing our $\ell=0,1$ invariants to those used in Ref.~\cite{Thompson:2016fxe}. We also thank Leor Barack, Jeremy Miller, and Niels Warburton for numerous helpful discussions. AP and AS acknowledge the support of a Royal Society University Research Fellowship Enhancement Award. AP is additionally supported by a Royal Society University Research Fellowship and a UKRI Frontier Research Grant (as selected by the ERC) under the Horizon Europe
Guarantee scheme [grant number EP/Y008251/1]. AP's early work on this paper was supported by the Natural Sciences and Engineering Research Council of Canada and by the European Research Council under the European Union's Seventh Framework Programme (FP7/2007-2013)/ERC Grant No. 304978. AS acknowledges partial support from STFC Consolidated Grant no. ST/V005596/1.

\appendix

\section{Transformations of $\tilde \xi^\alpha_{(n)}$ and $\widetilde h^{(n)}_{\alpha\beta}$}\label{transformation of xi tilde}

In Sec.~\ref{gauge fixing} we discuss the construction of invariant perturbations $\widetilde{h}^{(n)}_{\mu\nu}$ through a procedure of gauge fixing. These invariants depend on vector fields $\tilde\xi^\alpha_{(n)}$. Here we show how $\tilde\xi^\alpha_{(n)}$ and $\widetilde{h}^{(n)}_{\mu\nu}$ transform under a gauge transformation in the case of a fully fixed gauge and in the case of an only partially fixed gauge.

\subsection{Fully fixed gauge}

First we review the derivation of the standard transformation rules for metric perturbations, as displayed in Eq.~\eqref{Deltah}. Perturbations $h^{(n)}_{\mu\nu}$ are defined by an identification between the background spacetime manifold and the perturbed spacetime manifold. An identification is specified by the flow of a vector field $X$ through the one-parameter family of spacetimes, where $\e$ is the parameter; see Fig. 1 of Ref.~\cite{Pound:15b} for an illustration. The $n$th-order metric perturbation is then the $n$th-order term in the Taylor expansion along this flow, 
\beq
h^{(n)}_{\mu\nu}=\frac{1}{n!}\Lie^n_X{\sf g}_{\mu\nu}, 
\eeq
evaluated on the background manifold. If we instead work in a gauge specified by a vector field $Y$, then $h^{(n)}_{\mu\nu}$ is instead $\frac{1}{n!}\Lie^n_Y{\sf g}_{\mu\nu}$. The two quantities differ by 
\beq
\Delta h^{(n)}_{\mu\nu} = \frac{1}{n!}\Lie^n_Y{\sf g}_{\mu\nu} - \frac{1}{n!}\Lie^n_X{\sf g}_{\mu\nu}.
\eeq
Basic manipulations of the Lie derivatives put $\Delta h^{(n)}_{\mu\nu}$ in the form~\eqref{Deltah} with the definitions%
\begin{align}
\xi^\mu_{(1)} := Y^\mu-X^\mu,\label{xi1 def}\\
\xi^\mu_{(2)} := \frac{1}{2}[X,Y]^\mu.\label{xi2 def}
\end{align}

Now suppose $\widetilde{h}^{(n)}_{\mu\nu}$ is the $n$th-order metric perturbation in a gauge specified by a vector field $Z$, and $h^{(n)}_{\mu\nu}$ is in the gauge specified by $X$. Then 
\beq
\widetilde{h}^{(n)}_{\mu\nu}=h^{(n)}_{\mu\nu}+\Delta_{X\to Z} h^{(n)}_{\mu\nu}, 
\eeq
where
\beq
\Delta_{X\to Z} h^{(n)}_{\mu\nu} = \frac{1}{n!}\Lie^n_Z{\sf g}_{\mu\nu} - \frac{1}{n!}\Lie^n_X{\sf g}_{\mu\nu}.
\eeq
This can be written in the form~\eqref{tilde h} with definitions analogous to \eqref{xi1 def} and \eqref{xi2 def}: 
\begin{align}
\tilde \xi^\mu_{(1)} := Z^\mu-X^\mu,\\
\tilde \xi^\mu_{(2)} := \frac{1}{2}[X,Z]^\mu.
\end{align}
If $h^{(n)}_{\mu\nu}$ is instead in a gauge specified by $Y$, then the vectors $\tilde \xi^\mu_{(n)}$ become $\tilde \xi^\mu_{(1)} := Z^\mu-Y^\mu$ and $\tilde \xi^\mu_{(2)} := \frac{1}{2}[Y,Z]^\mu$. Therefore, under the transformation from $X$ to $Y$, they transform as
\begin{align}
\Delta\tilde \xi^\mu_{(1)} &:= (Z^\mu-Y^\mu) - (Z^\mu-X^\mu),\\
\Delta\tilde \xi^\mu_{(2)} &:= \frac{1}{2}[Y,Z]^\mu - \frac{1}{2}[X,Z]^\mu.
\end{align}
Expressing these in terms of $\xi^\mu_{(n)}$ and $\tilde\xi^\mu_{(n)}$, we obtain
\begin{align}
\Delta\tilde \xi^\mu_{(1)} &= -\xi^\mu_{(1)},\label{D tilde xi1}\\
\Delta\tilde \xi^\mu_{(2)} &= -\xi^\mu_{(2)}-\frac{1}{2}[\tilde\xi_{(1)},\xi_{(1)}]^\mu.\label{D tilde xi2}
\end{align}

We now show that these transformation rules for $\tilde\xi^\mu_{(n)}$ imply the invariance of $\widetilde{h}^{(n)}_{\mu\nu}$. Referring to the definition of $\widetilde{h}^{(n)}_{\mu\nu}$ in Eq.~\eqref{tilde h}, we see that at first order we have 
\beq\label{D h1 tilde}
\Delta\widetilde{h}^{(1)}_{\mu\nu} = \Delta h^{(1)}_{\mu\nu} + \Lie_{\Delta\tilde\xi_{(1)}}g_{\mu\nu} = \Lie_{\xi_{(1)}+\Delta\tilde\xi_{(1)}}g_{\mu\nu},
\eeq
where we used $\Delta h^{(1)}_{\mu\nu}=\Lie_{\xi_{(1)}}g_{\mu\nu}$. If the gauge of $\widetilde{h}^{(1)}_{\mu\nu}$ is fully fixed, then $\Lie_{\xi_{(1)}+\Delta\tilde\xi_{(1)}}g_{\mu\nu}$ trivially vanishes because in that case $\Delta\tilde\xi^\mu_{(1)}=-\xi^\mu_{(1)}$. 

Again referring to the definition of $\widetilde{h}^{(n)}_{\mu\nu}$ in Eq.~\eqref{tilde h}, we see that at second order, the gauge transformation is
\beq
\Delta\widetilde{h}^{(2)}_{\mu\nu} = \Delta h^{(2)}_{\mu\nu} +  \Lie_{\Delta\tilde\xi_{(2)}}g_{\mu\nu} + \Delta \widetilde{H}_{\mu\nu},
\eeq
where
\begin{multline}
\Delta \widetilde{H}_{\mu\nu} = \Lie_{\tilde \xi_{(1)}+\Delta\tilde\xi_{(1)}}\!\!\left(h^{(1)}_{\mu\nu}+\Delta h^{(1)}_{\mu\nu}+\tfrac{1}{2}\Lie_{\tilde \xi_{(1)}+\Delta \tilde \xi_{(1)}}g_{\mu\nu}\right) \\
	- \Lie_{\tilde \xi_{(1)}}\!\!\left(h^{(1)}_{\mu\nu}+\tfrac{1}{2}\Lie_{\tilde \xi_{(1)}}g_{\mu\nu}\right).
\end{multline}
This can be manipulated into the form
\begin{multline}\label{D h2 tilde}
\Delta\widetilde{h}^{(2)}_{\mu\nu} = \Lie_{(\xi_{(2)}+\Delta\tilde\xi_{(2)}+\frac{1}{2}[\tilde\xi_{(1)},\xi_{(1)}])}g_{\mu\nu}\\
					+\Lie_{\xi_{(1)}+\Delta\tilde\xi_{(1)}}\widetilde{h}^{(1)}_{\mu\nu}+\frac{1}{2}\Lie^2_{\xi_{(1)}+\Delta\tilde \xi_{(1)}}g_{\mu\nu}\\
					+\frac{1}{2}\Lie_{[\tilde\xi_{(1)}-\xi_{(1)},\xi_{(1)}+\Delta\tilde\xi_{(1)}]}g_{\mu\nu}.
\end{multline}
If the gauge of $\widetilde{h}^{(n)}_{\mu\nu}$ is fully fixed, then this immediately vanishes by virtue of Eqs.~\eqref{D tilde xi1} and \eqref{D tilde xi2} for $\Delta\tilde \xi^\mu_{(n)}$. 

\subsection{Partially fixed gauge}

In many instances, the gauge is only partially (or ``mostly'') fixed. The most pervasive such case, discussed in Sec.~\ref{gauge fixing}, is when the partially fixed gauge is specified up to transformations generated by a Killing vector of the background. The gauge of $\widetilde h_{\mu\nu}^{(n)}$ in this scenario depends on the gauge of $h^{(n)}_{\mu\nu}$. 

In the geometrical description from the previous section, the vector field $Z$ itself now depends on the gauge of $h^{(n)}_{\mu\nu}$. We can label it $Z_X$ in the $X$ gauge and $Z_Y$ in the $Y$ gauge. A short calculation leads to the following modified versions of Eqs.~\eqref{D tilde xi1} and \eqref{D tilde xi2}:
\begin{align}
\Delta\tilde \xi^\mu_{(1)} &= -\xi^\mu_{(1)} + \eta^\mu_{(1)},\\
\Delta\tilde \xi^\mu_{(2)} &= -\xi^\mu_{(1)} -\frac{1}{2}[\tilde\xi_{(1)},\xi_{(1)}]^\mu \nonumber\\
&\quad + \eta^\mu_{(2)} +\frac{1}{2}[\xi_{(1)}-\tilde\xi_{(1)},\eta_{(1)}]^\mu,
\end{align}
where $\eta^\mu_{(1)}:=Z^\mu_Y-Z^\mu_X$ and $\eta^\mu_{(2)}:=\frac{1}{2}[Z_X,Z_Y]^\mu$ are the generators of the transformation from the gauge defined by $Z_X$ to the gauge defined by $Z_Y$. If the gauge conditions on $\widetilde h^{(n)}_{\mu\nu}$ specify $Z$ up to an isometry of the background, then $\eta^\mu_{(1)}$ and $\eta^\mu_{(2)}$ must be Killing vectors of the background.

Given these results, we can now assess the transformations of $\widetilde h^{(n)}_{\mu\nu}$ using Eqs.~\eqref{D h1 tilde} and \eqref{D h2 tilde}, which remain valid in this scenario. A trivial calculation shows
\begin{align}
\Delta\widetilde{h}^{(1)}_{\mu\nu} &= \Lie_{\eta_{(1)}}g_{\mu\nu},\label{D h1 tilde partial fixing}\\
\Delta\widetilde{h}^{(2)}_{\mu\nu} &= \Lie_{\eta_{(2)}}g_{\mu\nu}
					+\Lie_{\eta_{(1)}}\widetilde{h}^{(1)}_{\mu\nu}
					+\frac{1}{2}\Lie^2_{\eta_{(1)}}g_{\mu\nu}.\label{D h2 tilde partial fixing}
\end{align}
This is simply the expected transformation from the $Z_X$ gauge to the $Z_Y$ gauge. If $\eta^\mu_{(n)}$ are Killing vectors of the background, then%
\beq
\Delta\widetilde{h}^{(1)}_{\mu\nu}=0
\eeq
because $\Lie_{\eta_{(1)}}g_{\mu\nu}=0$. Therefore $\widetilde{h}^{(1)}_{\mu\nu}$ is invariant even if the gauge fixing is only specified up to background Killing symmetries. On the other hand, Eq.~\eqref{D h2 tilde} reduces to
\begin{align}
\Delta\widetilde{h}^{(2)}_{\mu\nu} &= \Lie_{\eta_{(1)}} \widetilde{h}^{(1)}_{\mu\nu}.
\end{align}
Here we see that $\widetilde{h}^{(2)}_{\mu\nu}$ is only invariant if (i) the first-order metric perturbation possesses the same Killing symmetries as $g_{\mu\nu}$ and (ii)  the gauge of $\widetilde{h}^{(1)}_{\mu\nu}$ respects that symmetry. 

The same considerations apply for more generic unfixed low ($\ell=0,1$) modes. If we fix the gauge of the $\ell>1$ modes of $\widetilde h^{(1)}_{\mu\nu}$ but leave the gauge of the low modes entirely unfixed, then $\eta^\mu_{(n)}$ are generic $\ell=0,1$ vector fields. More concretely, if $\tilde\xi_{(1),\ell\leq 1}^{\mu}=0$, then 
\beq
\Delta\tilde\xi_{(1)}^{\mu} = - \xi^\mu_{(1),\ell>1} = - \xi^\mu_{(1)}+\xi^\mu_{(1),\ell\leq1};
\eeq
that is, $\eta^\mu_{(1)}$ is simply the $\ell=0,1$ piece of $\xi^\mu_{(1)}$. The $\ell>1$ modes of $\widetilde{h}^{(1)}_{\mu\nu}$ are invariant because, by virtue of Eq.~\eqref{D h1 tilde partial fixing}, $\Delta\widetilde{h}^{(1)}_{\mu\nu}$ is confined to $\ell=0,1$ modes. But the $\ell>1$ modes of $\widetilde{h}^{(2)}_{\mu\nu}$ are \emph{not} invariant because the term $\Lie_{\eta_{(1)}}\widetilde{h}^{(1)}_{\mu\nu}=\Lie_{\xi_{(1),\ell\leq1}}\widetilde{h}^{(1)}_{\mu\nu}$ in Eq.~\eqref{D h2 tilde partial fixing} generically contributes nonzero amounts to all modes of $\Delta\widetilde{h}^{(2)}_{\mu\nu}$.


\section{Decomposition of curvature tensors into tensors on ${\cal M}^2\times S^2$ }\label{M2S2-decomposition}

In this appendix we provide the $2+2$D decompositions of the linear and quadratic quantities appearing in the Einstein equations~\eqref{EFE1} and \eqref{EFE2}, following the decomposition procedure described in Sec.~\ref{covariant 2+2 decomposition}.

\subsection{Linear terms}
The linear quantities in Eqs.~\eqref{EFE1} and \eqref{EFE2} are ${\cal E}_{\mu\nu}[h]$ and ${\cal F}_{\mu\nu}[h]$, defined in Eqs.~\eqref{E} and \eqref{F}. For a generic perturbation $h_{\mu\nu}$ with components $h_{ab}$, $h_{aA}$, and $h_{AB}$, the $2+2D$ decomposition of ${\cal E}_{\mu\nu}$ is%
\begin{widetext}
\begin{subequations}\label{E-M2S2}
\begin{align}
{\cal E}_{ab}[h] &= \delta_{c}\delta^{c}h_{ab}+ \frac{1}{r^2}\D{A}D^{A}h_{ab}+\frac{2}{r^4} h^{A}{}_{\!A} r_{a} r_{b}  + \frac{2}{r} r^{c} \delta_{c}h_{ab} -  \frac{4}{r^3}\D{A}h_{(a}{}^{\!A}r_{b)}  -  \frac{4}{r^2} h_{c(a} r_{b)} r^{c}   \nonumber\\
			&\quad +2 R[\delta]_{a}{}^{c}{}_{b}{}^{d}h_{cd} -\frac{2M}{r^5}h^A{}_{\!A} q_{ab},\\
{\cal E}_{aA}[h] &= \delta_{b}\delta^{b}h_{aA} + \frac{1}{r^2} \D{B}D^{B}h_{aA} + \frac{2}{r} r^{b} \D{A}h_{ab}  -  \frac{2}{r^3} r_{a} \D{B}h_{A}{}^{\!B} -  \frac{f h_{aA} + 4 h^{b}{}_{\!A} r_{a} r_{b} }{r^2},\\
{\cal E}_{AB}[h] &=  \delta_{a}\delta^{a}h_{AB} + \frac{1}{r^2}\D{F}D^{F}h_{AB} - \frac{4 M h_{AB}}{r^3}  + 2 h_{ab} r^{a} r^{b} \Omega_{AB} + \frac{2}{r} r^{a} (2\D{(A}h_{B)a}- \delta_{a}h_{AB})  \nonumber\\
			&\quad + \frac{2}{r^2} R[D]_{A}{}^{F}{}_{B}{}^{G}h_{FG}  + \frac{2}{r^2} f h_{AB}  -  \frac{2}{r^2} f h_{F}{}^{\!F} \Omega_{AB}-\frac{2M}{r}h_a{}^a\Omega_{AB}.
\end{align}
\end{subequations}
Recall that $h_{ab}$, $h_{aA}$, and $h_{AB}$ are defined with indices down, and their indices are raised with $g^{ab}$ and $\Omega^{AB}$, such that $h^{aA}:=g^{ab}\Omega^{AB}h_{bB}$, for example.

${\cal F}_{\mu\nu}$ is defined in Eq.~\eqref{F}. Its components are found to be%
\begingroup
\allowdisplaybreaks
\begin{subequations}\label{F-M2S2}
\begin{align}
{\cal F}_{ab}[h] &= \frac{2 M}{r^5} \bar h^{A}{}_{\!A} g_{ab} -  \frac{6}{r^4}\bar h^{A}{}_{\!A} r_{a} r_{b} -  2\delta_{(a}\delta^{c}h_{b)c} 
			 - \frac{4}{r} r^{c} \delta_{(a}\bar h_{b)c} + \frac{2}{r^2}\left(2 \bar h_{c(a} r_{b)}r^{c} - \delta_{(a}D^{A}\bar h_{b)A}\right) \nonumber\\*
			&\quad -  \frac{4 M}{r^3} \bar h_{ab} +  \frac{2}{r^3}r_{(a} \left(\delta_{b)}\bar h^{A}{}_{\!A} + 2 D^{A}\bar h_{b)A}\right)  ,\\
{\cal F}_{aA}[h] &= - \delta_{a}\delta_{b}\bar h^{b}{}_{\!A} -  \frac{2}{r} \bigl[r^{b} (\delta_{a}\bar h_{bA} + \D{A}\bar h_{ab})- r_{a} \delta_{b}\bar h^{b}{}_{\!A}\bigr] 
			 -  \D{A}\delta_{b}\bar h_{a}{}^{b} + \frac{6}{r^2} \bar h^{b}{}_{\!A} r_{a} r_{b} - \frac{1}{r^2} \delta_{a}\D{B}\bar h_{A}{}^{\!B}  \nonumber\\*
			&\quad - \frac{1}{r^2} \D{A}\D{B}\bar h_{a}{}^{\!B} -  \frac{2 M}{r^3} \bar h_{aA} + \frac{1}{r^3} r_{a} (\D{A}\bar h^{B}{}_{\!B} + 4 \D{B}\bar h_{A}{}^{\!B}),\\
{\cal F}_{AB}[h] &= -4 \bar h_{ab} r^{a} r^{b} \Omega_{AB} - 2 r^{a} r \Omega_{AB} \delta_{b}\bar h_{a}{}^{b} - 2\D{(A}\delta^{a}\bar h_{B)a}
			 + \frac{2 f}{r^2}\bar h^{F}{}_{\!F} \Omega_{AB} -  \frac{2}{r^2}\D{(A}D^{F}\bar h_{B)F} \nonumber\\*
			&\quad -  \frac{2}{r} r^{a} \left(2\D{(A}\bar h_{B)a} + \Omega_{AB} \D{F}\bar h_{a}{}^{\!F}\right),
\end{align}
\end{subequations}
\endgroup
where we have opted to not explicitly express $\bar h_{\mu\nu}$ in terms of $h_{\mu\nu}$, and parentheses around indices indicate symmetrization.

\subsection{Quadratic terms}
The quadratic quantities in Eq.~\eqref{EFE2} are ${\cal A}_{\mu\nu}[h]$, ${\cal B}_{\mu\nu}[h]$, and ${\cal C}_{\mu\nu}[h]$, defined in Eqs.~\eqref{A}--\eqref{C}. Decomposing them in the same manner as we did the linear terms, we find for ${\cal A}_{\mu\nu}$,
\begin{subequations}\label{A-M2S2}\allowdisplaybreaks
\begin{align}
{\cal A}_{ab}[h] &= 2r^{-6} h_{AB} h^{AB} r_{a} r_{b} + \tfrac{1}{2} \delta_{a}h^{cd} \delta_{b}h_{cd}  -  2r^{-5}h^{AB} r_{(a} \delta_{b)}h_{AB} + 2\delta_{[d}h_{c]b}\delta^{d}h_{a}{}^{c} + 2r^{-4}\D{[B}h_{A]b} D^{B}h_{a}{}^{\!A} \nonumber\\*
				& \quad  -  2r^{-3}h^{cA}r_{(a} \left(\delta_{b)}h_{cA} -  \delta_{|c|}h_{b)A} + \D{|A|}h_{b)c}\right) + r^{-2}\delta_{a}h^{cA} \delta_{b}h_{cA} + r^{-2}\delta_{c}h_{bA} \delta^{c}h_{a}{}^{\!A} \nonumber\\*
				& \quad - 2r^{-2}\delta^{c}h_{(a}{}^{\!A} \D{|A|}h_{b)c}  + r^{-2}\D{A}h_{bc} D^{A}h_{a}{}^{c} + 2r^{-4} h_{cA} h^{cA} r_{a} r_{b} + \tfrac{1}{2}r^{-4} \delta_{a}h^{AB} \delta_{b}h_{AB},\\
{\cal A}_{aA}[h] &= - \delta_{b}h_{ac} \delta^{c}h^{b}{}_{\!A} + \delta_{c}h_{ab} \delta^{c}h^{b}{}_{\!A} + \tfrac{1}{2} \delta_{a}h^{bc} \D{A}h_{bc}  - r^{-1}h^{b}{}_{\!A} r^{c} (2\delta_{[a}h_{b]c} + \delta_{c}h_{ab}) + r^{-1}h_{b}{}^{c} r^{b} \D{A}h_{ac}   \nonumber\\*
				& \quad + r^{-1}h_{bc} r^{b} (\delta_{a}h^{c}{}_{\!A} -  \delta^{c}h_{aA})  -  r^{-5}h^{BF} r_{a} \D{A}h_{BF}  - r^{-3}h_{AB} r^{b} (\delta_{a}h_{b}{}^{\!B} + \delta_{b}h_{a}{}^{\!B}) +  r^{-3}h_{A}{}^{\!B} r^{b} \D{B}h_{ab} \nonumber\\*
				& \quad +  2r^{-3}h^{bB}(r_{(b} \delta_{a)}h_{AB} + \D{A}h_{B[a}r_{b]}  - \D{B}h_{A(a}r_{b)})  - 2r^{-2} h^{b}{}_{\!A} h_{bc} r_{a} r^{c} +  r^{-2}\delta_{b}h_{AB} \delta^{b}h_{a}{}^{\!B} \nonumber\\*
				& \quad +  r^{-2}\delta_{a}h^{bB} \D{A}h_{bB} - r^{-2}\delta^{b}h_{a}{}^{\!B} \D{B}h_{bA}  - r^{-2}\delta^{b}h_{A}{}^{\!B} \D{B}h_{ab} +  r^{-2}\D{B}h_{ab} D^{B}h^{b}{}_{\!A} \nonumber\\*
				& \quad - 2 r^{-4} h^{bB} h_{AB} r_{a} r_{b} +  \tfrac{1}{2} r^{-4} \delta_{a}h^{BF} \D{A}h_{BF} + 2r^{-4}\D{[F}h_{B]A}D^{F}h_{a}{}^{\!B},\\
{\cal A}_{AB}[h] &= 2 h_{a}{}^{c} h_{bc} r^{a} r^{b} \Omega_{AB} -  \delta_{a}h_{bB} \delta^{b}h^{a}{}_{\!A} + \delta_{b}h_{aB} \delta^{b}h^{a}{}_{\!A}  + r^{-1}h^{a}{}_{\!B} r^{b} (2\delta_{[a}h_{b]A} -  \D{A}h_{ab})   + 2 fr^{-4} h_{A}{}^{\!F} h_{BF}\nonumber\\*
				&\quad + 2r^{-1} r^{a} \bigl(2h_{ab} \D{(A}h^{b}{}_{\!B)}- h_{a}{}^{b} \delta_{b}h_{AB}\bigr)  + r^{-1}h^{a}{}_{\!A} r^{b} (2\delta_{[a}h_{b]B} -  \D{B}h_{ab}) + \tfrac{1}{2} \D{A}h^{ab} \D{B}h_{ab}  \nonumber\\*
				&\quad -  2r^{-3}h_{(A}{}^{\!F} r^{a} \delta_{|a|}h_{B)F}  - 2r^{-3}h_{F(A} r^{a} \D{B)}h_{a}{}^{\!F}  + 4r^{-3} h^{aF} r_{a} \D{(A}h_{B)F} - 2r^{-3} h^{aF} r_{a} \D{F}h_{AB} \nonumber\\*
				&\quad + 2 r^{-3} h_{F(A} r^{a} D^{F}h_{B)a} +2 fr^{-2} h_{aB} h^{a}{}_{\!A}  + 2r^{-2} h^{aF} h^{b}{}_{\!F} r_{a} r_{b} \Omega_{AB} + r^{-2}\delta_{a}h_{BF} \delta^{a}h_{A}{}^{\!F} \nonumber\\*
				&\quad + r^{-2}\D{A}h^{aF} \D{B}h_{aF} -  2r^{-2}\delta_{a}h_{F(A} D^{F}h^{a}{}_{\!B)}  + r^{-2}\D{F}h_{aB} D^{F}h^{a}{}_{\!A} + \tfrac{1}{2}r^{-4} \D{A}h^{FG} \D{B}h_{FG} \nonumber\\*
				&\quad  +2r^{-4}\D{[G}h_{F]B} D^{G}h_{A}{}^{\!F},
\end{align}
\end{subequations}
where square brackets indicate antisymmetrization and vertical bars indicate that the enclosed indices are excluded from the symmetrization or antisymmetrization.  For ${\cal B}_{\mu\nu}$,
\begin{subequations}\label{B-M2S2}\allowdisplaybreaks
\begin{align}
{\cal B}_{ab}[h] &= - 2 Mr^{-7} h_{AB} h^{AB} g_{ab} + 2r^{-6} h_{AB} h^{AB} r_{a} r_{b}  -  2 Mr^{-5} h_{cA} h^{cA} g_{ab} - 2r^{-5}h^{AB} r_{(a} \delta_{b)}h_{AB}  \nonumber\\*
				&\quad + h^{cd} (\delta_{b}\delta_{a}h_{cd} -  \delta_{d}\delta_{a}h_{bc} -  \delta_{d}\delta_{b}h_{ac} + \delta_{d}\delta_{c}h_{ab})   - r^{-3}h^{A}{}_{\!A} r^{c} (2\delta_{(a}h_{b)c} -  \delta_{c}h_{ab}) \nonumber\\*
				&\quad + r^{-3} h^{cA} \bigl[2 r_{c} (2\delta_{(a}h_{b)A} -  \D{A}h_{ab})  - 2 r_{(a} (\delta_{b)}h_{cA} + \delta_{|c|}h_{b)A} -  \D{|A|}h_{b)c})\bigr]  + r^{-2}h^{cA} (2 \delta_{b}\delta_{a}h_{cA} - 2 \delta_{c}\delta_{(a}h_{b)A}  \nonumber\\*
				&\quad  - 2 \D{A}\delta_{(a}h_{b)c}+ 2 \D{A}\delta_{c}h_{ab}) + r^{-4}h^{AB} (\delta_{b}\delta_{a}h_{AB} + \D{B}\D{A}h_{ab})  - 2r^{-4} h_{AB} D^{B}\delta_{(a}h_{b)}{}^{\!A},\\
{\cal B}_{aA}[h] &= h_{bc} (\delta^{c}\delta^{b}h_{aA}- \delta^{c}\delta_{a}h^{b}{}_{\!A}) + 2 Mr^{-3} h^{b}{}_{A} h_{ab} + r^{-1}h^{b}{}_{\!A} r^{c} (2\delta_{(a}h_{b)c} - \delta_{c}h_{ab})  + r^{-3}h_{AB} r^{b} \delta_{a}h_{b}{}^{\!B} \nonumber\\*
				&\quad - r^{-1} h_{bc} \bigl[r^{b} (\delta_{a}h^{c}{}_{\!A} + \delta^{c}h_{aA}) - 2 r_{a} \delta^{c}h^{b}{}_{\!A}\bigr]  + r^{-1}h_{b}{}^{c} r^{b} \D{A}h_{ac} -  r^{-1}h^{bc} r_{a} \D{A}h_{bc} \nonumber\\*
				&\quad + 2h^{bc} \D{A}\delta_{[a}h_{c]b}  + 2r^{-3}h^{bB}r_{(a} \delta_{b)}h_{AB}  -  r^{-3}h_{AB} r^{b} \delta_{b}h_{a}{}^{\!B} + 6r^{-3} h^{bB} r_{[b} \D{|A|}h_{a]B}  \nonumber\\*
				&\quad -  r^{-3}h^{B}{}_{\!B} r^{b} (2\delta_{[a}h_{b]A} + \D{A}h_{ab}) + 6r^{-3} h^{bB} r_{[a} \D{|B|}h_{b]A}  + r^{-3}h_{A}{}^{\!B} r^{b} \D{B}h_{ab} \nonumber\\*
				&\quad - r^{-2}h^{bB} (\delta_{b}\delta_{a}h_{AB} + 2\delta_{b}\D{[A}h_{B]a} - 2 \D{A}\delta_{a}h_{bB} + 2\D{B}\delta_{[a}h_{b]A} + \D{B}\D{A}h_{ab}) + 2Mr^{-5} h_{a}{}^{\!B} h_{AB} \nonumber\\*
				&\quad + 4r^{-5}h^{BF} r_{a} \D{[F}h_{A]B}  - 4r^{-4} h^{bB} h_{AB} r_{a} r_{b}  + 2r^{-4} h^{b}{}_{\!A} h^{B}{}_{\!B} r_{a} r_{b}  + r^{-4} h^{BF} \D{A}\delta_{a}h_{BF} \nonumber\\*
				&\quad - r^{-4}h^{BF} \D{F}\delta_{a}h_{AB} - r^{-4}h_{BF} D^{F}\D{A}h_{a}{}^{\!B} + r^{-4} h_{BF} D^{F}D^{B}h_{aA},\\
{\cal B}_{AB}[h] &= - 2h_{ab} \delta^{b}\D{(A}h^{a}{}_{\!B)} + 2r^{-3}  h_{F(A} r^{a} D^{F}h_{B)a}+ h^{bc} r^{a} r \Omega_{AB} (\delta_{a}h_{bc} - 2 \delta_{c}h_{ab}) \nonumber\\*
				&\quad + h^{ab} (\delta_{b}\delta_{a}h_{AB} + \D{B}\D{A}h_{ab}) -  2 Mr^{-1} h_{ab} h^{ab} \Omega_{AB} + 2r^{-1} h^{a}{}_{\!A} r^{b} \delta_{[a}h_{b]B}  \nonumber\\*
				&\quad - 4 r^{-1}h^{aF} r^{b} \Omega_{AB} \delta_{[a}h_{b]F}  - 2r^{-1} h_{a}{}^{b} r^{a} \delta_{b}h_{AB}  + r^{-1} h^{a}{}_{\!B} r^{b} (2\delta_{[a}h_{b]A} + \D{A}h_{ab})  \nonumber\\*
				&\quad + r^{-1} h^{a}{}_{\!A} r^{b} \D{B}h_{ab} - 2r^{-1} h^{aF} r^{b} \Omega_{AB} \D{F}h_{ab}  - 4r^{-2} h^{a}{}_{A} h^{b}{}_{\!B} r_{a} r_{b} + 4r^{-2} h_{ab} h_{AB} r^{a} r^{b}  \nonumber\\*
				&\quad + 4r^{-2} h^{aF} h^{b}{}_{\!F} r_{a} r_{b} \Omega_{AB} - 2r^{-2} h_{ab} h^{F}{}_{\!F} r^{a} r^{b} \Omega_{AB} + 2r^{-2} f (h_{aB} h^{a}{}_{\!A} -  h_{aF} h^{aF} \Omega_{AB})  \nonumber\\*
				&\quad - 2r^{-2}h^{aF} \delta_{a}\D{(A}h_{B)F} - 4r^{-3} h^{aF} r_{a} \D{F}h_{AB}   + r^{-2} h^{aF} \delta_{a}\D{F}h_{AB} + 2r^{-2} h^{aF} \D{B}\D{A}h_{aF}  \nonumber\\*
				&\quad + r^{-2} h^{aF} \D{F}\delta_{a}h_{AB} - 2r^{-2}h^{aF} \D{F}\D{(A}h_{B)a}    - 2r^{-3} h_{FG} r^{a} \Omega_{AB} D^{G}h_{a}{}^{\!F} \nonumber\\*
				&\quad + 2 f r^{-4} (h_{A}{}^{\!F} h_{BF} -  h_{FG} h^{FG} \Omega_{AB})  +r^{-4} h^{FG} (\D{B}\D{A}h_{FG} -  2\D{G}\D{(A}h_{B)F}  + \D{G}\D{F}h_{AB}) \nonumber\\*
				&\quad -  2 Mr^{-3} h_{aF} h^{aF} \Omega_{AB}  - 2r^{-3} h_{(A}{}^{\!F} r^{a} \delta_{|a|}h_{B)F} + r^{-3}  h^{FG} r^{a} \Omega_{AB} \delta_{a}h_{FG} \nonumber\\*
				&\quad + 2r^{-3} h_{F(A} r^{a} \D{B)}h_{a}{}^{\!F} + 4r^{-3} h^{aF} r_{a} \D{(A}h_{B)F} + r^{-3} h^{F}{}_{\!F} r^{a} (\delta_{a}h_{AB} -  2\D{(A}h_{B)a}).
\end{align}
\end{subequations}
For ${\cal C}_{\mu\nu}$,
\begin{subequations}\label{C-M2S2}\allowdisplaybreaks
\begin{align}
{\cal C}_{ab}[h] &=- 2r^{-1} h_{c}{}^{d} r^{c} (2\delta_{(a}h_{b)d} -  \delta_{d}h_{ab}) + \tfrac{1}{2} (2\delta_{(a}h_{b)}{}^{c} -  \delta^{c}h_{ab}) 
						(\delta_{c}h^{d}{}_{d} - 2 \delta_{d}h_{c}{}^{d}) \nonumber\\*
				&\quad-  2r^{-3} h^{cA} r_{c} (2\delta_{(a}h_{b)A} -  \D{A}h_{ab}) + \tfrac{1}{2} r^{-2}\bigl[2\delta_{(a}h_{b)}{}^{c} \delta_{c}h^{A}{}_{\!A} -  \delta_{c}h^{A}{}_{\!A} \delta^{c}h_{ab} \nonumber\\*
				&\quad- 4 \delta_{(a}h_{b)c} \D{A}h^{cA} 
						+ 2 \delta_{c}h_{ab} \D{A}h^{cA} + 2 \delta_{c}h^{cA} \D{A}h_{ab} + 2\delta_{(a}h_{b)}{}^{\!A} (\D{A}h^{c}{}_{c}-2 \delta_{c}h^{c}{}_{\!A}) \nonumber\\*
				&\quad - \D{A}h^{c}{}_{c} D^{A}h_{ab}\bigr] + \tfrac{1}{2}r^{-4}(2\delta_{(a}h_{b)}{}^{\!A} -  D^{A}h_{ab}) (\D{A}h^{B}{}_{\!B} - 2 \D{B}h_{A}{}^{\!B}),\\
{\cal C}_{aA}[h] &= - r^{-1}r_{a}h^{b}{}_{\!A}  (\delta_{b}h^{c}{}_{c} - 2 \delta_{c}h_{b}{}^{c})   - 2r^{-1}r^{b} h_{bc}  (\delta_{a}h^{c}{}_{\!A} - \delta^{c}h_{aA}) - r^{-5}h_{A}{}^{\!B} r_{a} (\D{B}h^{F}{}_{\!F} - 2 \D{F}h_{B}{}^{\!F}) \nonumber\\*
				&\quad- 2r^{-1}r^{b} h_{b}{}^{c}  \D{A}h_{ac} + \tfrac{1}{2} \bigl[\delta_{a}h^{b}{}_{\!A} (\delta_{b}h^{c}{}_{c} - 2 \delta_{c}h_{b}{}^{c})  - \delta_{b}h^{c}{}_{c} \delta^{b}h_{aA} + 2 \delta^{b}h_{aA} \delta_{c}h_{b}{}^{c} - 2 \delta_{b}h^{bc} \D{A}h_{ac} \nonumber\\*
				&\quad+ \delta^{c}h^{b}{}_{b} \D{A}h_{ac}\bigr] - 2r^{-3} h^{bB} r_{b} (\delta_{a}h_{AB} + 2\D{[A}h_{B]a})  +  r^{-3}r_a\bigl[2 h_{AB} \delta_{b}h^{bB} -  h^{b}{}_{\!A} (\delta_{b}h^{B}{}_{\!B} - 2 \D{B}h_{b}{}^{\!B}) \nonumber\\*
				&\quad-  h_{A}{}^{\!B} \D{B}h^{b}{}_{b}\bigr] 
					+ \tfrac{1}{2} r^{-2}\bigl[8 h^{b}{}_{\!A} h_{bc} r_{a} r^{c} - 2 \delta_{a}h_{AB} \delta_{b}h^{bB} + \delta_{a}h^{b}{}_{\!A} \delta_{b}h^{B}{}_{\!B} -  \delta_{b}h^{B}{}_{\!B} \delta^{b}h_{aA}  \nonumber\\*
					&\quad- 2 \delta_{b}h^{bB} \D{A}h_{aB} 
					+ \delta^{b}h^{B}{}_{\!B} \D{A}h_{ab} + 2 \delta_{b}h^{bB} \D{B}h_{aA}  - 2 \delta_{a}h^{b}{}_{\!A} \D{B}h_{b}{}^{\!B} + 2 \delta^{b}h_{aA} \D{B}h_{b}{}^{\!B}  \nonumber\\*
					&\quad- 2 \D{A}h_{ab} \D{B}h^{bB} + \delta_{a}h_{A}{}^{\!B} \D{B}h^{b}{}_{b}  + \D{A}h_{a}{}^{\!B} \D{B}h^{b}{}_{b} 
					-  \D{B}h^{b}{}_{b} D^{B}h_{aA}\bigr] \nonumber\\*
					&\quad+ 4r^{-4} h^{bB} h_{AB} r_{a} r_{b} + \tfrac{1}{2} r^{-4}(\delta_{a}h_{A}{}^{\!B} + \D{A}h_{a}{}^{\!B} - D^{B}h_{aA}) (\D{B}h^{F}{}_{\!F} - 2 \D{F}h_{B}{}^{\!F}),\\
{\cal C}_{AB}[h] &= h_{a}{}^{b} r^{a} r \Omega_{AB} (\delta_{b}h^{c}{}_{c} - 2 \delta_{c}h_{b}{}^{c}) + \tfrac{1}{2} (2 \delta_{a}h^{ab} \delta_{b}h_{AB} 
					-8 h_{a}{}^{c} h_{bc} r^{a} r^{b} \Omega_{AB} -  \delta_{b}h_{AB} \delta^{b}h^{a}{}_{a} + 2\delta_{a}h^{b}{}_{b} \D{(A}h^{a}{}_{\!B)} \nonumber\\*
					&\quad - 4 \delta_{b}h_{a}{}^{b} \D{(A}h^{a}{}_{\!B)} ) + r^{-1}r^a\bigl[h_{a}{}^{b} (2 \delta_{b}h_{AB} + \Omega_{AB} \delta_{b}h^{F}{}_{\!F}) - 2 h_{ab} (2\D{(A}h^{b}{}_{\!B)} + \Omega_{AB} \D{F}h^{bF})\bigr]\nonumber\\*
					&\quad   + r^{-1}h^{aF} r_{a} \Omega_{AB} (\D{F}h^{b}{}_{b}-2 \delta_{b}h^{b}{}_{\!F}) + \tfrac{1}{2}r^{-4}(2\D{(A}h_{B)}{}^{\!F} -  D^{F}h_{AB}) (\D{F}h^{G}{}_{\!G} - 2 \D{G}h_{F}{}^{\!G})\nonumber\\*
					&\quad - \tfrac{1}{2} r^{-2}\bigl[8 h^{aF} h^{b}{}_{\!F} r_{a} r_{b} \Omega_{AB} + 4 \delta_{a}h^{aF} \D{(A}h_{B)F} -  \delta_{a}h^{F}{}_{\!F} (2D_{(A}h^{a}{}_{\!B)}
					- \delta^{a}h_{AB}) \nonumber\\*
					&\quad + 4 \D{(A}h^{a}{}_{\!B)} \D{F}h_{a}{}^{\!F} - 2 \delta_{a}h_{AB} \D{F}h^{aF} - 2 \delta_{a}h^{aF} \D{F}h_{AB} - 2 \D{(A}h_{B)F} D^{F}h^{a}{}_{a}+ \D{F}h_{AB} D^{F}h^{a}{}_{a}\bigr] \nonumber\\*
					&\quad  + r^{-3}r_{a}h^{aF} (2 \D{F}h_{AB} -4 \D{(A}h_{B)F}  + \Omega_{AB} \D{F}h^{G}{}_{\!G} - 2 \Omega_{AB} \D{G}h_{F}{}^{\!G}).
\end{align}
\end{subequations}

\section{Covariant derivatives of scalar harmonics}\label{Covariant derivs of Y}

Our method of decomposing the Einstein equation requires expressing covariant derivatives of $Y_{\ell m}$ in terms of spin-weighted harmonics. We do so by noting that each covariant derivative in $\D{A_1}\cdots \D{A_s} Y_{\ell m}$ acts on a tensor of spin weight 0, such that Eqs.~\eqref{eth} and \eqref{ethbar} imply $\D{A} = \frac{1}{2}(\tilde m_A \edth'+\tilde m^*_A \edth)$. Using this along with Eq.~\eqref{eth m}, we find
\begin{subequations}\label{D...DY}
\begin{align}
\D{A_1}\cdots\D{A_s} Y_{\ell m} &= \frac{1}{2^s}\left(\tilde m_{A_1} \edth'+\tilde m^*_{A_1} \edth\right)\cdots\left(\tilde m_{A_s} \edth'+\tilde m^*_{A_s} \edth\right) Y_{\ell m}\\
&= \frac{1}{2^s}\sum_{\sigma(s)} \prod_{j=1}^s\alpha^s_{i,j} (\tilde m_{A_j}\edth')Y_{\ell m} \\
&= \frac{1}{2^s}\left[\tilde m_{A_1}\cdots\tilde m_{A_s}\edth'^sY_{\ell m}+\tilde m^*_{A_1}\cdots\tilde m^*_{A_s}\edth^sY_{\ell m}+\cdots\right].
\end{align}
\end{subequations}
The sum runs over all of the $2^s$ distinct products of $\tilde m_{A_j}\edth'$ and $\tilde m^*_{A_j}\edth$ (with $s$ total factors) acting on $Y_{\ell m}$. This is made precise in the second line, where $\sigma(s)$ is the set of $s$-tuples with all elements either 0 or 1, $\sigma(s)_{i,j}$ is the $j$th element of the $i$th $s$-tuple, $\alpha^s_{i,j}(\tilde m_{A_j}\edth')= \tilde m_{A_j}\edth'$ if $\sigma(s)_{i,j}=0$, and $\alpha^s_{i,j} (\tilde m_{A_j}\edth') = \tilde m^*_{A_j}\edth$ if $\sigma(s)_{i,j}=1$. 

This sum is straightforwardly written in terms of spin-weighted harmonics using the definition~\eqref{sYlm definition} and the identities~\eqref{sYlm identities}. For $s=1$ and $s=2$, the results are Eqs.~\eqref{DY} and \eqref{DDY}. For $s=3$ and $s=4$, the results are
\begin{align}
\D{A}\D{B}\D{C}Y_{\ell m} &= \frac{1}{8}\lam{\ell}{3}\left({}_{-3}Y_{\ell m}\tilde m_A \tilde m_B \tilde m_C-{}_3Y_{\ell m}\tilde m^*_A \tilde m^*_B \tilde m^*_C\right)\nonumber\\
&\quad +\frac{1}{8}\lam{\ell}{2}\sqrt{( l+2)( l-1)}\left({}_{1}Y_{\ell m}\tilde m_A \tilde m^*_B\tilde m^*_C-{}_{-1}Y_{\ell m}\tilde m^*_A\tilde m_B\tilde m_C\right) \nonumber\\
&\quad -\frac{1}{4}\lam{\ell}{1}^3\left({}_{-1}Y_{\ell m}\tilde m_A-{}_1Y_{\ell m}\tilde m^*_A\right)\Omega_{BC},\label{DDDY}
\end{align}
and
\begin{align}
\D{A}\D{B}\D{C}\D{D}Y_{\ell m} &= \frac{1}{16}\lam{\ell}{4}\left({}_{-4}Y_{\ell m}\tilde m_A\tilde m_B\tilde m_C\tilde m_D+{}_4Y_{\ell m}\tilde m^*_A\tilde m^*_B\tilde m^*_C\tilde m^*_D\right)\nonumber\\
&\quad -\frac{1}{16}\lam{\ell}{2}\left\{ {}_{2}Y_{\ell m}\left[( l+3)( l-2)\tilde m_A\tilde m^*_B+( l+2)( l-1)\tilde m^*_A\tilde m_B\right]\tilde m^*_C\tilde m^*_D\right.\nonumber\\
&\quad +\left.{}_{-2}Y_{\ell m}\left[( l+3)( l-2)\tilde m^*_A\tilde m_B+( l+2)( l-1)\tilde m_A\tilde m^*_B\right]\tilde m_C\tilde m_D\right\} \nonumber\\
&\quad -\frac{1}{8}\lam{\ell}{1}^3\sqrt{( l+2)( l-1)}\left({}_{-2}Y_{\ell m}\tilde m_A\tilde m_B+{}_2Y_{\ell m}\tilde m^*_A\tilde m^*_B\right)\Omega_{CD}\nonumber\\
&\quad +\frac{1}{4}\lam{\ell}{1}^4\,\Omega_{AB}\Omega_{CD}Y_{\ell m}
+\frac{1}{16} \lam{\ell}{2}^2(\tilde m_A\tilde m_B\tilde m^*_C\tilde m^*_D+\tilde m^*_A\tilde m^*_B\tilde m_C\tilde m_D)Y_{\ell m}.\label{DDDDY}
\end{align}

Equation~\eqref{D...DY} can also be used to derive the relationships~\eqref{tensor vs spin-weighted harmonics} between tensor and spin-weighted harmonics. $Y^{\ell m}_{ A_1\cdots A_s}$ is the symmetric trace-free piece of Eq.~\eqref{D...DY}, which picks out the two terms that contain only $\tilde m$'s or only $\tilde m^*$'s. Equation~\eqref{tensor Y vs sYlm} then immediately follows from the definition~\eqref{sYlm definition}. Given the identity \eqref{epsm=im}, the definition~\eqref{rank-s harmonics} of $X^{\ell m}_{A_1\cdots A_s}$ likewise picks out the two terms that contain only $\tilde m$'s or only $\tilde m^*$'s in Eq.~\eqref{D...DY}, and Eq.~\eqref{tensor X vs sYlm} then immediately follows.

\section{Quadratic coupling functions}\label{coupling functions}

In this appendix we list the coupling functions appearing in the decompositions of quadratic quantities. We only provide expressions for even-parity coupling functions: for example, ${\cal A}^{\ell'm's'\ell''m''s''}_{a+}$ and ${\cal A}^{\ell'm's'\ell''m''s''}_{+}$. Odd-parity analogues can be obtained using the rule
\beq
\begin{array}{r l}
{\cal A}^{\ell'm's'\ell''m''s''}_{a-} &= -i{\cal A}^{\ell'm's'\ell''m''s''}_{a+}\\[5pt]
{\cal A}^{\ell'm's'\ell''m''s''}_{-} &= -i{\cal A}^{\ell'm's'\ell''m''s''}_{+}
\end{array}
\quad\text{with }\sigma_\pm \to -\sigma_\mp.
\eeq
The quantities $\sigma:=(-1)^{\ell+\ell'+\ell''}$ and $\sigma_\pm:=\sigma\pm1$ arise from use of Eq.~\eqref{C symmetries}.

\subsection{Gauge transformation}

The coupling functions appearing in Eq.~\eqref{H decomposition} are
\begin{subequations}\label{H coupling functions}
\allowdisplaybreaks%
\begin{align}
H^{\ell'm'0\ell''m''0}_{ab} &= 2h^{\ell^{\prime \prime}{} \mathit{m}^{\prime \prime}{}}_{c(a} \delta_{b)}\zeta_{\ell^{\prime}{} \mathit{m}^{\prime}{}}^{c} + \zeta_{\ell^{\prime \prime}{} \mathit{m}^{\prime \prime}{}}^{c} \delta_{c}h^{\ell^{\prime}{} \mathit{m}^{\prime}{}}_{ab}
+\delta_{(a}\zeta_{\ell^{\prime}{}
\mathit{m}^{\prime}{}}^{c} \delta_{b)}\zeta^{\ell^{\prime \prime}{}\mathit{m}^{\prime \prime}}_{c} 
+ \zeta_{\ell^{\prime \prime}{} \mathit{m}^{\prime \prime}{}}^{c}\delta_{c}\delta_{(a}\zeta^{\ell^{\prime}{} \mathit{m}^{\prime}}_{b)} 
+ \delta_{(a}\zeta^{\ell^{\prime}{} \mathit{m}^{\prime}}_{|c|} \delta^{c}\zeta^{\ell^{\prime \prime}{}\mathit{m}^{\prime \prime}}_{b)},\\
H^{\ell'm'1\ell''m''-1}_{ab} &= - \frac{i}{2} \Bigl[h^{\ell^{\prime \prime}m^{\prime\prime}}_{ab} (Z^{-}_{\ell^{\prime}m^{\prime}} - i Z^{+}_{\ell^{\prime}m^{\prime}}) - 
h^{\ell^{\prime}m^{\prime}}_{ab} 
(Z^{-}_{\ell^{\prime \prime}{} \mathit{m}^{\prime \prime}{}} + i 
Z^{+}_{\ell^{\prime \prime}{} \mathit{m}^{\prime \prime}{}}) + 2(h_{(a+}^{\ell^{\prime \prime}{} \mathit{m}^{\prime \prime}{}}-i h_{(a-}^{\ell^{\prime \prime}{} \mathit{m}^{\prime \prime}{}} ) \
(\delta_{b)}Z^{-}_{\ell^{\prime}{} \mathit{m}^{\prime}{}} - i 
\delta_{b)}Z^{+}_{\ell^{\prime}{} \mathit{m}^{\prime}{}}) \nonumber\\*
&\qquad\quad -2(h_{(a+}^{\ell^{\prime}{} \mathit{m}^{\prime}{}} + i h_{(a-}^{\ell^{\prime}{} \mathit{m}^{\prime}{}}) (\delta_{b)}Z^{-}_{\ell^{\prime \prime}\mathit{m}^{\prime \prime}{}}+i \delta_{b)}Z^{+}_{\ell^{\prime \prime}{} \mathit{m}^{\prime\prime}{}})\Bigr] 
+ r^2 (\delta_{(a}Z^{-}_{\ell^{\prime \prime}{} \mathit{m}^{\prime \prime}{}} + i \delta_{(a}Z^{+}_{\ell^{\prime \prime}{} \mathit{m}^{\prime \prime}{}}) ( i \delta_{b)}Z^{+}_{\ell^{\prime}{} \mathit{m}^{\prime}{}}-\delta_{b)}Z^{-}_{\ell^{\prime}{} \mathit{m}^{\prime}{}}) \nonumber\\*
&\quad -  \frac{i}{2} \Bigl[\zeta^{\ell^{\prime \prime}{} \mathit{m}^{\prime \prime}}_{(a} (\delta_{b)}Z^{-}_{\ell^{\prime}{} \mathit{m}^{\prime}{}} - i \delta_{b)}Z^{+}_{\ell^{\prime}{} \mathit{m}^{\prime}{}}) 
 - \zeta^{\ell^{\prime}{} \mathit{m}^{\prime}}_{(a} (\delta_{b)}Z^{-}_{\ell^{\prime \prime}{} \mathit{m}^{\prime \prime}{}} + i \delta_{b)}Z^{+}_{\ell^{\prime \prime}{} \mathit{m}^{\prime \prime}{}}) 
 - (Z^{-}_{\ell^{\prime \prime}{} \mathit{m}^{\prime \prime}{}} + i Z^{+}_{\ell^{\prime \prime}{} \mathit{m}^{\prime \prime}{}}) \delta_{(a}\zeta^{\ell^{\prime}{} \mathit{m}^{\prime}}_{b)}\nonumber\\*
&\qquad\quad+ (Z^{-}_{\ell^{\prime}{} \mathit{m}^{\prime}{}} - i Z^{+}_{\ell^{\prime}{} \mathit{m}^{\prime}{}}) \delta_{(a}\zeta^{\ell^{\prime \prime}{} \mathit{m}^{\prime \prime}}_{b)}\Bigr],\\
H^{\ell'm'1\ell''m''0}_{a+} &= \frac{1}{4} \Bigl\{2 \sigma_{+}{} h^{\ell^{\prime \prime}{} \mathit{m}^{\prime \prime}{}}_{ac} \zeta_{\ell^{\prime}{} \mathit{m}^{\prime}{}}^{c} 
+\lam{\ell^{\prime \prime}}{1}^2\Bigl[ i \sigma_{-}{} h_{a+}^{\ell^{\prime \prime}{} \mathit{m}^{\prime \prime}{}} Z^{-}_{\ell^{\prime}{} \mathit{m}^{\prime}{}}  
-  \sigma_{+}{} h_{a-}^{\ell^{\prime}{} \mathit{m}^{\prime}{}} Z^{-}_{\ell^{\prime \prime}{} \mathit{m}^{\prime \prime}{}} 
- i\sigma_- h_{a+}^{\ell^{\prime}{} \mathit{m}^{\prime}{}} Z^{-}_{\ell^{\prime \prime}{} \mathit{m}^{\prime \prime}{}}  \nonumber\\*
&\qquad\ 
-  \sigma_{+}{} h_{a+}^{\ell^{\prime \prime}{} \mathit{m}^{\prime \prime}{}} Z^{+}_{\ell^{\prime}{} \mathit{m}^{\prime}{}} 
+ h_{a-}^{\ell^{\prime \prime}{} \mathit{m}^{\prime \prime}{}} (\sigma_{+}{} Z^{-}_{\ell^{\prime}{} \mathit{m}^{\prime}{}} + i \sigma_{-}{} Z^{+}_{\ell^{\prime}{} \mathit{m}^{\prime}{}}) 
+ i \sigma_{-}{} h_{a-}^{\ell^{\prime}{} \mathit{m}^{\prime}{}} Z^{+}_{\ell^{\prime \prime}{} \mathit{m}^{\prime \prime}{}} 
-  \sigma_+ h_{a+}^{\ell^{\prime}{} \mathit{m}^{\prime}{}} Z^{+}_{\ell^{\prime \prime}{} \mathit{m}^{\prime \prime}{}}\Bigr]\nonumber\\*
&\qquad\  
- 2i \sigma_- h_{\trAB}^{\ell^{\prime \prime}{} \mathit{m}^{\prime \prime}{}} \delta_{a}Z^{-}_{\ell^{\prime}{} \mathit{m}^{\prime}{}} 
+ 2 \sigma_+ h_{\trAB}^{\ell^{\prime \prime}{} \mathit{m}^{\prime \prime}{}} \delta_{a}Z^{+}_{\ell^{\prime}{} \mathit{m}^{\prime}{}} 
- 2i \sigma_- h_{c-}^{\ell^{\prime}{} \mathit{m}^{\prime}{}} \delta_{a}\zeta_{\ell^{\prime \prime}{} \mathit{m}^{\prime \prime}{}}^{c} 
+ 2 \sigma_+ h_{c+}^{\ell^{\prime}{} \mathit{m}^{\prime}{}} \delta_{a}\zeta_{\ell^{\prime \prime}{} \mathit{m}^{\prime \prime}}^{c} \nonumber\\*
&\qquad\  
- 2i \sigma_- \zeta_{\ell^{\prime \prime}{} \mathit{m}^{\prime \prime}{}}^{c} \delta_{c}h_{a-}^{\ell^{\prime}{} \mathit{m}^{\prime}{}}
+ 2 \sigma_{+}{} \zeta_{\ell^{\prime \prime}{} \mathit{m}^{\prime \prime}{}}^{c} \delta_{c}h_{a+}^{\ell^{\prime}{} \mathit{m}^{\prime}{}}\Bigr\}
-r r_{c}\zeta_{\ell^{\prime \prime}{} \mathit{m}^{\prime \prime}}^{c} 
\bigl(i \sigma_- \delta_{a}Z^{-}_{\ell^{\prime}{} \mathit{m}^{\prime}{}} 
- \sigma_{+} \delta_{a}Z^{+}_{\ell^{\prime}{} \mathit{m}^{\prime}{}}\bigr) \nonumber\\*
&\quad 
+ \frac{1}{8} \Bigl[\lam{\ell^{\prime \prime}}{1}^2\bigl(-i \sigma_{-}{} Z^{-}_{\ell^{\prime \prime}{} \mathit{m}^{\prime \prime}{}} \zeta^{\ell^{\prime}{} \mathit{m}^{\prime}}_{a}
 -  \sigma_{+}{} Z^{+}_{\ell^{\prime \prime}{} \mathit{m}^{\prime \prime}{}} \zeta^{\ell^{\prime}{} \mathit{m}^{\prime}}_{a} 
+ i \sigma_{-}{} Z^{-}_{\ell^{\prime}{} \mathit{m}^{\prime}{}} \zeta^{\ell^{\prime \prime}{} \mathit{m}^{\prime \prime}}_{a}  
-  \sigma_{+}{} Z^{+}_{\ell^{\prime}{} \mathit{m}^{\prime}{}} \zeta^{\ell^{\prime \prime}{} \mathit{m}^{\prime \prime}}_{a} \bigr) \nonumber\\*
&\qquad\quad
+ 2\sigma_+ \zeta_{\ell^{\prime}{} \mathit{m}^{\prime}{}}^{c} \delta_{a}\zeta^{\ell^{\prime \prime}{} \mathit{m}^{\prime \prime}}_{c} 
+ 2 \sigma_{+}{} \zeta_{\ell^{\prime}{} \mathit{m}^{\prime}{}}^{c} \delta_{a}\zeta^{\ell^{\prime \prime}{} \mathit{m}^{\prime \prime}}_{c} 
+ 2 \sigma_{+}{} \zeta_{\ell^{\prime \prime}{} \mathit{m}^{\prime \prime}{}}^{c} \delta_{c}\zeta^{\ell^{\prime}{} \mathit{m}^{\prime}}_{ a} 
+ 2 \sigma_{+}{} \zeta_{\ell^{\prime}{} \mathit{m}^{\prime}{}}^{c} \delta_{c}\zeta^{\ell^{\prime \prime}{} \mathit{m}^{\prime \prime}}_{a}\Bigr] \nonumber\\*
&\quad 
+ \frac{1}{8} r^2 \Bigl\{- Z^{-}_{\ell^{\prime \prime}{} \mathit{m}^{\prime \prime}{}} \lam{\ell^{\prime \prime}}{1}^2 (\sigma_{+}{} \delta_{a}Z^{-}_{\ell^{\prime}{} \mathit{m}^{\prime}{}} + i \sigma_{-}{} \delta_{a}Z^{+}_{\ell^{\prime}{} \mathit{m}^{\prime}{}}) 
+ 3Z^{+}_{\ell^{\prime \prime}{} \mathit{m}^{\prime \prime}{}}\lam{\ell^{\prime \prime}}{1}^2 \bigl(i \sigma_{-} \delta_{a}Z^{-}_{\ell^{\prime}{} \mathit{m}^{\prime}{}} -  \sigma_{+}{} \delta_{a}Z^{+}_{\ell^{\prime}{} \mathit{m}^{\prime}{}}\bigr) \nonumber\\*
&\qquad\qquad + \lam{\ell^{\prime \prime}}{1}^2 \bigl[(\sigma_{+}{} Z^{-}_{\ell^{\prime}{} \mathit{m}^{\prime}{}} + i \sigma_{-}{} Z^{+}_{\ell^{\prime}{} \mathit{m}^{\prime}{}}) \delta_{a}Z^{-}_{\ell^{\prime \prime}{} \mathit{m}^{\prime \prime}{}} + (i \sigma_{-}{} Z^{-}_{\ell^{\prime}{} \mathit{m}^{\prime}{}} -  \sigma_{+}{} Z^{+}_{\ell^{\prime}{} \mathit{m}^{\prime}{}}) \delta_{a}Z^{+}_{\ell^{\prime \prime}{} \mathit{m}^{\prime \prime}{}}\bigr] \nonumber\\*
&\qquad\qquad
+ 2 \sigma_{+}{} \zeta_{\ell^{\prime \prime}{} \mathit{m}^{\prime \prime}{}}^{c} \delta_{c}\delta_{a}Z^{+}_{\ell^{\prime}{} \mathit{m}^{\prime}{}}  
- 2i \sigma_-\delta_{a} (\zeta_{\ell^{\prime \prime}{} \mathit{m}^{\prime \prime}{}}^{c} \delta_{c}Z^{-}_{\ell^{\prime}{} \mathit{m}^{\prime}{}}) 
+ 2\sigma_+ \delta_{a}\zeta_{\ell^{\prime \prime}{} \mathit{m}^{\prime \prime}}^{c} \delta_{c}Z^{+}_{\ell^{\prime}{} \mathit{m}^{\prime}{}}\Bigr\}
,\\
H^{\ell'm'2\ell''m''-1}_{a+} &= - \frac{1}{4}h_{a-}^{\ell^{\prime \prime}{} \mathit{m}^{\prime \prime}{}} (\sigma_{+}{} Z^{-}_{\ell^{\prime}{} \mathit{m}^{\prime}{}} + i \sigma_{-}{} Z^{+}_{\ell^{\prime}{} \mathit{m}^{\prime}{}}) 
+ \frac{i}{4}\Bigl[h_{a+}^{\ell^{\prime \prime}{} \mathit{m}^{\prime \prime}{}} (\sigma_{-}{} Z^{-}_{\ell^{\prime}{} \mathit{m}^{\prime}{}} + i \sigma_{+}{} Z^{+}_{\ell^{\prime}{} \mathit{m}^{\prime}{}})
- \sigma_{-}{} h_{a+}^{\ell^{\prime}{} \mathit{m}^{\prime}{}} Z^{-}_{\ell^{\prime \prime}{} \mathit{m}^{\prime \prime}{}}
 + i \sigma_{+}{} h_{a+}^{\ell^{\prime}{} \mathit{m}^{\prime}{}} Z^{+}_{\ell^{\prime \prime}{} \mathit{m}^{\prime \prime}{}} \nonumber\\*
&\quad + h_{a-}^{\ell^{\prime}{} \mathit{m}^{\prime}{}} (i \sigma_{+}{} Z^{-}_{\ell^{\prime \prime}{} \mathit{m}^{\prime \prime}{}} + \sigma_{-}{} Z^{+}_{\ell^{\prime \prime}{} \mathit{m}^{\prime \prime}{}}) 
+ i \sigma_{+}{} h_{-}^{\ell^{\prime}{} \mathit{m}^{\prime}{}} \delta_{a}Z^{-}_{\ell^{\prime \prime}{} \mathit{m}^{\prime \prime}{}} 
- \sigma_{-}{} h_{+}^{\ell^{\prime}{} \mathit{m}^{\prime}{}} \delta_{a}Z^{-}_{\ell^{\prime \prime}{} \mathit{m}^{\prime \prime}{}} \nonumber\\*
&\quad + (\sigma_{-}{} h_{-}^{\ell^{\prime}{} \mathit{m}^{\prime}{}} + i \sigma_{+}{} h_{+}^{\ell^{\prime}{} \mathit{m}^{\prime}{}}) \delta_{a}Z^{+}_{\ell^{\prime \prime}{} \mathit{m}^{\prime \prime}{}}\Bigr]
-\frac{1}{8} \Bigl[(i \sigma_{-}{} Z^{-}_{\ell^{\prime \prime}{} \mathit{m}^{\prime \prime}{}} + \sigma_{+}{} Z^{+}_{\ell^{\prime \prime}{} \mathit{m}^{\prime \prime}{}}) \zeta^{\ell^{\prime}{} \mathit{m}^{\prime}}_{a} - (i \sigma_{-}{} Z^{-}_{\ell^{\prime}{} \mathit{m}^{\prime}{}} -  \sigma_{+}{} Z^{+}_{\ell^{\prime}{} \mathit{m}^{\prime}{}}) \zeta^{\ell^{\prime \prime}{} \mathit{m}^{\prime \prime}}_{a}\Bigr] 
\nonumber\\*
&\quad - \frac{1}{8} r^2 \Bigl[3 (\sigma_{+}{} Z^{-}_{\ell^{\prime}{} \mathit{m}^{\prime}{}} + i \sigma_{-}{} Z^{+}_{\ell^{\prime}{} \mathit{m}^{\prime}{}}) \delta_{a}Z^{-}_{\ell^{\prime \prime}{} \mathit{m}^{\prime \prime}{}} 
+ Z^{-}_{\ell^{\prime \prime}{} \mathit{m}^{\prime \prime}{}} (\sigma_{+}{} \delta_{a}Z^{-}_{\ell^{\prime}{} \mathit{m}^{\prime}{}} + i \sigma_{-}{} \delta_{a}Z^{+}_{\ell^{\prime}{} \mathit{m}^{\prime}{}}) \nonumber\\*
&\qquad\qquad - Z^{+}_{\ell^{\prime \prime}{} \mathit{m}^{\prime \prime}{}} (i \sigma_{-}{} \delta_{a}Z^{-}_{\ell^{\prime}{} \mathit{m}^{\prime}{}} -  \sigma_{+}{} \delta_{a}Z^{+}_{\ell^{\prime}{} \mathit{m}^{\prime}{}}) 
-3i (\sigma_{-}{} Z^{-}_{\ell^{\prime}{} \mathit{m}^{\prime}{}} + i \sigma_{+}{} Z^{+}_{\ell^{\prime}{} \mathit{m}^{\prime}{}}) \delta_{a}Z^{+}_{\ell^{\prime \prime}{} \mathit{m}^{\prime \prime}{}}\Bigr]
,\\
H^{\ell'm'0\ell''m''0}_\trAB &= - h_{\trAB}^{\ell^{\prime \prime}{} \mathit{m}^{\prime \prime}{}} Z^{+}_{\ell^{\prime}{} \mathit{m}^{\prime}{}} \lam{\ell^{\prime}}{1}^2 
+ \zeta_{\ell^{\prime \prime}{} \mathit{m}^{\prime \prime}{}}^{c} \delta_{c}h_{\trAB}^{\ell^{\prime}{} \mathit{m}^{\prime}{}}
+r_{a} r_{c} \zeta_{\ell^{\prime}{} \mathit{m}^{\prime}}^{c} \zeta_{\ell^{\prime \prime}{} \mathit{m}^{\prime \prime}}^{a} 
+ \frac{M}{r} \zeta^{\ell^{\prime}{} \mathit{m}^{\prime}{}}_{c} \zeta_{\ell^{\prime \prime}{} \mathit{m}^{\prime \prime}{}}^{c}\nonumber\\*
&\quad - r_{c} r \bigl(2 Z^{+}_{\ell^{\prime}{} \mathit{m}^{\prime}{}} \zeta_{\ell^{\prime \prime}{} \mathit{m}^{\prime \prime}}^{c} \lam{\ell^{\prime}}{1}^2  
- \zeta_{\ell^{\prime \prime}{} \mathit{m}^{\prime \prime}{}}^{a} \delta_{a}\zeta_{\ell^{\prime}{} \mathit{m}^{\prime}}^{c}\bigr) 
+ \frac{1}{2} r^2 \lam{\ell^{\prime}}{1}^2  \bigl(Z^{+}_{\ell^{\prime}{} \mathit{m}^{\prime}{}} Z^{+}_{\ell^{\prime \prime}{} \mathit{m}^{\prime \prime}{}} \lam{\ell^{\prime\prime}}{1}^2  -  \zeta_{\ell^{\prime \prime}{} \mathit{m}^{\prime \prime}{}}^{c} \delta_{c}Z^{+}_{\ell^{\prime}{} \mathit{m}^{\prime}{}}\bigr)
,\\
H^{\ell'm'1\ell''m''-1}_\trAB &= \frac{i}{2} \Bigl[h_{\trAB}^{\ell^{\prime \prime}{} \mathit{m}^{\prime \prime}{}} (iZ^{+}_{\ell^{\prime}{} \mathit{m}^{\prime}{}}-Z^{-}_{\ell^{\prime}{} \mathit{m}^{\prime}{}}) 
+ h_{\trAB}^{\ell^{\prime}{} \mathit{m}^{\prime}{}} (Z^{-}_{\ell^{\prime \prime}{} \mathit{m}^{\prime \prime}{}} + i Z^{+}_{\ell^{\prime \prime}{} \mathit{m}^{\prime \prime}{}}) 
+ (h_{c-}^{\ell^{\prime \prime}{} \mathit{m}^{\prime \prime}{}} + i h_{c+}^{\ell^{\prime \prime}{} \mathit{m}^{\prime \prime}{}}) \zeta_{\ell^{\prime}{} \mathit{m}^{\prime}}^{c} \nonumber\\*
&\qquad -  (h_{c-}^{\ell^{\prime}{} \mathit{m}^{\prime}{}} - i h_{c+}^{\ell^{\prime}{} \mathit{m}^{\prime}{}}) \zeta_{\ell^{\prime \prime}{} \mathit{m}^{\prime \prime}}^{c}\Bigr]
- \tfrac{1}{2} \zeta_{\ell^{\prime}{} \mathit{m}^{\prime}{}}^{c} \zeta^{\ell^{\prime \prime}{} \mathit{m}^{\prime \prime}}_{c} 
+ \frac{i}{2}r_{c} r \bigl[(Z^{-}_{\ell^{\prime \prime}{} \mathit{m}^{\prime \prime}{}} + i Z^{+}_{\ell^{\prime \prime}{} \mathit{m}^{\prime \prime}{}}) \zeta_{\ell^{\prime}{} \mathit{m}^{\prime}}^{c} 
-  (Z^{-}_{\ell^{\prime}{} \mathit{m}^{\prime}{}} - i Z^{+}_{\ell^{\prime}{} \mathit{m}^{\prime}{}}) \zeta_{\ell^{\prime \prime}{} \mathit{m}^{\prime \prime}}^{c}\bigr] 
\nonumber\\*
&\quad + \frac{1}{4} r^2 \Bigl\{-i Z^{-}_{\ell^{\prime \prime}{} \mathit{m}^{\prime \prime}{}} Z^{+}_{\ell^{\prime}{} \mathit{m}^{\prime}{}} \lam{\ell^{\prime}}{1}^2 
+ Z^{+}_{\ell^{\prime \prime}{} \mathit{m}^{\prime \prime}{}} \bigl[i Z^{-}_{\ell^{\prime}{} \mathit{m}^{\prime}{}} \lam{\ell^{\prime \prime}}{1}^2 + Z^{+}_{\ell^{\prime}{} \mathit{m}^{\prime}{}} (\lam{\ell^{\prime}}{1}^2 + \lam{\ell^{\prime \prime}}{1}^2)\bigr] 
- \zeta_{\ell^{\prime \prime}{} \mathit{m}^{\prime \prime}{}}^{c} (i \delta_{c}Z^{-}_{\ell^{\prime}{} \mathit{m}^{\prime}{}} + \delta_{c}Z^{+}_{\ell^{\prime}{} \mathit{m}^{\prime}{}}) 
\nonumber\\*
&\qquad\qquad
+ i \zeta_{\ell^{\prime}{} \mathit{m}^{\prime}{}}^{c} (\delta_{c}Z^{-}_{\ell^{\prime \prime}{} \mathit{m}^{\prime \prime}{}} + i \delta_{c}Z^{+}_{\ell^{\prime \prime}{} \mathit{m}^{\prime \prime}{}})\Bigr\}
,\\ 
H^{\ell'm'2\ell''m''-2}_\trAB &= \frac{1}{4} \Bigl[(h_{-}^{\ell^{\prime \prime}{} \mathit{m}^{\prime \prime}{}} + i h_{+}^{\ell^{\prime \prime}{} \mathit{m}^{\prime \prime}{}}) (Z^{-}_{\ell^{\prime}{} \mathit{m}^{\prime}{}} - i Z^{+}_{\ell^{\prime}{} \mathit{m}^{\prime}{}}) 
+ (h_{-}^{\ell^{\prime}{} \mathit{m}^{\prime}{}} - i h_{+}^{\ell^{\prime}{} \mathit{m}^{\prime}{}}) (Z^{-}_{\ell^{\prime \prime}{} \mathit{m}^{\prime \prime}{}} + i Z^{+}_{\ell^{\prime \prime}{} \mathit{m}^{\prime \prime}{}})\Bigr]\nonumber\\*
&\quad
+\frac{1}{2} r^2 (Z^{-}_{\ell^{\prime}{} \mathit{m}^{\prime}{}} - i Z^{+}_{\ell^{\prime}{} \mathit{m}^{\prime}{}}) (Z^{-}_{\ell^{\prime \prime}{} \mathit{m}^{\prime \prime}{}} + i Z^{+}_{\ell^{\prime \prime}{} \mathit{m}^{\prime \prime}{}}),\\
H^{\ell'm'1\ell''m''1}_{+} &= \frac{1}{4} \mu_{\ell^{\prime}}^2 \bigl[(\sigma_{+}{} h_{-}^{\ell^{\prime}{} \mathit{m}^{\prime}{}} + i \sigma_{-}{} h_{+}^{\ell^{\prime}{} \mathit{m}^{\prime}{}}) Z^{-}_{\ell^{\prime \prime}{} \mathit{m}^{\prime \prime}{}} 
+ (i \sigma_{-}{} h_{-}^{\ell^{\prime}{} \mathit{m}^{\prime}{}} -  \sigma_{+}{} h_{+}^{\ell^{\prime}{} \mathit{m}^{\prime}{}}) Z^{+}_{\ell^{\prime \prime}{} \mathit{m}^{\prime \prime}{}}\bigr] 
- (i \sigma_{-}{} h_{c-}^{\ell^{\prime \prime}{} \mathit{m}^{\prime \prime}} - \sigma_{+}{} h_{c+}^{\ell^{\prime \prime}{} \mathit{m}^{\prime \prime}}) \zeta_{\ell^{\prime}{} \mathit{m}^{\prime}}^{c}
\nonumber\\*
&\quad+\frac{1}{2} \sigma_{+}{} \zeta_{\ell^{\prime}{} \mathit{m}^{\prime}{}}^{c} \zeta^{\ell^{\prime \prime}{} \mathit{m}^{\prime \prime}}_{c} 
+ \frac{1}{4} r^2 \Bigl\{\mu_{\ell^{\prime}}^2 \bigl[Z^{-}_{\ell^{\prime}{} \mathit{m}^{\prime}{}} (\sigma_{+}{} Z^{-}_{\ell^{\prime \prime}{} \mathit{m}^{\prime \prime}{}} + i \sigma_{-}{} Z^{+}_{\ell^{\prime \prime}{} \mathit{m}^{\prime \prime}{}}) 
+ Z^{+}_{\ell^{\prime}{} \mathit{m}^{\prime}{}} (i \sigma_{-}{} Z^{-}_{\ell^{\prime \prime}{} \mathit{m}^{\prime \prime}{}} -  \sigma_{+}{} Z^{+}{}_{\ell^{\prime \prime}{} \mathit{m}^{\prime \prime}{}})\bigr] 
\nonumber\\*
&\qquad\qquad\qquad\qquad\qquad\qquad - 2 \zeta_{\ell^{\prime}{} \mathit{m}^{\prime}{}}^{c} (i \sigma_{-}{} \delta_{c}Z^{-}_{\ell^{\prime \prime}{} \mathit{m}^{\prime \prime}{}} - \sigma_{+}{} \delta_{c}Z^{+}_{\ell^{\prime \prime}{} \mathit{m}^{\prime \prime}{}})\Bigr\},\\
H^{\ell'm'2\ell''m''0}_{+} &= - \frac{i}{2}\Bigl[2 h_{\trAB}^{\ell^{\prime \prime}{} \mathit{m}^{\prime \prime}{}} (\sigma_{-}{} Z^{-}_{\ell^{\prime}{} \mathit{m}^{\prime}{}} + i \sigma_{+}{} Z^{+}_{\ell^{\prime}{} \mathit{m}^{\prime}{}}) 
+ \sigma_{-}{} h_{+}^{\ell^{\prime}{} \mathit{m}^{\prime}{}} Z^{-}_{\ell^{\prime \prime}{} \mathit{m}^{\prime \prime}{}} \lam{\ell^{\prime \prime}}{1}^2 
- i \sigma_{+}{} h_{+}^{\ell^{\prime}{} \mathit{m}^{\prime}{}} Z^{+}_{\ell^{\prime \prime}{} \mathit{m}^{\prime \prime}{}} \lam{\ell^{\prime \prime}}{1}^2 \nonumber\\*
&\qquad\quad
- h_{-}^{\ell^{\prime}{} \mathit{m}^{\prime}{}} (i \sigma_{+}{} Z^{-}_{\ell^{\prime \prime}{} \mathit{m}^{\prime \prime}{}} +  \sigma_{-}{} Z^{+}_{\ell^{\prime \prime}{} \mathit{m}^{\prime \prime}{}}) \lam{\ell^{\prime \prime}}{1}^2 
+ \sigma_{-}{} \zeta_{\ell^{\prime \prime}{} \mathit{m}^{\prime \prime}{}}^{c} \delta_{c}h_{-}^{\ell^{\prime}{} \mathit{m}^{\prime}{}} 
+ i \sigma_{+}{} \zeta_{\ell^{\prime \prime}{} \mathit{m}^{\prime \prime}{}}^{c} \delta_{c}h_{+}^{\ell^{\prime}{} \mathit{m}^{\prime}{}}\Bigr]\nonumber\\*
&\quad -2 r_{c} r (i \sigma_{-}{} Z^{-}_{\ell^{\prime}{} \mathit{m}^{\prime}{}} - \sigma_{+}{} Z^{+}_{\ell^{\prime}{} \mathit{m}^{\prime}{}}) \zeta_{\ell^{\prime \prime}{} \mathit{m}^{\prime \prime}}^{c} 
- \frac{1}{2} r^2 \Bigl[i \sigma_{-}{} Z^{-}_{\ell^{\prime \prime}{} \mathit{m}^{\prime \prime}{}} Z^{+}_{\ell^{\prime}{} \mathit{m}^{\prime}{}} \lam{\ell^{\prime \prime}}{1}^2 
+ 2 \sigma_{+}{} Z^{+}_{\ell^{\prime}{} \mathit{m}^{\prime}{}} Z^{+}_{\ell^{\prime \prime}{} \mathit{m}^{\prime \prime}{}} \lam{\ell^{\prime \prime}}{1}^2 \nonumber\\*
&\quad+ Z^{-}_{\ell^{\prime}{} \mathit{m}^{\prime}{}} (\sigma_{+}{} Z^{-}_{\ell^{\prime \prime}{} \mathit{m}^{\prime \prime}{}} - 2i \sigma_{-}{} Z^{+}_{\ell^{\prime \prime}{} \mathit{m}^{\prime \prime}{}}) \lam{\ell^{\prime \prime}}{1}^2 
+ i \sigma_{-}{} \zeta_{\ell^{\prime \prime}{} \mathit{m}^{\prime \prime}{}}^{c} \delta_{c}Z^{-}_{\ell^{\prime}{} \mathit{m}^{\prime}{}} - \sigma_{+}{} \zeta_{\ell^{\prime \prime}{} \mathit{m}^{\prime \prime}{}}^{c} \delta_{c}Z^{+}_{\ell^{\prime}{} \mathit{m}^{\prime}{}}\Bigr]
,\\
H^{\ell'm'3\ell''m''-1}_{+} &= -\frac{1}{4} \bigl[(\sigma_{+}{} h_{-}^{\ell^{\prime}{} \mathit{m}^{\prime}{}} + i \sigma_{-}{} h_{+}^{\ell^{\prime}{} \mathit{m}^{\prime}{}}) Z^{-}_{\ell^{\prime \prime}{} \mathit{m}^{\prime \prime}{}} 
- (i \sigma_{-}{} h_{-}^{\ell^{\prime}{} \mathit{m}^{\prime}{}} -  \sigma_{+}{} h_{+}^{\ell^{\prime}{} \mathit{m}^{\prime}{}}) Z^{+}_{\ell^{\prime \prime}{} \mathit{m}^{\prime \prime}{}}\bigr]\nonumber\\
&\quad-\frac{1}{4} r^2 \bigl[ Z^{-}_{\ell^{\prime \prime}{} \mathit{m}^{\prime \prime}{}} (\sigma_{+}{} Z^{-}_{\ell^{\prime}{} \mathit{m}^{\prime}{}} + i \sigma_{-}{} Z^{+}_{\ell^{\prime}{} \mathit{m}^{\prime}{}}) 
- (i \sigma_{-}{} Z^{-}_{\ell^{\prime}{} \mathit{m}^{\prime}{}} -  \sigma_{+}{} Z^{+}_{\ell^{\prime}{} \mathit{m}^{\prime}{}}) Z^{+}_{\ell^{\prime \prime}{} \mathit{m}^{\prime \prime}{}}\bigr].
\end{align}
\end{subequations}

\subsection{Ricci tensor}

The quantities appearing in Eq.~\eqref{Ailm} are
\begin{subequations}\label{A coupling functions}
\allowdisplaybreaks%
\begin{align}
{\cal A}^{\ell'm'0\ell''m''0}_{ab} &= \frac{4 \widetilde{h}_{\trAB}^{\ell^{\prime} m^{\prime}} \widetilde{h}_{\trAB}^{\ell^{\prime \prime}m^{\prime \prime}} r_{a} r_{b}}{r^6} + \frac{1}{2} \delta_{a}\widetilde{h}^{\ell^{\prime}m^{\prime}}_{cd} \delta_{b}\widetilde{h}^{\ell^{\prime \prime} m^{\prime \prime}}_{ef}g^{ce}g^{df} - \frac{4 \widetilde{h}_{\trAB}^{\ell^{\prime \prime}m^{\prime \prime}} r_{(a} \delta_{b)}\widetilde{h}_{\trAB}^{\ell^{\prime} m^{\prime}}}{r^5}+ \frac{\widetilde{h}_{b-}^{\ell^{\prime}{} m^{\prime}{}} \widetilde{h}_{a-}^{\ell^{\prime \prime}{} m^{\prime \prime}{}} \lam{\ell^{\prime}}{1}^2 \lam{\ell^{\prime \prime}}{1}^2}{r^4}\nonumber\\*
&\quad + \frac{\delta_{a}\widetilde{h}_{\trAB}^{\ell^{\prime}{} m^{\prime}{}} \delta_{b}\widetilde{h}_{\trAB}^{\ell^{\prime \prime}{} m^{\prime \prime}{}}}{r^4} +2\delta_{[d}\widetilde{h}^{\ell^{\prime}{} m^{\prime}{}}_{c]b} \delta^{d}\widetilde{h}^{\ell^{\prime \prime}{} m^{\prime \prime}{}}_{ae}g^{ce},\\
{\cal A}^{\ell'm'1\ell''m''-1}_{ab} &= - \frac{2 \widetilde{h}_{c-}^{\ell^{\prime} m^{\prime}} \widetilde{h}_{d-}^{\ell^{\prime \prime}m^{\prime \prime}}g^{cd} r_{a} r_{b}}{r^4} - \frac{r_{(a}g^{cd}\left(i \widetilde{h}^{\ell^{\prime} m^{\prime}}_{b)c} - \delta_{b)}\widetilde{h}_{c-}^{\ell^{\prime}m^{\prime}}+\delta_{|c|}\widetilde{h}_{b)-}^{\ell^{\prime} m^{\prime}}\right)\widetilde{h}_{d-}^{\ell^{\prime \prime}m^{\prime \prime}}}{r^3} \nonumber\\*
&\quad +  \frac{r_{(a}g^{cd}\left(i \widetilde{h}^{\ell^{\prime \prime}m^{\prime \prime}}_{b)c} + \delta_{b)}\widetilde{h}_{c-}^{\ell^{\prime \prime}{} m^{\prime \prime}}-\delta_{|c|}\widetilde{h}_{b)-}^{\ell^{\prime \prime}m^{\prime \prime}}\right)\widetilde{h}_{d-}^{\ell^{\prime}m^{\prime}}}{r^3} -  \frac{g^{cd}\left(\widetilde{h}^{\ell^{\prime}m^{\prime}}_{c(a} \widetilde{h}^{\ell^{\prime \prime}m^{\prime \prime}}_{b)d} + \delta_{(a}\widetilde{h}_{|c|-}^{\ell^{\prime \prime} m^{\prime \prime}} \delta_{b)}\widetilde{h}_{d-}^{\ell^{\prime} m^{\prime}}\right)}{ r^2} \nonumber\\*
&\quad - \frac{\delta_{c}\widetilde{h}_{(a-}^{\ell^{\prime} m^{\prime}} \delta^{c}\widetilde{h}_{b)-}^{\ell^{\prime \prime} m^{\prime \prime}} - i \bigl(\widetilde{h}^{\ell^{\prime \prime} m^{\prime \prime}}_{c(a}\delta^{c}\widetilde{h}_{b)-}^{\ell^{\prime}m^{\prime}} -  \widetilde{h}^{\ell^{\prime} m^{\prime}}_{c(a} \delta^{c}\widetilde{h}_{b)-}^{\ell^{\prime \prime} m^{\prime \prime}}\bigr)}{ r^2},\\
{\cal A}^{\ell'm'1\ell''m''0}_{a+} &= - \frac{\sigma_{+}{} \widetilde{h}_{\trAB}^{\ell^{\prime}{} m^{\prime}{}} \widetilde{h}_{\trAB}^{\ell^{\prime \prime}{} m^{\prime \prime}{}} r_{a}}{r^5} + \frac{i \sigma_{-}{} \widetilde{h}_{b-}^{\ell^{\prime}{} m^{\prime}{} } \widetilde{h}_{\trAB}^{\ell^{\prime \prime}{} m^{\prime \prime}{}} r_{a} r^{b}}{r^4} + \frac{1}{2r^4} \widetilde{h}_{\trAB}{}^{\ell^{\prime}{} m^{\prime}{}} \bigl(i \sigma_{-}{} \widetilde{h}_{a-}^{\ell^{\prime \prime}{} m^{\prime \prime}{}} \lam{\ell''}{1}^2 + \sigma_{+}{} \delta_{a}\widetilde{h}_{\trAB}^{\ell^{\prime \prime}{} m^{\prime \prime}{}}\bigr) \nonumber\\*
&\quad -  \frac{1}{2r^3} \widetilde{h}_{b-}^{\ell^{\prime}{} m^{\prime}{}} r^{b} \bigl(\sigma_{+}{} \widetilde{h}_{a-}^{\ell^{\prime \prime}{} m^{\prime \prime}{}} \lam{1}{\ell^{\prime \prime}}^2 + i \sigma_{-}{} \delta_{a}\widetilde{h}_{\trAB}{}^{\ell^{\prime \prime}{} m^{\prime \prime}{}}\bigr) +  \frac{1}{2r^3} \widetilde{h}_{\trAB}^{\ell^{\prime \prime}{} m^{\prime \prime}{}} r^{b} \bigl(\sigma_{+}{} \widetilde{h}^{\ell^{\prime}{} m^{\prime}{}}_{ab} + 2i \sigma_{-}{} \delta_{(a}\widetilde{h}_{b)-}^{\ell^{\prime}{} m^{\prime}{}} \bigr) \nonumber\\*
&\quad - \frac{1}{2r^3}i \sigma_{-}{} \widetilde{h}_{b-}^{\ell^{\prime}{} m^{\prime}{}} r_{a} \delta^{b}\widetilde{h}_{\trAB}^{\ell^{\prime \prime}{} m^{\prime \prime}{}} + \frac{1}{r^2}i \sigma_{-}{} \widetilde{h}^{\ell^{\prime \prime}{} m^{\prime \prime}{}}_{bc} \widetilde{h}_{d-}^{\ell^{\prime}{} m^{\prime}{}} r_{a}g^{bd} r^{c} - \frac{1}{2r^2} \sigma_{+}{} \widetilde{h}_{b-}^{\ell^{\prime \prime}{} m^{\prime \prime}{} } \lam{\ell^{\prime \prime}}{1}^2 g^{bc}\delta_{(a}\widetilde{h}_{c)-}^{\ell^{\prime}{} m^{\prime}{}} \nonumber\\*
&\quad -\frac{1}{4r^2} \left[2i\sigma_{-}{} \delta_{b}\widetilde{h}_{\trAB}^{\ell^{\prime \prime}{} m^{\prime \prime}{}} \delta^{b}\widetilde{h}_{a-}^{\ell^{\prime}{} m^{\prime}{}} - \widetilde{h}^{\ell^{\prime}{} m^{\prime}{}}_{ab} g^{bc}\left(i\sigma_{-}{} \widetilde{h}_{c-}^{\ell^{\prime \prime}{} m^{\prime \prime}{}} \lam{\ell^{\prime \prime}}{1}^2 - 2 \sigma_{+}{} \delta_{c}\widetilde{h}_{\trAB}^{\ell^{\prime \prime}{} m^{\prime \prime}{}}\right)\right] \nonumber\\*
&\quad + \frac{1}{2r}i \sigma_{-}{} \widetilde{h}_{b-}^{\ell^{\prime}{} m^{\prime}{}} g^{bd}r^{c} \left(\delta_{a}\widetilde{h}^{\ell^{\prime \prime}{} m^{\prime \prime}{}}_{cd} -  2\delta_{[d}\widetilde{h}^{\ell^{\prime \prime}{} m^{\prime \prime}{}}_{c]a}\right) + \frac{1}{2r} \widetilde{h}^{\ell^{\prime \prime}{} m^{\prime \prime}{}}_{bc} r^{b} g^{cd}\left(\sigma_{+}{} \widetilde{h}^{\ell^{\prime}{} m^{\prime}{}}_{ad} - 2i \sigma_{-}{} \delta_{[a}\widetilde{h}_{d]-}^{\ell^{\prime}{} m^{\prime}}\right) \nonumber\\*
&\quad + \frac{1}{4}  g^{bd}\bigl(\sigma_{+}{} \widetilde{h}^{\ell^{\prime}{} m^{\prime}}_{bc} \delta_{a}\widetilde{h}^{\ell^{\prime \prime}{} m^{\prime \prime}{}}_{de}g^{ce} + 4i \sigma_{-}{} \delta_{[b}\widetilde{h}^{\ell^{\prime \prime}{} m^{\prime \prime}{}}_{c]a}\delta^{c}\widetilde{h}_{d-}^{\ell^{\prime}{} m^{\prime}}\bigr),\\
{\cal A}^{\ell'm'2\ell''m''-1}_{a+} &= \frac{\sigma_{+}{} \widetilde{h}_{b-}^{\ell^{\prime}{} m^{\prime}{} b} \widetilde{h}_{c-}^{\ell^{\prime \prime}{} m^{\prime \prime}{}}g^{bc} r_{a}}{2 r^3} + \frac{\widetilde{h}_{b-}^{\ell^{\prime}{} m^{\prime}}g^{bc} \bigl(i \sigma_{-}{} \widetilde{h}^{\ell^{\prime \prime}{} m^{\prime \prime}{}}_{ac} + 2\sigma_{+}{}\delta_{[c}\widetilde{h}_{a]-}^{\ell^{\prime \prime}{} m^{\prime \prime}{}}\bigr)}{4 r^2},\\
{\cal A}^{\ell'm'0\ell''m''0}_\trAB &= 2 \widetilde{h}^{\ell^{\prime}{} m^{\prime}{}}_{ad} \widetilde{h}^{\ell^{\prime \prime}{} m^{\prime \prime}{}}_{bc}g^{cd} r^{a} r^{b} + \frac{2 f \widetilde{h}_{\trAB}^{\ell^{\prime}{} m^{\prime}{}} \widetilde{h}_{\trAB}^{\ell^{\prime \prime}{} m^{\prime \prime}{}}}{r^4} -  \frac{2 \widetilde{h}_{\trAB}^{\ell^{\prime \prime}{} m^{\prime \prime}{}} r^{a} \delta_{a}\widetilde{h}_{\trAB}^{\ell^{\prime}{} m^{\prime}{}}}{r^3} +\frac{\widetilde{h}_{a-}^{\ell^{\prime}{} m^{\prime}{}} \widetilde{h}_{b-}^{\ell^{\prime \prime}{} m^{\prime \prime}{}}g^{ab} \lam{\ell^{\prime}}{1}^2 \lam{\ell^{\prime \prime}}{1}^2}{2r^2} \nonumber\\*
&\quad + \frac{\delta_{a}\widetilde{h}_{\trAB}^{\ell^{\prime \prime}{} m^{\prime \prime}{}} \delta^{a}\widetilde{h}_{\trAB}^{\ell^{\prime}{} m^{\prime}{}}}{r^2} -  \frac{2 \widetilde{h}^{\ell^{\prime \prime}{} m^{\prime \prime}{}}_{ab} r^{a} \delta^{b}\widetilde{h}_{\trAB}^{\ell^{\prime}{} m^{\prime}{}}}{r},\\
{\cal A}^{\ell'm'1\ell''m''-1}_\trAB &= - \frac{\widetilde{h}_{\trAB}^{\ell^{\prime}{} m^{\prime}{}} \widetilde{h}_{\trAB}^{\ell^{\prime \prime}{} m^{\prime \prime}{}}}{r^4} - \frac{(fg^{ab}+2r^ar^b) \widetilde{h}_{a-}^{\ell^{\prime}{} m^{\prime}} \widetilde{h}_{b-}^{\ell^{\prime \prime}{} m^{\prime \prime}{}}}{r^2} - \frac{i}{2r}r^{b}g^{ac} \widetilde{h}_{c-}^{\ell^{\prime \prime}m^{\prime \prime}} \left(\widetilde{h}^{\ell^{\prime}m^{\prime}}_{ab} - 2i \delta_{[a}\widetilde{h}_{b]-}^{\ell^{\prime}m^{\prime}}\right) \nonumber\\*
&\quad +\frac{i}{2r} r^{b}g^{ac} \widetilde{h}_{c-}^{\ell^{\prime}m^{\prime}} \left(\widetilde{h}^{\ell^{\prime \prime}{} m^{\prime \prime}{}}_{ab} + 2i \delta_{[a}\widetilde{h}_{b]-}^{\ell^{\prime \prime} m^{\prime \prime}}\right) - \tfrac{1}{4}g^{ac}\left(\widetilde{h}^{\ell^{\prime}{} m^{\prime}}_{ab} \widetilde{h}^{\ell^{\prime \prime}{} m^{\prime \prime}{}}_{cd}g^{bd} -4\delta_{[a}\widetilde{h}_{b]-}^{\ell^{\prime \prime} m^{\prime \prime}} \delta^{b}\widetilde{h}_{c-}^{\ell^{\prime}m^{\prime}} \right),\\
{\cal A}^{\ell'm'2\ell''m''-2}_\trAB &= \frac{g^{ab}\widetilde{h}_{a-}^{\ell^{\prime}{} m^{\prime}} \widetilde{h}_{b-}^{\ell^{\prime \prime}{} m^{\prime \prime}{}}}{2 r^2},\\
{\cal A}^{\ell'm'1\ell''m''1}_{+} &= - \frac{2i \sigma_{-}{} \widetilde{h}_{a-}^{\ell^{\prime \prime}{} m^{\prime \prime}{}} \widetilde{h}_{\trAB}^{\ell^{\prime}{} m^{\prime}{}} r^{a}}{r^3} -  \frac{\sigma_{+}{} f \widetilde{h}_{a-}^{\ell^{\prime}{} m^{\prime}{}} \widetilde{h}_{b-}^{\ell^{\prime \prime}{} m^{\prime \prime}{}}g^{ab}}{r^2} + \frac{\widetilde{h}_{a-}^{\ell^{\prime \prime}{} m^{\prime \prime}{}} g^{ac}r^{b} \bigl(i \sigma_{-}{} \widetilde{h}^{\ell^{\prime}{} m^{\prime}{}}_{bc} + 2\sigma_{+}{}\delta_{[b}\widetilde{h}_{c]-}^{\ell^{\prime}{} m^{\prime}{}}\bigr)}{r} \nonumber\\*
&\quad + \frac{1}{4} \sigma_{+}{}g^{ac} \bigl(\widetilde{h}^{\ell^{\prime}{} m^{\prime}}_{ab} \widetilde{h}^{\ell^{\prime \prime}{} m^{\prime \prime}{}}_{cd}g^{bd} + 4 \delta_{[a}\widetilde{h}_{b]-}^{\ell^{\prime \prime}{} m^{\prime \prime}{}} \delta^{b}\widetilde{h}_{c-}^{\ell^{\prime}{} m^{\prime}{}}\bigr),\\
{\cal A}^{\ell'm'2\ell''m''0}_{+} &= - \frac{2i \sigma_{-}{} \widetilde{h}^{\ell^{\prime \prime}{} m^{\prime \prime}{}}_{ab} \widetilde{h}_{c-}^{\ell^{\prime}{} m^{\prime}{}}r^{b}g^{ac}}{r} + \frac{i \sigma_{-}{} \widetilde{h}_{a-}^{\ell^{\prime}{}m^{\prime}{}} \delta^{a}\widetilde{h}_{\trAB}^{\ell^{\prime\prime}{} m^{\prime \prime}{}}}{r^2}.
\end{align}
\end{subequations}
The quantities appearing in Eq.~\eqref{Bilm} are
\begin{subequations}\label{B coupling functions}
\allowdisplaybreaks%
\begin{align}
{\cal B}^{\ell'm'0\ell''m''0}_{ab} &=
  \frac{4(r r_{a} r_{b}- M g_{ab})\widetilde{h}_{\trAB}^{\ell^{\prime}{} m^{\prime}{}} \widetilde{h}_{\trAB}^{\ell^{\prime \prime}{} m^{\prime \prime}{}}}{r^7}
  -  \frac{4 \widetilde{h}_{\trAB}^{\ell^{\prime \prime}{} m^{\prime \prime}{}} r_{(a} \delta_{b)}\widetilde{h}_{\trAB}^{\ell^{\prime}{} m^{\prime}{}}}{r^5}
  + \frac{\widetilde{h}_{\trAB}^{\ell^{\prime \prime}{} m^{\prime \prime}{}} (2 \delta_{a}\delta_{b}\widetilde{h}_{\trAB}^{\ell^{\prime}{} m^{\prime}{}} - \lam{\ell'}{1}^2 \widetilde{h}^{\ell^{\prime}{} m^{\prime}{}}_{ab} )}{r^4} \nonumber \\
  &\quad -  \frac{2 \widetilde{h}_{\trAB}^{\ell^{\prime \prime}{} m^{\prime \prime}{}} r^{c} (2\delta_{(a}\widetilde{h}^{\ell^{\prime}{} m^{\prime}{}}_{b)c} -  \delta_{c}\widetilde{h}^{\ell^{\prime}{} m^{\prime}{}}_{ab})}{r^3} + \widetilde{h}^{\ell^{\prime \prime}{} m^{\prime \prime}}_{ef} g^{ce}g^{df} (\delta_{b}\delta_{a}\widetilde{h}^{\ell^{\prime}{} m^{\prime}{}}_{cd} + \delta_{d}\delta_{c}\widetilde{h}^{\ell^{\prime}{} m^{\prime}{}}_{ab} -  2\delta_{d}\delta_{(a}\widetilde{h}^{\ell^{\prime}{} m^{\prime}{}}_{b)c}),\\
{\cal B}^{\ell'm'1\ell''m''-1}_{ab} &= \frac{2 M g^{cd} \widetilde{h}_{c-}^{\ell^{\prime}{} m^{\prime}{}} \widetilde{h}_{d-}^{\ell^{\prime \prime}{} m^{\prime \prime}{}} g_{ab}}{r^5}
  - \frac{r^{c} \bigl[\widetilde{h}_{c-}^{\ell^{\prime}{} m^{\prime}{}} (2\delta_{(a}\widetilde{h}_{b)-}^{\ell^{\prime \prime}{} m^{\prime \prime}}-i \widetilde{h}^{\ell^{\prime \prime}{} m^{\prime \prime}{}}_{ab})+\widetilde{h}_{c-}^{\ell^{\prime \prime}{} m^{\prime \prime}} (2 \delta_{(a}\widetilde{h}_{b)-}^{\ell^{\prime}{} m^{\prime}{}}+i \widetilde{h}^{\ell^{\prime}{} m^{\prime}{}}_{ab})\bigr]}{r^3} \nonumber \\
  & \quad
  + \frac{g^{cd} r_{(a} \bigl[(\delta_{b)}\widetilde{h}_{d-}^{\ell^{\prime}{} m^{\prime}{}} + \delta_{|d|}\widetilde{h}_{b)-}^{\ell^{\prime}{} m^{\prime}{}} + i \widetilde{h}^{\ell^{\prime}{} m^{\prime}{}}_{b)d} )\widetilde{h}_{c-}^{\ell^{\prime \prime}{} m^{\prime \prime}{}}  + (\delta_{b)}\widetilde{h}_{d-}^{\ell^{\prime \prime}{} m^{\prime \prime}{}}+\delta_{|d|}\widetilde{h}_{b)-}^{\ell^{\prime \prime}{} m^{\prime \prime}{}}-i \widetilde{h}^{\ell^{\prime \prime}{} m^{\prime \prime}{}}_{b)d})\widetilde{h}_{c-}^{\ell^{\prime}{} m^{\prime}{}} \bigr]}{r^3} \nonumber \\
  & \quad
  + \frac{g^{cd} \widetilde{h}_{c-}^{\ell^{\prime \prime}{} m^{\prime \prime}{}} \bigl[2\delta_{d}\delta_{(a}\widetilde{h}_{b)-}^{\ell^{\prime}{} m^{\prime}{}}- 2 \delta_{b}\delta_{a}\widetilde{h}_{d-}^{\ell^{\prime}{} m^{\prime}{}}  + 2i (\delta_{d}\widetilde{h}^{\ell^{\prime}{} m^{\prime}{}}_{ab} - \delta_{(a}\widetilde{h}^{\ell^{\prime}{} m^{\prime}{}}_{b)d})\bigr]}{2 r^2} \nonumber \\
  & \quad + \frac{g^{cd} \widetilde{h}_{c-}^{\ell^{\prime}{} m^{\prime}{}} \bigl[2\delta_{d}\delta_{(a}\widetilde{h}_{b)-}^{\ell^{\prime \prime}{} m^{\prime \prime}{}} - 2 \delta_{b}\delta_{a}\widetilde{h}_{d-}^{\ell^{\prime \prime}{} m^{\prime \prime}{}} - 2i (\delta_{d}\widetilde{h}^{\ell^{\prime \prime}{} m^{\prime \prime}{}}_{ab} - \delta_{(a}\widetilde{h}^{\ell^{\prime \prime}{} m^{\prime \prime}{}}_{b)d}) \bigr]}{2 r^2},\\
{\cal B}^{\ell'm'1\ell''m''0}_{a+} &= 
  - \frac{(\sigma_{+}{} \widetilde{h}_{\trAB}^{\ell^{\prime}{} m^{\prime}{}} r_{a}+ i M \sigma_{-}{} \widetilde{h}_{a-}^{\ell^{\prime}{} m^{\prime}{}}) \widetilde{h}_{\trAB}^{\ell^{\prime \prime}{} m^{\prime \prime}{}}}{r^5}
  + \frac{( \sigma_{+}{} \delta_{a}\widetilde{h}_{\trAB}^{\ell^{\prime}{} m^{\prime}{}} + i \sigma_{-}{} \lam{\ell'}{1}^2 \widetilde{h}_{a-}^{\ell^{\prime}{} m^{\prime}{}})\widetilde{h}_{\trAB}^{\ell^{\prime \prime}{} m^{\prime \prime}{}} }{2 r^4} \nonumber \\
  & \quad + \frac{\sigma_{+}{} \bigl[3  \lam{\ell''}{1}^2  \widetilde{h}_{b-}^{\ell^{\prime}{} m^{\prime}{}} (g^{bc} \widetilde{h}_{c-}^{\ell^{\prime \prime}{} m^{\prime \prime}{}}r_{a}  -  r^b \widetilde{h}_{a-}^{\ell^{\prime \prime}{} m^{\prime \prime}{}} ) - r^{b} \widetilde{h}^{\ell^{\prime}{} m^{\prime}{}}_{ab} \widetilde{h}_{\trAB}^{\ell^{\prime \prime}{} m^{\prime \prime}{}} \bigr]}{2r^3}
  + \sigma_{+}{}  g^{bd} g^{ce} \widetilde{h}^{\ell^{\prime \prime}{} m^{\prime \prime}{}}_{bc} \delta_{[a}\widetilde{h}^{\ell^{\prime}{} m^{\prime}{}}_{d]e} \nonumber \\
  & \quad
  + \frac{i \sigma_{-}{} \bigl[\widetilde{h}_{\trAB}^{\ell^{\prime \prime}{} m^{\prime \prime}{}} r^{b} \delta_{[a}\widetilde{h}_{b]-}^{\ell^{\prime}{} m^{\prime}{}} - g^{bc}\widetilde{h}_{b-}^{\ell^{\prime}{} m^{\prime}{}} (M \widetilde{h}^{\ell^{\prime \prime}{} m^{\prime \prime}{}}_{ac}
  +r_{(a} \delta_{c)}\widetilde{h}_{\trAB}^{\ell^{\prime \prime}{} m^{\prime \prime}{}})\bigr]}{r^3}
  + i \sigma_{-}{} g^{cd} \delta^{b} \delta_{[a}\widetilde{h}_{c]-}^{\ell^{\prime}{} m^{\prime}{}}\widetilde{h}^{\ell^{\prime \prime}{} m^{\prime \prime}{}}_{bd} \nonumber \\
  & \quad
  + \frac{6 \sigma_{+}{} \lam{\ell''}{1}^2 g^{bc} \widetilde{h}_{b-}^{\ell^{\prime}{} m^{\prime}{}} \delta_{[c}\widetilde{h}_{a]-}^{\ell^{\prime \prime}{} m^{\prime \prime}{}} 
  -  i \sigma_{-}{} g^{bc} \widetilde{h}_{b-}^{\ell^{\prime}{} m^{\prime}{}} (\lam{\ell''}{1}^2\widetilde{h}^{\ell^{\prime \prime}{} m^{\prime \prime}{}}_{ac}  - 2 \delta_{c}\delta_{a}\widetilde{h}_{\trAB}^{\ell^{\prime \prime}{} m^{\prime \prime}{}})}{4 r^2} 
  + \frac{\sigma_{+}{} g^{bd} g^{ce} r_{[b} \widetilde{h}^{\ell^{\prime}{} m^{\prime}{}}_{a]c}\widetilde{h}^{\ell^{\prime \prime}{} m^{\prime \prime}{}}_{de}}{r} \nonumber \\
  & \quad
  + \frac{i \sigma_{-}{} g^{cd} \bigl[2(r^{b} \delta_{(a}\widetilde{h}_{c)-}^{\ell^{\prime}{} m^{\prime}{}} -  r_{a} \delta^{b}\widetilde{h}_{c-}^{\ell^{\prime}{} m^{\prime}{}})\widetilde{h}^{\ell^{\prime \prime}{} m^{\prime \prime}{}}_{bd}- \widetilde{h}_{c-}^{\ell^{\prime}{} m^{\prime}{}} r^{b} (2\delta_{(a}\widetilde{h}^{\ell^{\prime \prime}{} m^{\prime \prime}{}}_{d)b} -  \delta_{b}\widetilde{h}^{\ell^{\prime \prime}{} m^{\prime \prime}{}}_{ad})\bigr]}{2r}   ,\\
{\cal B}^{\ell'm'2\ell''m''-1}_{a+} &= \frac{g^{bc} \widetilde{h}_{b-}^{\ell^{\prime \prime}{} m^{\prime \prime}{}} \bigl(i\sigma_{-}{} \widetilde{h}^{\ell^{\prime}{} m^{\prime}{}}_{ac} - 2 \sigma_{+}{} \delta_{(a}\widetilde{h}_{c)-}^{\ell^{\prime}{} m^{\prime}{}}\bigr)}{4 r^2},\\
{\cal B}^{\ell'm'0\ell''m''0}_\trAB &= 
- \tfrac{1}{2} \lam{\ell'}{1}^2 g^{ac} g^{bd} \widetilde{h}^{\ell^{\prime}{} m^{\prime}}_{ab} \widetilde{h}^{\ell^{\prime \prime}{} m^{\prime \prime}{}}_{cd}  
-  \frac{\widetilde{h}_{\trAB}^{\ell^{\prime}{} m^{\prime}{}} \widetilde{h}_{\trAB}^{\ell^{\prime \prime}{} m^{\prime \prime}{}} (2 f + \lam{\ell'}{1}^2)}{r^4}
+ \frac{2 \widetilde{h}_{\trAB}^{\ell^{\prime \prime}{} m^{\prime \prime}{}} r^{a} \delta_{a}\widetilde{h}_{\trAB}^{\ell^{\prime}{} m^{\prime}{}}}{r^3}
+ \widetilde{h}^{\ell^{\prime \prime}{} m^{\prime \prime}}_{ab} \delta^{b}\delta^{a}\widetilde{h}_{\trAB}^{\ell^{\prime}{} m^{\prime}{}}\nonumber \\
& \quad -  \frac{2 M g^{ac} g^{bd} \widetilde{h}^{\ell^{\prime}{} m^{\prime}}_{ab} \widetilde{h}^{\ell^{\prime \prime}{} m^{\prime \prime}{}}_{cd} 
+ 2 \widetilde{h}^{\ell^{\prime \prime}{} m^{\prime \prime}{}}_{ab} r^{a} \delta^{b}\widetilde{h}_{\trAB}^{\ell^{\prime}{} m^{\prime}{}}}{r}
+ r r^{a} g^{bd} g^{ce} \widetilde{h}^{\ell^{\prime \prime}{} m^{\prime \prime}}_{bc}(\delta_{a}\widetilde{h}^{\ell^{\prime}{} m^{\prime}{}}_{de} - 2 \delta_{e}\widetilde{h}^{\ell^{\prime}{} m^{\prime}{}}_{ad}),\\
{\cal B}^{\ell'm'1\ell''m''-1}_\trAB &= \frac{2 M g^{ab}\widetilde{h}_{a-}^{\ell^{\prime}{} m^{\prime}{}} \widetilde{h}_{b-}^{\ell^{\prime \prime}{} m^{\prime \prime}} - i  r^{a}(\widetilde{h}_{a-}^{\ell^{\prime \prime}{} m^{\prime \prime}{}} \widetilde{h}_{\trAB}^{\ell^{\prime}{} m^{\prime}{}} - \widetilde{h}_{a-}^{\ell^{\prime}{} m^{\prime}{}} \widetilde{h}_{\trAB}^{\ell^{\prime \prime}{} m^{\prime \prime}{}})}{r^3}
-  \frac{2 r^{a} r^{b} \widetilde{h}_{a-}^{\ell^{\prime}{} m^{\prime}} \widetilde{h}_{b-}^{\ell^{\prime \prime}{} m^{\prime \prime}}}{r^2} \nonumber \\
& \quad - \frac{2i (\widetilde{h}_{a-}^{\ell^{\prime}{} m^{\prime}{}}\delta^{a}\widetilde{h}_{\trAB}^{\ell^{\prime \prime}{} m^{\prime \prime}{}}-\widetilde{h}_{a-}^{\ell^{\prime \prime}{} m^{\prime \prime}} \delta^{a}\widetilde{h}_{\trAB}^{\ell^{\prime}{} m^{\prime}{}})
-  (4 f + \mu_{\ell'}^2 + \mu_{\ell''}^2 + \lam{\ell'}{1}^2 + \lam{\ell''}{1}^2)g^{ab}\widetilde{h}_{a-}^{\ell^{\prime}{} m^{\prime}{}} \widetilde{h}_{b-}^{\ell^{\prime \prime}{} m^{\prime \prime}{}}}{4r^2} \nonumber \\
& \quad + \frac{g^{ab}r^{c} \bigl[\widetilde{h}_{a-}^{\ell^{\prime \prime}{} m^{\prime \prime}}(2\delta_{[b}\widetilde{h}_{c]-}^{\ell^{\prime}{} m^{\prime}{}}-i \widetilde{h}^{\ell^{\prime}{} m^{\prime}{}}_{bc})
  +  \widetilde{h}_{a-}^{\ell^{\prime}{} m^{\prime}{}} (2\delta_{[b}\widetilde{h}_{c]-}^{\ell^{\prime \prime}{} m^{\prime \prime}{}} + i \widetilde{h}^{\ell^{\prime \prime}{} m^{\prime \prime}{}}_{bc})\bigr]}{2 r},\\ 
{\cal B}^{\ell'm'1\ell''m''1}_{+} &= - \frac{2i \sigma_{-}{}  r^{a} \widetilde{h}_{a-}^{\ell^{\prime \prime}{} m^{\prime \prime}{}} \widetilde{h}_{\trAB}^{\ell^{\prime}{} m^{\prime}{}}}{r^3} 
   + \frac{\sigma_{+}{} \bigl[4 r^{a} r^{b} -  g^{ab} (2 f + \mu_{\ell'}^2 + \lam{\ell'}{1}^2)\bigr]\widetilde{h}_{a-}^{\ell^{\prime}{} m^{\prime}{}} \widetilde{h}_{b-}^{\ell^{\prime \prime}{} m^{\prime \prime}{}}  
   + 2 i \sigma_{-}{} \widetilde{h}_{a-}^{\ell^{\prime \prime}{} m^{\prime \prime}{}} \delta^{a}\widetilde{h}_{\trAB}^{\ell^{\prime}{} m^{\prime}{}}}{2r^2} \nonumber \\
   & \quad
   +  \frac{r^{a} g^{bc} \widetilde{h}_{c-}^{\ell^{\prime \prime}{} m^{\prime \prime}{}} (2\sigma_{+}{} \delta_{[a}\widetilde{h}_{b]-}^{\ell^{\prime}{} m^{\prime}{}} -i \sigma_{-}{} \widetilde{h}^{\ell^{\prime}{} m^{\prime}{}}_{ab})}{r},\\
{\cal B}^{\ell'm'2\ell''m''0}_{+} &= \tfrac{1}{2} \widetilde{h}^{\ell^{\prime \prime}{} m^{\prime \prime}{}}_{ab} g^{ac}(\sigma_{+}{}g^{bd}\widetilde{h}^{\ell^{\prime}{} m^{\prime}{}}_{cd} + 2i \sigma_{-}{} \delta^{b}\widetilde{h}_{c-}^{\ell^{\prime}{} m^{\prime}{}}).
\end{align}
\end{subequations}
The quantities appearing in Eq.~\eqref{Cilm} are
\begin{subequations}\label{C coupling functions}
\allowdisplaybreaks%
\begin{align}
{\cal C}^{\ell'm'0\ell''m''0}_{ab} &= 2\delta_{(a}\widetilde{h}^{\ell^{\prime}{} m^{\prime}{}}_{b)c} \delta^{c}\widetilde{h}_{\trab}^{\ell^{\prime \prime}{} m^{\prime \prime}{}} - \delta_{c}\widetilde{h}^{\ell^{\prime \prime}{} m^{\prime \prime}{}}_{ab} \delta^{c}\widetilde{h}_{\trab}^{\ell^{\prime}{} m^{\prime}{}}
-  \frac{\delta_{c}\widetilde{h}^{\ell^{\prime \prime}{} m^{\prime \prime}{}}_{ab} \delta^{c}\widetilde{h}_{\trAB}^{\ell^{\prime}{} m^{\prime}{}} -  2\delta_{(a}\widetilde{h}^{\ell^{\prime}{} m^{\prime}{}}_{b)c} \delta^{c}\widetilde{h}_{\trAB}^{\ell^{\prime \prime}{} m^{\prime \prime}{}}}{r^2} \nonumber \\
& \quad
 -  \frac{2 r^{c} g^{de}\widetilde{h}^{\ell^{\prime \prime}{} m^{\prime \prime}{}}_{cd} (2\delta_{(a}\widetilde{h}^{\ell^{\prime}{} m^{\prime}{}}_{b)e} -  \delta_{e}\widetilde{h}^{\ell^{\prime}{} m^{\prime}{}}_{ab})}{r}
 -  g^{cd} (2\delta_{(a}\widetilde{h}^{\ell^{\prime}{} m^{\prime}{}}_{b)c} -  \delta_{c}\widetilde{h}^{\ell^{\prime}{} m^{\prime}{}}_{ab}) \delta^{e}\widetilde{h}^{\ell^{\prime \prime}{} m^{\prime \prime}{}}_{de},\\
{\cal C}^{\ell'm'1\ell''m''-1}_{ab} &= \frac{r^{c} \bigl[
  \widetilde{h}_{c-}^{\ell^{\prime \prime}{} m^{\prime \prime}} (2\delta_{(a}\widetilde{h}_{b)-}^{\ell^{\prime}{} m^{\prime}{}}+i \widetilde{h}^{\ell^{\prime}{} m^{\prime}{}}_{ab}) 
+ \widetilde{h}_{c-}^{\ell^{\prime}{} m^{\prime}} ( 2\delta_{(a}\widetilde{h}_{b)-}^{\ell^{\prime \prime}{} m^{\prime \prime}{}}-i \widetilde{h}^{\ell^{\prime \prime}{} m^{\prime \prime}{}}_{ab})\bigr]}{r^3}
\nonumber \\
& \quad
 + \frac{\bigl(\widetilde{h}^{\ell^{\prime \prime}{} m^{\prime \prime}{}}_{ab} + 2i\delta_{(a}\widetilde{h}_{b)-}^{\ell^{\prime \prime}{} m^{\prime \prime}{}}\bigr) (\widetilde{h}_{\trab}^{\ell^{\prime}{} m^{\prime}{}} - i \delta^{c}\widetilde{h}_{c-}^{\ell^{\prime}{} m^{\prime}})
 + (\widetilde{h}^{\ell^{\prime}{} m^{\prime}{}}_{ab} - 2i \delta_{(a}\widetilde{h}_{b)-}^{\ell^{\prime}{} m^{\prime}{}}) (\widetilde{h}_{\trab}^{\ell^{\prime \prime}{} m^{\prime \prime}{}} + i \delta^{c}\widetilde{h}_{c-}^{\ell^{\prime \prime}{} m^{\prime \prime}})}{2 r^2},\\
{\cal C}^{\ell'm'1\ell''m''0}_{a+} &= 
- \frac{2i \sigma_{-}{}  r_{a} r^{b}\widetilde{h}_{b-}^{\ell^{\prime}{} m^{\prime}{}} \widetilde{h}_{\trAB}^{\ell^{\prime \prime}{} m^{\prime \prime}{}}}{r^4} 
+ \frac{\sigma_{+}{} ( \lam{\ell''}{1}^2 r^{b} \widetilde{h}_{b-}^{\ell^{\prime}{} m^{\prime}{}} \widetilde{h}_{a-}^{\ell^{\prime \prime}{} m^{\prime \prime}{}}- r_{a}\widetilde{h}_{\trab}^{\ell^{\prime}{} m^{\prime}{}} \widetilde{h}_{\trAB}^{\ell^{\prime \prime}{} m^{\prime \prime}{}})}{r^3}\nonumber \\
& \quad
+ \frac{i \sigma_{-}{} \bigl[r^{b} \widetilde{h}_{b-}^{\ell^{\prime}{} m^{\prime}{}} \delta_{a}\widetilde{h}_{\trAB}^{\ell^{\prime \prime}{} m^{\prime \prime}{}}
- r_{a} (\widetilde{h}_{\trAB}^{\ell^{\prime \prime}{} m^{\prime \prime}{}} \delta^{b}\widetilde{h}_{b-}^{\ell^{\prime}{} m^{\prime}} -  \widetilde{h}_{b-}^{\ell^{\prime}{} m^{\prime}{}} \delta^{b}\widetilde{h}_{\trAB}^{\ell^{\prime \prime}{} m^{\prime \prime}{}})\bigr]}{r^3}
+ \tfrac{1}{2} \bigl(\sigma_{+} \widetilde{h}^{\ell^{\prime}{} m^{\prime}{}}_{ab} + i \sigma_{-} \delta_{b}\widetilde{h}_{a-}^{\ell^{\prime}{} m^{\prime}{}}
\bigr) \delta^{b}\widetilde{h}_{\trab}^{\ell^{\prime \prime}{} m^{\prime \prime}{}} \nonumber \\
& \quad
+ \frac{i\sigma_{-}{} \bigl[\delta_{a}\widetilde{h}_{\trAB}^{\ell^{\prime \prime}{} m^{\prime \prime}{}} \delta^{b}\widetilde{h}_{b-}^{\ell^{\prime}{} m^{\prime}{}} + 2 \delta^{b}\widetilde{h}_{\trAB}^{\ell^{\prime \prime}{} m^{\prime \prime}{}} \delta_{[b}\widetilde{h}_{a]-}^{\ell^{\prime}{} m^{\prime}{}}-4 r_{a} r^{b} g^{cd} \widetilde{h}^{\ell^{\prime \prime}{} m^{\prime \prime}{}}_{bc} \widetilde{h}_{d-}^{\ell^{\prime}{} m^{\prime}{}} - \lam{\ell''}{1}^2 \widetilde{h}_{a-}^{\ell^{\prime \prime}{} m^{\prime \prime}{}} \widetilde{h}_{\trab}^{\ell^{\prime}{} m^{\prime}{}} 
\bigr]}{2 r^2} \nonumber \\
& \quad
+ \frac{\sigma_{+}{} \bigl[\widetilde{h}_{\trab}^{\ell^{\prime}{} m^{\prime}{}} \delta_{a}\widetilde{h}_{\trAB}^{\ell^{\prime \prime}{} m^{\prime \prime}{}} + \lam{\ell''}{1}^2 \widetilde{h}_{a-}^{\ell^{\prime \prime}{} m^{\prime \prime}{}}  \delta^{b}\widetilde{h}_{b-}^{\ell^{\prime}{} m^{\prime}{}} + \widetilde{h}^{\ell^{\prime}{} m^{\prime}{}}_{ab} \delta^{b}\widetilde{h}_{\trAB}^{\ell^{\prime \prime}{} m^{\prime \prime}{}}\bigr]}{2r^2} \nonumber \\
& \quad
-  \tfrac{1}{2} \sigma_{+}{} g^{bc} \widetilde{h}^{\ell^{\prime}{} m^{\prime}{}}_{ab} \delta^{d}\widetilde{h}^{\ell^{\prime \prime}{} m^{\prime \prime}{}}_{cd} 
-  \tfrac{1}{2}i \sigma_{-}{} \bigl(\delta_{a}\widetilde{h}_{b-}^{\ell^{\prime}{} m^{\prime}{}} \delta^{b}\widetilde{h}_{\trab}^{\ell^{\prime \prime}{} m^{\prime \prime}{}} + 2 g^{bc}\delta_{[b}\widetilde{h}_{a]-}^{\ell^{\prime}{} m^{\prime}{}} \delta^{d}\widetilde{h}^{\ell^{\prime \prime}{} m^{\prime \prime}{}}_{cd}\bigr) \nonumber \\
& \quad
-  \frac{\sigma_{+}{} g^{bc} r^{d} \widetilde{h}^{\ell^{\prime}{} m^{\prime}{}}_{ab} \widetilde{h}^{\ell^{\prime \prime}{} m^{\prime \prime}{}}_{cd}
-  i \sigma_{-}{} \bigl[r_{a} \widetilde{h}_{b-}^{\ell^{\prime}{} m^{\prime}{}} (\delta^{b}\widetilde{h}_{\trab}^{\ell^{\prime \prime}{} m^{\prime \prime}{}} -  g^{bc}\delta^{d}\widetilde{h}^{\ell^{\prime \prime}{} m^{\prime \prime}{}}_{cd}) 
+  2g^{bc}r^{d} \delta_{[a}\widetilde{h}_{b]-}^{\ell^{\prime}{} m^{\prime}{}}\widetilde{h}^{\ell^{\prime \prime}{} m^{\prime \prime}{}}_{cd} \bigr]}{r},\\
{\cal C}^{\ell'm'0\ell''m''0}_\trAB &= \delta^{a}\widetilde{h}_{\trAB}^{\ell^{\prime \prime}{} m^{\prime \prime}{}} \delta^{b}\widetilde{h}^{\ell^{\prime}{} m^{\prime}{}}_{ab}
 -4 r^{a} r^{b} g^{cd} \widetilde{h}^{\ell^{\prime}{} m^{\prime}{}}_{ac} \widetilde{h}^{\ell^{\prime \prime}{} m^{\prime \prime}{}}_{bd}
 -  \delta_{a}\widetilde{h}_{\trAB}^{\ell^{\prime}{} m^{\prime}{}} \delta^{a}\widetilde{h}_{\trab}^{\ell^{\prime \prime}{} m^{\prime \prime}{}}
 \nonumber \\
 & \quad
 + 2 r r^{a} (\widetilde{h}^{\ell^{\prime \prime}{} m^{\prime \prime}{}}_{ab} \delta^{b}\widetilde{h}_{\trab}^{\ell^{\prime}{} m^{\prime}{}} - g^{bc} \widetilde{h}^{\ell^{\prime \prime}{} m^{\prime \prime}{}}_{ab} \delta^{d}\widetilde{h}^{\ell^{\prime}{} m^{\prime}{}}_{cd})
  -  \frac{\delta_{a}\widetilde{h}_{\trAB}^{\ell^{\prime \prime}{} m^{\prime \prime}{}} \delta^{a}\widetilde{h}_{\trAB}^{\ell^{\prime}{} m^{\prime}{}}}{r^2}
 + \frac{4 \widetilde{h}^{\ell^{\prime \prime}{} m^{\prime \prime}{}}_{ab} r^{a} \delta^{b}\widetilde{h}_{\trAB}^{\ell^{\prime}{} m^{\prime}{}}}{r},\\
{\cal C}^{\ell'm'1\ell''m''-1}_\trAB &= \frac{4 r^{a} r^{b} \widetilde{h}_{a-}^{\ell^{\prime}{} m^{\prime}{}} \widetilde{h}_{b-}^{\ell^{\prime \prime}{} m^{\prime \prime}{}}}{r^2}
 + \frac{r^{a} \bigl[\widetilde{h}_{a-}^{\ell^{\prime \prime}{} m^{\prime \prime}{}} (\delta^{b}\widetilde{h}_{b-}^{\ell^{\prime}{} m^{\prime}{}}+ i \widetilde{h}_{\trab}^{\ell^{\prime}{} m^{\prime}{}}) + \widetilde{h}_{a-}^{\ell^{\prime}{} m^{\prime}{}} (\delta^{b}\widetilde{h}_{b-}^{\ell^{\prime \prime}{} m^{\prime \prime}{}}-i \widetilde{h}_{\trab}^{\ell^{\prime \prime}{} m^{\prime \prime}{}})\bigr]}{r},\\ 
{\cal C}^{\ell'm'1\ell''m''1}_{+} &= \frac{2i \sigma_{-}{} r^{a}\widetilde{h}_{a-}^{\ell^{\prime \prime}{} m^{\prime \prime}{}} \widetilde{h}_{\trAB}^{\ell^{\prime}{} m^{\prime}{}}}{r^3}
 + \frac{\widetilde{h}_{\trAB}^{\ell^{\prime}{} m^{\prime}{}} (\sigma_{+}{} \widetilde{h}_{\trab}^{\ell^{\prime \prime}{} m^{\prime \prime}{}} + i \sigma_{-}{} \delta^{a}\widetilde{h}_{a-}^{\ell^{\prime \prime}{} m^{\prime \prime}})}{r^2},\\
{\cal C}^{\ell'm'2\ell''m''0}_{+} &= \frac{2i \sigma_{-}{} r^{a} g^{bc} \widetilde{h}^{\ell^{\prime \prime}{} m^{\prime \prime}{}}_{ab} \widetilde{h}_{c-}^{\ell^{\prime}{} m^{\prime}{}}}{r}
 -  \frac{i \sigma_{-}{} \widetilde{h}_{a-}^{\ell^{\prime}{} m^{\prime}{}} \delta^{a}\widetilde{h}_{\trAB}^{\ell^{\prime \prime}{} m^{\prime \prime}{}}}{r^2} 
 - i \sigma_{-}{} \widetilde{h}_{a-}^{\ell^{\prime}{} m^{\prime}{}} (\delta^{a}\widetilde{h}_{\trab}^{\ell^{\prime \prime}{} m^{\prime \prime}{}} - g^{ab} \delta^{c}\widetilde{h}^{\ell^{\prime \prime}{} m^{\prime \prime}{}}_{bc}).
\end{align}
\end{subequations}

\subsection{Stress-energy terms}

The mode decomposition of Eq.~\eqref{calT2} is
\begin{subequations}\label{calT2 modes}
\allowdisplaybreaks
\begin{align}
\widetilde{\cal T}^{(2)\ell m}_{ab} &= \widetilde{T}^{(2)\ell m}_{ab} - g_{ab}\left(\widetilde{T}^{(2)\ell m}_{\trab}+r^{-2}\widetilde{T}^{(2)\ell m}_{\trAB}\right)
					+ \sum_{\substack{\ell'm'\\\ell''m''}}\sum_{s'=0,1} \lam{\ell^{\prime}}{s'} \lam{\ell^{\prime \prime}}{s'}C^{\ell m 0}_{\ell'm's'\ell''m'',-s'}\ {\cal T}^{\ell'm's'\ell''m'',-s'}_{ab},\\
\widetilde{\cal T}^{(2)\ell m}_{a\pm} &= \widetilde{T}^{(2)\ell m}_{a\pm} 
					 + \sum_{\substack{\ell'm'\\\ell''m''}}\frac{\lam{\ell^{\prime}}{1}}{\lam{\ell}{1}}C^{\ell m 1}_{\ell'm'1\ell''m''0}\ {\cal T}^{\ell'm'1\ell''m''0}_{a\pm},\\
\widetilde{\cal T}^{(2)\ell m}_{\trAB} &= -r^2\widetilde{T}^{(2)\ell m}_{\trab} 
					 + \sum_{\substack{\ell'm'\\\ell''m''}}\sum_{s'=0,1}\lam{\ell^{\prime}}{s'} \lam{\ell^{\prime \prime}}{s'}C^{\ell m 0}_{\ell'm's'\ell''m'',-s'}\ {\cal T}^{\ell'm's'\ell''m'',-s'}_{\trAB},\\
\widetilde{\cal T}^{(2)\ell m}_{\pm} &= \widetilde{T}^{(2)\ell m}_{\pm}. 
\end{align}
\end{subequations}
The coupling functions ${\cal T}^{\ell'm's'\ell''m'',s''}_{\boldsymbol{\cdot}}$ are
\begin{subequations}
\allowdisplaybreaks
\begin{align}
{\cal T}^{\ell'm'0\ell''m''0}_{ab} &= \frac{1}{2} g_{ab}g^{ce}g^{df}\widetilde h^{\ell^{\prime}{} \mathit{m}^{\prime}}_{cd}  T^{\ell^{\prime \prime}{} \mathit{m}^{\prime \prime}{}}_{ef} 
-  \frac{1}{2} \widetilde h^{\ell^{\prime}{} \mathit{m}^{\prime}{}}_{ab}g^{cd} T^{\ell^{\prime \prime}{} \mathit{m}^{\prime \prime}}_{cd} 
+ r^{-4}g_{ab}\widetilde h_{\trAB}^{\ell^{\prime \prime}{} \mathit{m}^{\prime \prime}{}} T_{\trAB}^{\ell^{\prime}{} \mathit{m}^{\prime}{}} - r^{-2}\widetilde h^{\ell^{\prime \prime}{} \mathit{m}^{\prime \prime}{}}_{ab} T_{\trAB}^{\ell^{\prime}{} \mathit{m}^{\prime}{}},\\ 
{\cal T}^{\ell'm'1\ell''m'',-1}_{ab} &= - \frac{1}{2r^2}g_{ab}g^{cd}\Bigl[\widetilde h_{c-}^{\ell^{\prime \prime}{} \mathit{m}^{\prime \prime}}\bigl(T_{d-}^{\ell^{\prime}{} \mathit{m}^{\prime}{}} - i T_{d+}^{\ell^{\prime}{} \mathit{m}^{\prime}{}}\bigr) + \widetilde h_{c-}^{\ell^{\prime}{} \mathit{m}^{\prime}}\bigl(T_{d-}^{\ell^{\prime \prime}{} \mathit{m}^{\prime \prime}{}} + i T_{d+}^{\ell^{\prime \prime}{} \mathit{m}^{\prime \prime}{}}\bigr)\Bigr],\\
%
%
%
%
{\cal T}^{\ell'm'1\ell''m''0}_{a+} &= \frac{i}{4}\sigma_{-}{}\widetilde h_{a-}^{\ell^{\prime}{} \mathit{m}^{\prime}{}}g^{bc} T^{\ell^{\prime \prime}{} \mathit{m}^{\prime \prime}{}}_{bc} 
+ \frac{i }{2 r^2 }\sigma_{-}{}\widetilde h_{a-}^{\ell^{\prime}{} \mathit{m}^{\prime}{}} T_{\trAB}^{\ell^{\prime \prime}{} \mathit{m}^{\prime \prime}{}},\\
%
%
{\cal T}^{\ell'm'0\ell''m''0}_{\trAB} &= \frac{1}{2} r^2 g^{ac}g^{bd}\widetilde h^{\ell^{\prime}{} \mathit{m}^{\prime}}_{ab} T^{\ell^{\prime \prime}{} \mathit{m}^{\prime \prime}{}}_{cd} 
-  \frac{1}{2}\widetilde h_{\trAB}^{\ell^{\prime}{} \mathit{m}^{\prime}{}} g^{ab}T^{\ell^{\prime \prime}{} \mathit{m}^{\prime \prime}}_{ab} 
+ r^{-2}\widetilde h_{\trAB}^{\ell^{\prime \prime}{} \mathit{m}^{\prime \prime}{}} T_{\trAB}^{\ell^{\prime}{} \mathit{m}^{\prime}{}} -  r^{-2}\widetilde h_{\trAB}^{\ell^{\prime}{} \mathit{m}^{\prime}{}} T_{\trAB}^{\ell^{\prime \prime}{} \mathit{m}^{\prime \prime}{}},\\
{\cal T}^{\ell'm'1\ell''m'',-1}_{\trAB} &= - \frac{1}{2}g^{ab}\Bigl[\widetilde h_{a-}^{\ell^{\prime \prime}{} \mathit{m}^{\prime \prime}{}}\bigl(T_{b-}^{\ell^{\prime}{} \mathit{m}^{\prime}{}} - i T_{b+}^{\ell^{\prime}{} \mathit{m}^{\prime}{}}\bigr) 
+ \widetilde h_{a-}^{\ell^{\prime}{} \mathit{m}^{\prime}{}}\bigl(T_{b-}^{\ell^{\prime \prime}{} \mathit{m}^{\prime \prime}{}} + i T_{b+}^{\ell^{\prime \prime}{} \mathit{m}^{\prime \prime}{}}\bigr)\Bigr].
\end{align}
\end{subequations}
\end{widetext}

\section{Field equations for low multipoles}

In this appendix we summarize the special cases of the field equations~\eqref{EFEnlm fixed} for $\ell=0$ and $\ell=1$. We specifically discuss the $\ell=0$ and odd-parity $\ell=1$ equations, which we are able to relate to the evolution of mass and spin. We have not found a new or illuminating form for the even-parity $\ell=1$ equations, which are associated with a displacement of the center of mass~\cite{Martel-Poisson:05}.

\subsection{$\ell=0$}

The $\ell=0$ field equations, with $\ell m$ labels omitted for visual simplicity, are
\begin{subequations}%
\begin{align}
\dR_{ab}[\widetilde{h}^{(1)}_{\boldsymbol\cdot}] &= 8\pi{\cal T}^{(1)}_{ab},\\
\dR_{\trAB}[\widetilde{h}^{(1)}_{\boldsymbol\cdot}] &= 8\pi{\cal T}^{(1)}_{\trAB},
\end{align}
\end{subequations}
and their analogues at second order. These can be written entirely in terms of the invariant variables $\widetilde h_{rr}$ and $\varphi$ defined in Eqs.~\eqref{h00rr tilde} and \eqref{ell=0 invariant}. However, we opt to replace $\widetilde h_{rr}$ with an effective mass perturbation $\delta M$, defined from 
\beq
\widetilde h_{rr} = \frac{\partial g_{rr}}{\partial M}\delta M=\frac{2\delta M}{rf^2}.
\eeq
In terms of $\delta M$ and $\varphi$,
\begingroup%
\allowdisplaybreaks
\begin{subequations}%
\begin{align}
\dR_{tt} &= \frac{M \delta M}{fr^4} - \frac{\partial^2_t\delta M}{fr} -  \frac{M \partial_r\delta M}{r^3} \nonumber\\
&\quad - \frac{f (2r-M) \varphi}{2 r^2} - \tfrac{1}{2} f^2\partial_r\varphi ,\\
\dR_{tr} &=  \frac{2}{r^2f}\frac{\partial \delta M}{\partial t},\\
\dR_{rr} &= -\frac{2 r-3 M}{f^3 r^4} \delta M + \frac{\partial_t^2\delta M}{f^3 r} + \frac{2 r-3 M}{f^2 r^3} \partial_r\delta M \nonumber\\
&\quad + \frac{3 M \varphi}{2f^2 r^2} + \tfrac{1}{2} \partial_r\varphi ,\\
\dR_{\trAB} &= r^{-1}\partial_r(r\delta M) + \frac{2M}{r^2f}\delta M+ \frac{1}{2}rf\varphi.
\end{align}
\end{subequations}
\endgroup

We can further reduce these field equations by eliminating $\varphi$. Solving $\dR_{\trAB}=8\pi{\cal T}_\trAB$ for $\varphi$ yields 
\begin{align}\label{phi to dM}
\varphi = -\frac{2 \partial_r\delta M}{rf}-\frac{2 \delta M}{r^2 f^2}+\frac{16 \pi \mathcal{T}_\trAB}{rf},
\end{align}
which reduces the remaining three components of the field equations to equations for $\delta M$. The $tr$ component, $\dR_{tr}=8\pi{\cal T}_{tr}$, reads
\beq\label{Mdot}
\frac{\partial}{\partial t} \delta M = 4\pi r^2 f T_{tr},
\eeq
where we have used ${\cal T}_{tr}=T_{tr}$. Equation~\eqref{Mdot} is a flux-balance equation, equating the rate of change of the mass at radius $r$ to the flux of energy crossing the $r={\rm const.}$ surface. This  determines $\delta M $ up to a time-independent function of $r$. The function of $r$ can be determined up to a constant from the ``anti-trace'' piece of the field equations, 
\beq
f^{-1} \dR_{tt}+f \dR_{rr}=8\pi(f^{-1} {\cal T}_{tt}+f {\cal T}_{rr}),
\eeq
which can be simplified to
\beq\label{dM/dr}
\frac{\partial}{\partial r}\delta M = 4 \pi r^2 f^{-1} T_{tt}
\eeq
after using \eqref{phi to dM} and expressing ${\cal T}_{\alpha\beta}$ in terms of $T_{\alpha\beta}$. Equation~\eqref{dM/dr} relates the mass within a sphere of radius $r$ to the total energy within that sphere. Equations~\eqref{Mdot} and \eqref{dM/dr} together determine $\delta M $ up to a constant $\delta M_0$, corresponding to a trivial perturbation toward another Schwarzschild solution with mass $M+\e\delta M_0$. 

In this way, the entire invariant content of the $\ell=0$ solution is placed in $\delta M$, which satisfies physical energy-balance equations. The remaining piece of the field equations is the trace piece,
\beq
\frac{1}{2}g^{ab}\dR_{ab}=4\pi g^{ab}{\cal T}_{ab}:=8\pi{\cal T}_\trab.
\eeq
After using Eq.~\eqref{phi to dM}, we can simplify this to a wave equation for $\delta M$,
\beq\label{Box dM}
\Box_{{\cal M}^2}\delta M  = 8\pi \left(f \partial_r {\cal T}_\trAB+\frac{M}{r^2} {\cal T}_\trAB - rf  {\cal T}_{\trab}\right).
\eeq
This final equation is redundant due to the Bianchi identities, but it shows that the mass perturbation propagates causally according to a hyperbolic equation.

The same calculations apply at second order with the obvious replacements of source terms. As a final comment in this section, we note that at second order the quadratic source terms dramatically simplify for $\ell=0$ due to Eq.~\eqref{ell=0 coupling C}.  Equation~\eqref{Ailm}, for example, reduces to
\beq\label{restricted sums ell=0}
\sum_{s'=0}^{s'_{\rm max}}\ \sum_{\ell'=s'}^\infty\ \sum_{m'=-\ell'}^{+\ell'}\!\!\lam{\ell^{\prime}}{s'} \lam{\ell^{\prime \prime}}{s'}\frac{(-1)^{m'+s'}}{\sqrt{4\pi}} {\cal A}^{\ell' m' s' \ell',-m',-s'}_{\boldsymbol{\cdot}}.
\eeq


\subsection{$\ell=1$, odd parity}

The $\ell=1$, odd-parity field equations, with $\ell m$ labels omitted for visual simplicity, are
\begin{align}
\dR_{a-}[\widetilde{h}^{(1)}_{\boldsymbol\cdot}] &= 8\pi{\cal T}^{(1)}_{a-},
\end{align}
and thieir analogues at second order. These can be written entirely in terms of the invariant variable $\varphi_-$ defined in Eq.~\eqref{ell=1 odd invariant}. However, we opt to write it in terms of an effective angular momentum variable $\delta J$ defined by
\beq
\varphi_- = \frac{\delta J}{r^4}.
\eeq
Explicitly, the field equations reduce to 
\begin{subequations}\label{dJ eqns}%
\begin{align}
\frac{\partial}{\partial r}\delta J &=  - \frac{16\pi r^2}{f}T_{t-},\label{dJ/dr}\\
\frac{\partial}{\partial t}\delta J &= - 16\pi r^2 f T_{r-}.\label{Jdot}
\end{align}
\end{subequations}

In analogy with Eqs.~\eqref{Mdot} and \eqref{dM/dr}, Eq.~\eqref{Jdot} can be interpreted as the statement that the angular momentum within a sphere of radius $r$ changes at a rate equal to the instantaneous flux of angular momentum into the sphere, while Eq.~\eqref{dJ/dr} can be interpreted as the statement that the total angular momentum within the sphere is equal to the integrated angular momentum density within the sphere. These two equations determine $\delta J$ up to a constant. The constant represents a perturbation toward a Kerr solution with spin parameter $\frac{\delta J}{M}$.

In analogy with Eq.~\eqref{Box dM}, from Eq.~\eqref{dJ eqns} we can derive a wave equation for $\delta J$:
\beq\label{Box dJ}
\Box_{{\cal M}^2}\delta J  = -16\pi \epsilon^{ab}\delta _a\left(r^2T_{b-}\right).
\eeq

\section{Field equations in Barack--Lousto--Sago conventions}\label{BLS conventions}

For self-force computations in the Lorenz gauge, the most common set of conventions are those of Barack and Lousto~\cite{Barack-Lousto:05} as modified by Barack and Sago~\cite{Barack-Sago:07}.\footnote{The Barack--Lousto--Sago basis is related to the Barack--Lousto basis by $Y^{3\ell m(BLS)}_{\mu\nu}=fY^{3\ell m(BL)}_{\mu\nu}$, leading to coefficients related by $S^{(BL)}_{3\ell m}=f S^{(BLS)}_{3\ell m}$.} These conventions were used at first order in Refs.~\cite{Barack-Sago:07,Akcay-Warburton-Barack:13,Wardell-Warburton:15} (among others) and in all second-order calculations~\cite{Pound:2019lzj,Miller:2020bft,Warburton:2021kwk,Wardell:2021fyy}. In this appendix we describe the translation of our results into these conventions.

The Barack--Lousto--Sago conventions use a set of harmonics $Y^{i\ell m}_{\mu\nu}$, where $i$ runs from 1 to 10, with nonvanishing components given by%
\begingroup
\allowdisplaybreaks
\begin{subequations}
\begin{align}
Y^{1\ell m}_{ab} &= \frac{1}{\sqrt{2}}(t_a t_b + f^{-2}r_a r_b)Y^{\ell m},\\
Y^{2\ell m}_{ab} &= \frac{f^{-1}}{\sqrt{2}}(t_a r_b + r_a t_b)Y^{\ell m},\\
Y^{3\ell m}_{ab} &= -\frac{1}{\sqrt{2}}g_{ab}Y^{\ell m},\\
Y^{4\ell m}_{aA} &= \frac{r}{\sqrt{2}\lam{\ell}{1}}t_a Y_A^{\ell m},\\
Y^{5\ell m}_{aA} &= \frac{rf^{-1}}{\sqrt{2}\lam{\ell}{1}}r_a Y_A^{\ell m},\\
Y^{6\ell m}_{AB} &= \frac{r^2}{\sqrt{2}}\Omega_{AB}Y^{\ell m},\\
Y^{7\ell m}_{AB} &= \frac{\sqrt{2}r^2}{\lam{\ell}{2}}Y^{\ell m}_{AB},\\
Y^{8\ell m}_{aA} &= -\frac{r}{\sqrt{2}\lam{\ell}{1}}t_aX_A^{\ell m},\\
Y^{9\ell m}_{aA} &= -\frac{r f^{-1}}{\sqrt{2}\lam{\ell}{1}}r_aX_A^{\ell m},\\
Y^{10\ell m}_{AB} &= -\frac{\sqrt{2}r^2}{\lam{\ell}{2}}X^{\ell m}_{AB}.
\end{align}
\end{subequations}
\endgroup
These harmonics are orthogonal (but not orthonormal) with respect to the inner product $\<S_{\mu\nu},Q_{\mu\nu}\>:=\int \eta^{\alpha\mu}\eta^{\beta\nu}S^*_{\alpha\beta}Q_{\mu\nu}d\Omega$, where $\eta^{\mu\nu}:={\rm diag}(1,f^2,r^{-2}\Omega^{AB})$.\footnote{The definition of $\eta^{\mu\nu}$ corrects a typo in Ref.~\cite{Barack-Lousto:05}, as previously noted in Ref.~\cite{Wardell-Warburton:15}} If we expand a tensor $v_{\mu\nu}$ as
\beq\label{Yilm-expansion}
v_{\mu\nu}=\sum_{i \ell m}v_{i\ell m}Y^{i\ell m}_{\mu\nu},
\eeq
then the coefficients are given by\footnote{This corrects Eq.~(2.7) in Ref.~\cite{Wardell-Warburton:15}, which omitted the factor of $N_i$, as previously noted in Ref.~\cite{Miller:2020bft}.}
\begin{equation}
v_{i\ell m} = \kappa_i\int Y^{i\ell m*}_{\alpha\beta}\eta^{\alpha\mu}\eta^{\beta\nu}v_{\mu\nu}d\Omega,\label{coefficients}
\end{equation}
where $\kappa_3=f^{-2}$ and $\kappa_i=1$ for $i\neq3$.

An advantage of these harmonics is that they are well suited to assessing (or imposing) the regularity of a tensor at the future horizon. A tensor $v_{\mu\nu}=\sum_{i\ell m}v_{i\ell m}Y^{i\ell m}_{\mu\nu}$ has continuous components in ingoing Eddington--Finkelstein coordinates $(v,r)$ at $r=2M$ if and only if 
\begin{enumerate}
\item each coefficient $v_{i\ell m}$ is continuous there,
\item $v_{2\ell m} = v_{1\ell m}+\O(f^2)$,
\item $v_{i\ell m} = v_{i+1,\ell m}+\O(f)$ for $i=4,8$.
\end{enumerate}
If each $v_{i\ell m}$ is a smooth function of $v$ and $r$ at $r=2M$, then the above conditions are equivalent to smoothness of $v_{\mu\nu}$ at the future horizon.

Another advantage of this set of harmonics is that it makes trace reversals trivial. If we expand a tensor $v_{\mu\nu}$ as in Eq.~\eqref{Yilm-expansion} and its trace reverse as $\bar v_{\mu\nu}=\sum_{i\ell m}\bar v_{i\ell m}Y^{i\ell m}_{\mu\nu}$, then the coefficients in the two expansions are related by%
\begin{subequations}\label{BLS trace reversal}%
\begin{align}%
\bar v_{i\ell m} &= v_{i\ell m} \text{ if } i\neq3,6,\\
\bar v_{3\ell m} &= v_{6\ell m},\\
\bar v_{6\ell m} &= v_{3\ell m}.
\end{align}
\end{subequations}
Hence, a trace reversal is accomplished by the simple switch $i=3\leftrightarrow i=6$.

The coefficients $v_{i\ell m}$ are related to the tensor-harmonic coefficients in the body of the paper according to
\begin{subequations}\label{BLS coeffs from tensor coeffs}
\begin{align}
v_{1\ell m} &= \frac{1}{\sqrt{2}}(v^{\ell m}_{tt}+f^2v^{\ell m}_{rr}),\\
v_{2\ell m} &= \sqrt{2}f v^{\ell m}_{tr},\\
v_{3\ell m} &= -\frac{1}{\sqrt{2}}g^{ab}v^{\ell m}_{ab},\\
v_{4\ell m} &= \frac{\sqrt{2}\lam{\ell}{1}}{r}v^{\ell m}_{t+},\\
v_{5\ell m} &= \frac{\sqrt{2}\lam{\ell}{1} f}{r}v^{\ell m}_{r+},\\
v_{6\ell m} &= \frac{\sqrt{2}}{r^2}v^{\ell m}_{\trAB},\\
v_{7\ell m} &= \frac{\lam{\ell}{2}}{\sqrt{2}r^2}v^{\ell m}_+,\\
v_{8\ell m} &= -\frac{\sqrt{2}\lam{\ell}{1}}{r}v^{\ell m}_{t-},\\
v_{9\ell m} &= -\frac{\sqrt{2}\lam{\ell}{1} f}{r}v^{\ell m}_{r-},\\
v_{10\ell m} &= -\frac{\lam{\ell}{2}}{\sqrt{2}r^2}v^{\ell m}_-.
\end{align} 
\end{subequations}
These relations can be used to obtain the linear and quadratic quantities in the Lorenz-gauge field equations (${\cal E}_{i\ell m}$, ${\cal A}_{i\ell m}$, and ${\cal B}_{i\ell m}$) from their counterparts ${\cal E}^{\ell m}_{\boldsymbol\cdot}$, ${\cal A}^{\ell m}_{\boldsymbol\cdot}$, and ${\cal B}^{\ell m}_{\boldsymbol\cdot}$ given in the body of the paper. However, the results will be expressed in terms of the field variables $h^{\ell m}_{\boldsymbol\cdot}$, which must be translated to Barack--Lousto--Sago variables.

Instead of directly using coefficients in an expansion of $h^{(n)}_{\mu\nu}$ of the form~\eqref{Yilm-expansion}, the Barack--Lousto--Sago convention is to scale those coefficients by convenient factors. Specifically, the trace-reversed field is expanded as
\beq\label{BLS expansion}
\bar h^{(n)}_{\mu\nu} = \frac{1}{r}\sum_{i\ell m}a_{i\ell}\bar h^{(n)}_{i\ell m}Y^{i\ell m}_{\mu\nu},
\eeq
where 
\beq
a_{i\ell}:=\frac{1}{\sqrt{2}}\begin{cases}1 & \text{for } i=1,2,3,6,\\
\frac{1}{\lam{\ell}{1}} & \text{for } i=4,5,8,9,\\
\frac{1}{\lam{\ell}{2}} & \text{for } i=7,10.
\end{cases}
\eeq
The $n$th-order field variables are then the coefficients $\bar h^{(n)}_{i\ell m}$. We can express our field variables $h^{(n)\ell m}_{\boldsymbol\cdot}$ in terms of these by inverting the relations~\eqref{BLS coeffs from tensor coeffs}, accounting for the rescaling, and performing the trace reversal $i=3\leftrightarrow i=6$. The result is given in Table~\ref{table:conventions}.

If we begin from a field equation of the form~\eqref{EFEn dG} and specialize to the Lorenz gauge, we reduce the equation to Eq.~\eqref{EFEn Lorenz}, reproduced here for convencience: ${\cal E}_{\mu\nu}[\bar h^{(n)}] = -2 S^{(n)}_{\mu\nu}.$ Obtaining ${\cal E}_{i\ell m}$ from ${\cal E}^{\ell m}_{\boldsymbol\cdot}$ using Eq.~\eqref{BLS coeffs from tensor coeffs}, substituting Eq.~\eqref{BLS expansion} into ${\cal E}_{i\ell m}$, and then adding terms proportional to the gauge condition, as described in Ref.~\cite{Barack-Lousto:05}, leads to the Barack--Lousto--Sago formulation of the linearized Einstein equation in the Lorenz gauge, which is written as
\beq\label{BLS EFE order n}
\Box^{2d}_{\rm sc}\bar h^{(n)}_{i\ell m} + {\cal M}^{j}_{i\ell}\bar h^{(n)}_{j\ell m} = \frac{rf}{2a_{i\ell}}S^{(n)}_{i\ell m}.
\eeq
Here $\Box^{2d}_{\rm sc}=\partial_u\partial_v +V_\ell$ is a two-dimensional scalar wave operator with potential $V_\ell=\tfrac{1}{4}f\left[2M/r^3+\ell(\ell+1)/r^2\right]$, and the terms ${\cal M}^j_{i\ell}\bar h^{(1)}_{j\ell m}$ are given in Eqs.~(A1)--(A10) of Ref.~\cite{Barack-Sago:07}. $S^{(n)}_{i\ell m}$ is obtained from $S^{(n)\ell m}_{\boldsymbol{\cdot}}$ via Eq.~\eqref{BLS coeffs from tensor coeffs}.

At first order, Eq.~\eqref{BLS EFE order n} becomes
\beq
\Box^{2d}_{\rm sc}\bar h^{(1)}_{i\ell m} + {\cal M}^{j}_{i\ell}\bar h^{(1)}_{j\ell m} = \frac{4\pi rf}{a_{i\ell}}T^{(1)}_{i\ell m}.
\eeq
At second order, it becomes
\begin{align}
\Box^{2d}_{\rm sc}\bar h^{(2)}_{i\ell m} + {\cal M}^{j}_{i\ell}\bar h^{(2)}_{j\ell m} = \frac{4\pi rf}{a_{i\ell}}T^{(2)}_{i\ell m}- \frac{rf}{2a_{i\ell}}\delta^2G_{i\ell m}[h^{(1)}]
\end{align}
or, more explicitly, in terms of the sources appearing in Sec.~\ref{summary},
\begin{multline}
\Box^{2d}_{\rm sc}\bar h^{(2)}_{i\ell m} + {\cal M}^{j}_{i\ell}\bar h^{(2)}_{j\ell m} = \frac{4\pi rf}{a_{i\ell}}\overline{\cal T}^{(2)}_{i\ell m} \\
- \frac{rf}{4a_{i\ell}}(\bar{\cal A}_{i\ell m}+\bar{\cal B}_{i\ell m}).
\end{multline}
The source terms $\overline{\cal T}^{(2)}_{i\ell m}$, $\bar{\cal A}_{i\ell m}$, and $\bar{\cal B}_{i\ell m}$ can be obtained explicitly by (i) constructing ${\cal T}^{(2)}_{i\ell m}$, ${\cal A}_{i\ell m}$, and ${\cal B}_{i\ell m}$ from ${\cal T}^{(2)\ell m}_{\boldsymbol{\cdot}}$, ${\cal A}^{\ell m}_{\boldsymbol{\cdot}}$, and ${\cal B}^{\ell m}_{\boldsymbol{\cdot}}$ using Eq.~\eqref{BLS coeffs from tensor coeffs}, (ii) replacing the variables $h^{(1)\ell m}_{\boldsymbol\cdot}$ with the variables $\bar h^{(1)}_{i\ell m}$ using Table~\ref{table:conventions}, (iii) performing the trace reversal  $i=3\leftrightarrow i=6$ of ${\cal T}^{(2)}_{i\ell m}$, ${\cal A}_{i\ell m}$, and ${\cal B}_{i\ell m}$ to obtain $\overline{\cal T}^{(2)}_{i\ell m}$, $\bar{\cal A}_{i\ell m}$, and $\bar {\cal B}_{i\ell m}$. 

In self-force computations, $T^{(1)}_{\mu\nu}$ is the stress-energy tensor of a point mass. At second order, rather than working with a stress-energy tensor and solving directly for the physical retarded field, we instead use a puncture scheme. A singular piece of the metric perturbation, representing the particle's local self-field that diverges at the particle's position, is moved to the right-hand side of the field equation, and one solves for the regular residual field. The field equations in that case then become
\begin{multline}
\Box^{2d}_{sc}\bar h^{(2)\res}_{i\ell m} + {\cal M}^j_{i\ell}\bar h^{(2)\res}_{j\ell m} = - \left(\Box^{2d}_{sc}\bar h^{(2)\P}_{i\ell m} + {\cal M}^j_{i\ell}\bar h^{(2)\P}_{j\ell m}\right)\\
													 - \frac{rf}{4a_{i\ell}}(\bar{\cal A}_{i\ell m}+\bar{\cal B}_{i\ell m}),\label{puncture scheme}
\end{multline}
where $\bar h^{(2)\P}_{i\ell m}$ are the harmonic coefficients in the expansion of the 4D puncture field given in Ref.~\cite{Pound-Miller:14}, and $\bar h^{(2)\res}_{i\ell m}:=\bar h^{(2)}_{i\ell m}-\bar h^{(2)\P}_{i\ell m}$ are the residual field modes. No stress-energy terms appear in Eq.~\eqref{puncture scheme}, and the total source on the right-hand side is defined on the puncture's worldline by taking the limit from off the worldline; see the discussion around Eqs.~(13)--(17) in Ref.~\cite{Upton:2021oxf}. The field equations are also further modified using a two-timescale expansion, as described in Ref.~\cite{Miller:2020bft}.


\bibliography{bibfile}

\end{document}